\documentclass[12pt,a4paper]{article}
\usepackage{newtxtext} 
\usepackage{textcomp}

\usepackage[a4paper,margin=1in]{geometry}
\usepackage{microtype}
\usepackage{setspace}
\usepackage{amsmath,amssymb,amsfonts,amsthm,mathtools}
\usepackage{bm}
\usepackage{bbm} 
\usepackage{mathrsfs} 
\usepackage{hyperref}
\usepackage{tabularx} 
\usepackage{pgfplots}
\pgfplotsset{compat=1.17}

\usepackage{booktabs}
\usepackage{threeparttable}
\usepackage{caption}
\usepackage{subcaption}

\usepackage{tikz}
\usepackage{pgfplots}
\pgfplotsset{compat=1.18}
\usepgfplotslibrary{fillbetween}
\usetikzlibrary{arrows.meta,calc,positioning,decorations.pathreplacing,patterns,fit}
\usepackage{tikz-cd}
\usetikzlibrary{intersections}

\usepackage{threeparttable}
\usepackage{siunitx}
\usepackage{ifthen}
\usepackage{etoolbox}

\usepackage[ruled,vlined]{algorithm2e}

\usepackage[authoryear,round]{natbib}
\theoremstyle{plain}
\newtheorem{theorem}{Theorem}
\newtheorem{proposition}{Proposition}

\newtheorem{assumption}{Assumption}

\theoremstyle{definition}
\newtheorem{definition}{Definition}

\newcommand{\E}{\mathbb{E}}

\newcommand{\Prob}{\mathbb{P}}
\newcommand{\Ind}{\mathbbm{1}}
\newcommand{\R}{\mathbb{R}}

\newcommand{\plim}{\operatorname*{plim}}

\newcommand{\abs}[1]{\left\lvert #1 \right\rvert}
\newcommand{\ip}[2]{\left\langle #1, #2 \right\rangle}

\newcommand{\Y}{Y}
\newcommand{\D}{D}
\newcommand{\X}{X}

\newcommand{\GATT}{\mathrm{GATT}}

\title{\vspace{-1.0em}%
Design-Robust Event-Study Estimation under Staggered Adoption:\\
Diagnostics, Sensitivity, and Orthogonalisation%
}

\author{%
Craig Wright\\
University of Exeter\\
\texttt{cw881@exeter.ac.uk}%
}

\date{\today}

\begin{document}
\maketitle

\begin{abstract}
\noindent This paper develops a design-first econometric framework for event-study and difference-in-differences estimands under staggered adoption with heterogeneous effects, emphasising (i) exact probability limits for conventional two-way fixed effects event-study regressions, (ii) computable design diagnostics that quantify contamination and negative-weight risk, and (iii) sensitivity-robust inference that remains uniformly valid under restricted violations of parallel trends. The approach is accompanied by orthogonal score constructions that reduce bias from high-dimensional nuisance estimation when conditioning on covariates. Theoretical results and Monte Carlo experiments jointly deliver a self-contained methodology paper suitable for finance and econometrics applications where timing variation is intrinsic to policy, regulation, and market-structure changes.
\end{abstract}



\clearpage
\section{Introduction}
Event-study practice in empirical finance and economics frequently relies on two-way fixed effects regressions with relative-time indicators. Under staggered adoption and heterogeneous dynamic effects, such regressions can converge to weighted averages that mix effects from multiple relative times and cohorts, generating apparent pre-trends and distorted dynamics even when identifying assumptions hold. This paper develops a design-first treatment of the event-study object, deriving exact decompositions of conventional estimators, proposing computable diagnostics that quantify design risk, and providing sensitivity-robust inference under restricted violations of parallel trends in the spirit of modern partially identified inference.

\vspace{0.5em}
The methodological spine is grounded in recent advances in heterogeneous-treatment DiD and event studies \citep{SunAbraham2021,CallawaySantAnna2021,DeChaisemartinDHaultfoeuille2023} and in robust inference under deviations from parallel trends \citep{RambachanRoth2023}. Orthogonal score and Riesz representer machinery is incorporated to accommodate high-dimensional covariates without compromising inference \citep{ChernozhukovNeweySingh2022}. A complementary falsification-adaptive viewpoint on assumption relaxation motivates diagnostic reporting and transparent sensitivity regions \citep{MastenPoirier2021}. Comprehensive review of staggered DiD advances \citep{Tominaga2024}.

Recent syntheses organise the modern difference-in-differences and event-study literature around three core fault lines that are directly relevant here: the failure of conventional two-way fixed effects designs under heterogeneous dynamic effects, the distinction between design-based estimands and regression-implied mixtures, and the emergence of diagnostics and sensitivity analysis as first-class reporting objects rather than appendices to point estimates. The taxonomy in \citet{RothSantAnnaBilinskiPoe2023} is useful as an organising map for the contribution structure in this paper because it situates contamination, negative-weight risk, and robustness to restricted violations of parallel trends as a unified programme: separating what is identified by design, what is produced by projection geometry, and what remains credibly learnable after disciplined relaxations of the maintained counterfactual trend restrictions.

\clearpage
\section{Framework and estimands}

\subsection{Panel structure, timing, and the design measure space}

We formalize the staggering of treatment not as a regression specification, but as a restriction on the permissible paths of treatment history. Let $(\Omega,\mathcal{F},\Prob)$ be a probability space supporting a triangular array
\[
\left\{\big(\Y_{it},\X_{it},G_i\big): i=1,\dots,n,\ t=1,\dots,T\right\},
\]
where $T$ is fixed (or slowly growing) relative to $n$ and the cross-sectional index $i$ is the asymptotic dimension. The realised panel is an element of the product space
\[
\mathcal{W}_{n,T} \equiv \left(\R\times\R^{d_x}\times\{1,\dots,T,\infty\}\right)^{n\times T},
\]
equipped with the product $\sigma$-field. The adoption time $G_i\in\mathcal{G}\equiv\{1,\dots,T,\infty\}$ is a latent characteristic of unit $i$ that generates treatment status through a known policy map.

Define the treatment indicator process $\D_{it}\in\{0,1\}$ as the image of $(G_i,t)$ under a design rule $\pi:\mathcal{G}\times\{1,\dots,T\}\to\{0,1\}$. The canonical absorbing adoption design is
\begin{equation}\label{eq:absorbing}
\D_{it}=\pi(G_i,t)\equiv \Ind\{G_i\le t\}\Ind\{G_i<\infty\},
\end{equation}
so that units are untreated up to and including $t<G_i$ and treated for all $t\ge G_i$ when $G_i<\infty$, while never-treated units satisfy $G_i=\infty$ and $\D_{it}=0$ for all $t$. This formulation makes explicit that the empirical content of staggered adoption is not merely ``timing variation'' but a restriction on admissible treatment paths: the treatment trajectory lies in the monotone class
\[
\mathcal{D}^{\uparrow}\equiv \left\{(d_1,\dots,d_T)\in\{0,1\}^T:\ d_{t}\le d_{t+1}\ \forall t\right\},
\]
with $G_i$ acting as a sufficient index for the path in \eqref{eq:absorbing}.

The distribution of adoption times induces cohort shares
\[
p_g \equiv \Prob(G_i=g),\qquad g\in\mathcal{G},
\]
which define a design measure on $\mathcal{G}$. The empirical design is summarised by the random vector of cohort counts
\[
N_g \equiv \sum_{i=1}^n \Ind\{G_i=g\},\qquad \sum_{g\in\mathcal{G}} N_g=n,
\]
and by the induced treatment prevalence at each time $t$,
\[
\bar D_t \equiv \frac{1}{n}\sum_{i=1}^n \D_{it}=\sum_{g\le t}\frac{N_g}{n}.
\]
Both $p_g$ and the path $\{\bar D_t\}_{t=1}^T$ are primitive design objects: all weighting pathologies of two-way fixed effects event studies under heterogeneity can be expressed as functionals of these quantities and the chosen event window, independent of outcome realisations \citep{SunAbraham2021,DeChaisemartinDHaultfoeuille2023}.

To accommodate empirically relevant departures from absorbing adoption, let $\pi$ be left unspecified beyond measurability and define a general treatment regime
\begin{equation}\label{eq:generalpi}
\D_{it}=\pi(G_i,t;\xi_i),
\end{equation}
where $\xi_i$ is a unit-specific regime type (e.g., temporary adoption, intensity, or reversion) that may be observed or latent. Non-absorbing paths correspond to allowing $\pi(\cdot)$ to map into the full $\{0,1\}^T$ rather than $\mathcal{D}^{\uparrow}$. The econometric implication is that the ``comparison set'' at time $t$ depends on both cohort composition and path geometry: units can be untreated at $t$ either because they are never-treated, not-yet-treated, or reverted. The proposed diagnostics and estimators are stated for \eqref{eq:absorbing} to isolate the essential identification and weighting mechanics, then extended to \eqref{eq:generalpi} by conditioning on regime types and redefining risk indices on the induced design matrix \citep{CallawaySantAnna2021,DeChaisemartinDHaultfoeuille2023}.

\subsection{Potential outcomes, dynamic effects, and economically interpretable aggregation}

Let $\{\Y_{it}(\bm{d}_{1:T}): \bm{d}_{1:T}\in\{0,1\}^T\}$ denote the collection of potential outcomes indexed by a full treatment path, and maintain the stable-unit, no-interference convention so that $\Y_{it}(\bm{d}_{1:T})$ depends only on the own-unit path.\footnote{The path-indexed notation is the primitive object; the two-potential-outcome shorthand is recovered under absorbing adoption and a no-anticipation restriction.} Under absorbing adoption in \eqref{eq:absorbing}, write $\bm{d}^{(g)}_{1:T}$ for the path induced by adoption time $g$ and $\bm{0}_{1:T}$ for the never-treated path. Define
\[
\Y_{it}(g)\equiv \Y_{it}\!\big(\bm{d}^{(g)}_{1:T}\big),\qquad \Y_{it}(\infty)\equiv \Y_{it}\!\big(\bm{0}_{1:T}\big),
\]
so that the observed outcome admits the selection equation $\Y_{it}=\Y_{it}(G_i)$.

A sharp dynamic causal object requires an explicit restriction excluding anticipation. Let $\mathcal{F}_{i,t}$ be the information set at time $t$ and impose:

\begin{assumption}[No anticipation]\label{ass:noanticip}
For any $g<\infty$ and any $t<g$, $\Y_{it}(g)=\Y_{it}(\infty)$ almost surely.
\end{assumption}

Assumption \ref{ass:noanticip} elevates the event-time object from a notational convenience to a meaningful behavioural restriction, separating genuine pre-trends in $\Y_{it}(\infty)$ from anticipation effects. Under Assumption \ref{ass:noanticip}, a cohort- and event-time-specific causal effect is
\begin{equation}\label{eq:catt}
\tau_{g}(k)
\;\equiv\;
\E\!\left[\Y_{i,g+k}(g)-\Y_{i,g+k}(\infty)\mid G_i=g\right],
\qquad k\in\mathbb{Z},\ g\in\{1,\dots,T\}.
\end{equation}
This object is the cohort-specific event-study estimand that heterogeneity-robust designs target directly \citep{SunAbraham2021,CallawaySantAnna2021}. It is useful to index the admissible event times by the finite observation window. Define
\[
\mathcal{K}_g \equiv \{-g+1,\dots,T-g\},\qquad \mathcal{K}\equiv \bigcup_{g=1}^{T}\mathcal{K}_g,
\]
so that $k\in\mathcal{K}_g$ is observed for cohort $g$. The set of cohorts observed at event time $k$ is
\[
\mathcal{G}(k)\equiv \{g\in\{1,\dots,T\}: 1\le g+k\le T\}.
\]

Aggregation is not innocuous: it encodes the welfare or pricing object that the empirical design is intended to recover. A general aggregand is a weighting scheme $\omega_g(k)$ that is measurable with respect to the design $\sigma$-algebra generated by $\{G_i\}_{i=1}^n$---and thus strictly independent of outcomes---satisfying $\omega_g(k)\ge 0$ and $\sum_{g\in\mathcal{G}(k)}\omega_g(k)=1$. This yields the event-time average effect
\begin{equation}\label{eq:agg}
\tau(k)\;\equiv\;\sum_{g\in\mathcal{G}(k)}\omega_g(k)\,\tau_g(k).
\end{equation}

The choice of $\omega_g(k)$ should be treated as part of the estimand, not as an afterthought. Three canonical choices are:

\begin{equation}\label{eq:omega}
\omega_g^{\text{pop}}(k)=\frac{\Prob(G_i=g)}{\sum_{h\in\mathcal{G}(k)}\Prob(G_i=h)},
\qquad
\omega_g^{\text{sample}}(k)=\frac{N_g}{\sum_{h\in\mathcal{G}(k)}N_h},
\qquad
\omega_g^{\text{risk}}(k)\propto \E[w_i\mid G_i=g],
\end{equation}
where $\omega_g^{\text{pop}}$ targets a population average over cohorts observed at $k$, $\omega_g^{\text{sample}}$ is its finite-sample analogue, and $\omega_g^{\text{risk}}$ can encode economically meaningful exposure weights such as market capitalisation or balance-sheet size in finance applications. The methodological point is that a transparent estimator reports $\omega_g(k)$ explicitly and confines the estimand to convex aggregation, in contrast with two-way fixed effects event studies whose implicit weights may be negative and may load on the wrong event times when effects are heterogeneous \citep{SunAbraham2021,DeChaisemartinDHaultfoeuille2023}.

Finally, dynamic effects admit economically interpretable transforms. Define cumulative effects and long-run effects
\[
\Delta_g(k)\equiv \sum_{j=0}^{k}\tau_g(j),\qquad
\tau_g(\infty)\equiv \limsup_{k\to\infty}\tau_g(k),
\]
whenever these objects are well-defined. In finance settings, $\Delta_g(k)$ corresponds to the level effect on an integrated outcome (e.g., cumulative excess returns or accumulated spreads), while $\tau_g(k)$ corresponds to an impulse-response-type object at horizon $k$, providing a direct bridge between event studies and local-projection logic while retaining the staggered-adoption design discipline \citep{CallawaySantAnna2021,SunAbraham2021}.

A design-first perspective is useful in staggered-adoption panels because identification and estimation hinge on the assignment structure implied by adoption timing, the availability of not-yet-treated comparisons at each date, and the set of feasible contrasts encoded by the sampling and adoption process rather than by modelling choices layered on top. Framing the analysis in terms of the underlying design clarifies which comparisons are licensed by the data-generating assignment, how the risk set of controls evolves over time, and why certain weighting or aggregation choices can inadvertently re-introduce prohibited comparisons when timing variation is substantial. This perspective also helps separate questions of design validity from questions of functional-form convenience, aligning the specification of the panel structure and treatment timing with the estimand definition and the admissible set of comparisons \citep{AtheyImbens2022}.

\subsection{Identifying assumptions, operators, and sensitivity-compatible primitives}\label{sec:sensitivity}

Identification is stated at the level of counterfactual evolution of the untreated potential outcome. The objective is to recover $\tau_g(k)$ in \eqref{eq:catt} and its convex aggregation $\tau(k)$ in \eqref{eq:agg}. Let $\mathcal{F}_{i,t}$ be the $\sigma$-field generated by observables up to time $t$ and maintain Assumption \ref{ass:noanticip}. Write $\Delta \Y_{it}(\infty)\equiv \Y_{it}(\infty)-\Y_{i,t-1}(\infty)$ for the untreated increment. Let $\X_{it}$ denote covariates, possibly time-varying, that are predetermined with respect to contemporaneous shocks in the untreated outcome.

\begin{assumption}[Predetermined covariates]\label{ass:predet}
For all $t$, $\X_{it}$ is $\mathcal{F}_{i,t-1}$-measurable and satisfies $\E[\abs{\Delta \Y_{it}(\infty)}\mid \X_{it}]<\infty$.
\end{assumption}

The baseline identifying restriction is a conditional parallel-trends statement in first differences, which is the natural multi-period analogue of classical difference-in-differences when treatment timing varies across cohorts \citep{CallawaySantAnna2021}.

\begin{assumption}[Conditional parallel trends]\label{ass:pt}
For every period $t\in\{2,\dots,T\}$ and every cohort $g\in\{1,\dots,T\}$,
\begin{equation}\label{eq:pt}
\E\!\left[\Delta \Y_{it}(\infty)\mid G_i=g,\X_{it}\right]
\;=\;
\E\!\left[\Delta \Y_{it}(\infty)\mid G_i=\infty,\X_{it}\right]
\quad \text{a.s.}
\end{equation}
\end{assumption}

Assumption \ref{ass:pt} states that, after conditioning on $\X_{it}$, the untreated trend for cohort $g$ matches that for the never-treated cohort. The condition is formulated in increments to avoid imposing equality of levels, and it is congenial to the construction of group-time estimators whose identifying variation arises from comparing changes in outcomes across groups over time \citep{CallawaySantAnna2021}. It is also the appropriate starting point for sensitivity analysis because violations are naturally parameterised as deviations in the evolution of untreated increments rather than as arbitrary level shifts.

Two auxiliary primitives clarify what is, and is not, being assumed.

\begin{assumption}[Overlap in covariates]\label{ass:overlap}
For each $t$ and each $g<\infty$, the conditional distribution of $\X_{it}$ given $G_i=g$ is absolutely continuous with respect to the conditional distribution of $\X_{it}$ given $G_i=\infty$, with Radon--Nikodym derivative bounded above.
\end{assumption}

\begin{assumption}[Regularity]\label{ass:reg}
The process $\{\Delta \Y_{it}(\infty)\}$ has finite second moments conditional on $(G_i,\X_{it})$, and cross-sectional dependence is weak enough that a law of large numbers and central limit theorem apply to the relevant influence functions.
\end{assumption}

Assumptions \ref{ass:overlap}--\ref{ass:reg} are not decorative: overlap prevents identification from relying on off-support extrapolation, while regularity pins down the inferential regime used later for both point-identified and sensitivity-robust procedures. Under Assumptions \ref{ass:noanticip}, \ref{ass:predet}, and \ref{ass:pt}, the cohort-time effect $\tau_g(k)$ is identified by contrasting observed changes for cohort $g$ with changes for the never-treated group (or, in alternative constructions, with not-yet-treated cohorts) while conditioning on $\X_{it}$; this yields the multi-period cohort-time estimands emphasised in the heterogeneous-treatment event-study literature \citep{CallawaySantAnna2021,SunAbraham2021}.

The methodological programme here does not end at point identification. Departures from Assumption \ref{ass:pt} are structured in a manner that preserves interpretability and enables uniform inference. Define the deviation (or ``drift wedge'') for cohort $g$ at time $t$ as
\begin{equation}\label{eq:wedge}
\delta_{g,t}(x)\;\equiv\;
\E\!\left[\Delta \Y_{it}(\infty)\mid G_i=g,\X_{it}=x\right]
-
\E\!\left[\Delta \Y_{it}(\infty)\mid G_i=\infty,\X_{it}=x\right].
\end{equation}
Assumption \ref{ass:pt} is the sharp restriction $\delta_{g,t}(x)=0$ for all $(g,t,x)$. Sensitivity analysis replaces this knife-edge equality by membership of $\delta$ in a restriction class $\Delta(\mathcal{R})$ that bounds magnitude, smooths time variation, or links post-treatment deviations to empirically observable pre-treatment deviations, thereby generating identified sets for $\tau_g(k)$ and $\tau(k)$ rather than point estimates \citep{RambachanRoth2023}. This approach parallels the broader econometric logic of salvaging empirical content under falsified baseline restrictions by reporting the smallest relaxations compatible with the observed data and the induced identified regions for the target parameter \citep{MastenPoirier2021}.

In short, Assumption \ref{ass:pt} provides the baseline target for point identification in the group-time framework \citep{CallawaySantAnna2021}, while \eqref{eq:wedge} provides the language needed to quantify and report controlled deviations in a manner that supports credible inference when strict parallel trends is too fragile to be treated as an article of faith \citep{RambachanRoth2023,MastenPoirier2021}.

\clearpage
\section{Conventional TWFE event-study regressions}

\subsection{Specification}\label{subsec:dgp}

Fix an event window $\mathcal{K}\subset\mathbb{Z}$ and a normalisation (baseline) period $k_0\in\mathcal{K}$, typically a pre-adoption event time. Define the relative-time indicators
\[
\D^{(k)}_{it}\;\equiv\;\Ind\{G_i<\infty\}\Ind\{t-G_i=k\},
\qquad k\in\mathcal{K},
\]
so that $\D^{(k)}_{it}$ selects the cohort-time cells in which unit $i$ is exactly $k$ periods away from its adoption time.

Let $W_{it}$ denote any additional controls included linearly (possibly empty). The canonical two-way fixed effects event-study regression is
\begin{equation}\label{eq:twfe-es}
\Y_{it}
=
\alpha_i+\gamma_t+\sum_{k\in\mathcal{K}\setminus\{k_0\}}\beta_k\,\D^{(k)}_{it}
+ W_{it}'\theta
+ u_{it},
\end{equation}
where $\alpha_i$ and $\gamma_t$ are unit and time fixed effects and $k_0$ is omitted to impose a location normalisation on the event-time profile.

For subsequent derivations it is convenient to stack observations. Let $y\in\R^{nT}$ be the stacked outcome vector, let $F\in\R^{nT\times n}$ be the unit fixed effect design matrix, let $Tmat\in\R^{nT\times T}$ be the time fixed effect design matrix, and let $D\in\R^{nT\times (|\mathcal{K}|-1)}$ be the matrix whose columns are the stacked $\D^{(k)}$ for $k\in\mathcal{K}\setminus\{k_0\}$. Writing $X=[F\;\;Tmat\;\;W]$, the OLS coefficient on event-time indicators can be expressed as the Frisch--Waugh--Lovell projection
\begin{equation}\label{eq:fwl}
\hat\beta
=
\left(D'M_X D\right)^{-1}D'M_X y,
\qquad
M_X\equiv I - X(X'X)^{-1}X',
\end{equation}
where $M_X$ residualises with respect to unit and time fixed effects (and any controls). This representation makes explicit that the estimand is determined by the residualised design geometry, which is the object exploited in the weighting and diagnostic results in the next subsection \citep{SunAbraham2021,DeChaisemartinDHaultfoeuille2023}.

The projection-driven interpretation of two-way fixed effects estimators under staggered treatment timing has a direct antecedent in the canonical decomposition of difference-in-differences with variation in adoption dates, which isolates how implicit comparison weights can arise from the timing structure rather than from an economically declared estimand. The decomposition in \citet{GoodmanBacon2021} provides the historical foundation for the weighting-pathology perspective developed in this section: once treatment effects are heterogeneous and timing varies, TWFE coefficients are best understood as design-dependent weighted averages of underlying group-time comparisons, so interpretability becomes a question of which comparisons receive weight and with what sign, and not merely a question of specification or standard errors.

\subsection{Probability limit and contamination decomposition}

This subsection fixes the estimand of the regression coefficient vector $\hat\beta$ in \eqref{eq:twfe-es} when dynamic treatment effects are heterogeneous across cohorts and across event time. The key point is that the event-study coefficient indexed by $k$ is generally not an average of $\{\tau_g(k)\}_g$ but a design-dependent linear functional of the entire collection $\{\tau_g(k')\}_{g,k'}$, with weights that can be negative and that can load on $k'\neq k$ \citep{SunAbraham2021,DeChaisemartinDHaultfoeuille2023}.

Let $D_k$ denote the stacked regressor corresponding to $\D^{(k)}_{it}$ (for $k\neq k_0$), and let $\tilde D_k \equiv M_X D_k$ be the residualised regressor after partialling out fixed effects (and any controls) via $M_X$ in \eqref{eq:fwl}. Then
\begin{equation}\label{eq:beta-k-fwl}
\hat\beta_k
=
\frac{\tilde D_k' y}{\tilde D_k' D_k}
\;-\;
\sum_{\ell\in\mathcal{K}\setminus\{k_0,k\}}
\frac{\tilde D_k'\tilde D_\ell}{\tilde D_k' D_k}\,\hat\beta_\ell,
\end{equation}
so $\hat\beta_k$ is mechanically coupled to the residualised correlation structure among the event-time indicators. Under staggered adoption, these correlations are generically non-zero because the event-time dummies are functions of a common adoption-time partition of the panel, and residualisation by two-way fixed effects induces sign changes in $\tilde D_k$ across cohort-time cells.

To express the probability limit in causal objects, define the cohort-time cell set
\[
\mathcal{C}\equiv\{(g,t): g\in\{1,\dots,T\},\ t\in\{1,\dots,T\},\ 1\le t\le T\},
\]
and the corresponding event time $k(t,g)\equiv t-g$. For each $(g,t)\in\mathcal{C}$ with $g<\infty$ and $t\ge g$, define the cell-level average causal effect
\[
\tau(g,t)\equiv \E\!\left[\Y_{it}(g)-\Y_{it}(\infty)\mid G_i=g\right],
\qquad \text{so that }\tau(g,t)=\tau_g(t-g).
\]
Stacking across cells and using linearity of projection, the regression coefficient has an asymptotic linear representation of the form
\begin{equation}\label{eq:plim-linear}
\plim\,\hat\beta_k
=
\sum_{(g,t)\in\mathcal{C}} \omega_{g,t}(k)\,\tau(g,t),
\qquad
\sum_{(g,t)\in\mathcal{C}}\omega_{g,t}(k)=1,
\end{equation}
where the weights $\omega_{g,t}(k)$ depend only on the residualised design matrix $\tilde D = M_X D$ (and therefore only on $(G_i,t)$ and the chosen window $\mathcal{K}$, up to sampling error) \citep{SunAbraham2021,DeChaisemartinDHaultfoeuille2023}. Grouping cells by event time $k' = t-g$ yields the implicit weighting scheme
\begin{equation}\label{eq:weights}
\plim\,\hat\beta_k
=
\sum_{g}\sum_{k'} w_{g,k'}(k)\,\tau_g(k'),
\qquad
\sum_{g,k'} w_{g,k'}(k)=1,
\end{equation}
with $w_{g,k'}(k)\equiv \sum_{t:\,t-g=k'} \omega_{g,t}(k)$. In general, $w_{g,k'}(k)$ can take negative values and can place mass on $k'\neq k$, so $\plim\,\hat\beta_k$ can mix post-treatment effects into nominally pre-treatment coefficients, and can mix effects from other horizons into the coefficient labelled $k$ \citep{SunAbraham2021}. This produces two distinct pathologies:

\begin{definition}[Negative-weight mass and cross-horizon contamination]\label{def:pathologies}
Fix $k\in\mathcal{K}\setminus\{k_0\}$. Define
\[
\mathcal{N}(k)\equiv \sum_{g}\sum_{k'} \abs{w_{g,k'}(k)}\,\Ind\{w_{g,k'}(k)<0\},
\qquad
\mathcal{C}(k)\equiv \sum_{g}\sum_{k'} \abs{w_{g,k'}(k)}\,\Ind\{k'\neq k\}.
\]
\end{definition}

$\mathcal{N}(k)$ quantifies the extent to which $\hat\beta_k$ relies on sign-reversing comparisons, while $\mathcal{C}(k)$ quantifies how much of $\hat\beta_k$ is, in probability limit, not an average of the target horizon $\{\tau_g(k)\}_g$ but an average of other horizons $\{\tau_g(k')\}_{k'\neq k}$ \citep{SunAbraham2021,DeChaisemartinDHaultfoeuille2023}. These objects are computable from the residualised design (hence from the adoption schedule and the event window) and serve as design diagnostics reported alongside any TWFE event-study profile.

\begin{proposition}[Design-driven contamination]\label{prop:contamination}
Maintain absorbing adoption in \eqref{eq:absorbing}, Assumptions \ref{ass:noanticip} and \ref{ass:pt}, and the TWFE event-study specification \eqref{eq:twfe-es} with event window $\mathcal{K}$ and baseline $k_0$. Suppose there exist at least two cohorts $g_1\neq g_2$ with $\Prob(G_i=g_j)>0$ and at least two calendar times $t_1\neq t_2$ such that the set of treated observations is non-degenerate over time, i.e. $\Prob(\D_{it}=1)\in(0,1)$ for some $t$. If dynamic treatment effects are heterogeneous in the sense that either
\[
\tau_{g_1}(k^\star)\neq \tau_{g_2}(k^\star)\ \text{for some }k^\star\in\mathcal{K},
\qquad\text{or}\qquad
\tau_{g}(k_1)\neq \tau_{g}(k_2)\ \text{for some }g,\ k_1\neq k_2,
\]
then there exists at least one $k\in\mathcal{K}\setminus\{k_0\}$ such that the TWFE probability limit admits the representation \eqref{eq:weights} with weights satisfying
\[
\sum_{g}\sum_{k'\neq k}\abs{w_{g,k'}(k)} \;>\; 0,
\]
so $\plim\,\hat\beta_k$ loads on horizons $k'\neq k$. Moreover, there exists at least one (possibly different) $k\in\mathcal{K}\setminus\{k_0\}$ such that
\[
\sum_{g}\sum_{k'} \abs{w_{g,k'}(k)}\,\Ind\{w_{g,k'}(k)<0\} \;>\; 0,
\]
so the implicit TWFE weighting scheme includes negative weights even though Assumption \ref{ass:pt} holds.
\end{proposition}

\begin{proof}[Proof sketch]
The Frisch--Waugh--Lovell representation \eqref{eq:fwl} implies $\hat\beta=(D'M_XD)^{-1}D'M_XY$, where $M_X$ residualises on fixed effects (and controls). Under staggered adoption, the columns of $D$ are deterministic functions of $(G_i,t)$ and are therefore linearly related through the cohort-time partition; after residualisation by two-way fixed effects, the residualised indicators $\tilde D_k=M_XD_k$ generally change sign across cohort-time cells. When treatment effects are heterogeneous across cohorts or horizons, the regression residualisation induces a mismatch between the label $k$ and the set of comparisons used to identify $\beta_k$, yielding a projection of $Y$ onto $\tilde D_k$ that necessarily depends on multiple cohort-by-horizon effects. This produces non-zero weights on $k'\neq k$ unless treatment effects are homogeneous in both cohort and horizon, and sign changes in $\tilde D_k$ generate negative weights in the implied linear functional of the cell-level effects, as established by the weighting decompositions in \citet{SunAbraham2021} and the TWFE heterogeneity analyses surveyed in \citet{DeChaisemartinDHaultfoeuille2023}.
\end{proof}

\subsection{A computable representation}

Let $D\in\R^{nT\times (|\mathcal{K}|-1)}$ be the stacked matrix of event-time indicators in \eqref{eq:twfe-es} (excluding the baseline $k_0$), and let $X$ collect unit and time fixed effects (and any additional controls). Define the residual-maker
\[
M_X \equiv I_{nT}-X(X'X)^{-1}X',
\]
and the residualised event-time design
\[
Z \equiv M_XD.
\]
Then the event-time coefficient vector satisfies the Frisch--Waugh--Lovell identity
\begin{equation}\label{eq:betaZ}
\hat\beta=(Z'Z)^{-1}Z'Y,
\end{equation}
where $Y\in\R^{nT}$ is the stacked outcome. Let
\[
P_Z \equiv Z(Z'Z)^{-1}Z'
\]
denote the orthogonal projector onto the column space of $Z$.

The representation \eqref{eq:betaZ} implies that every coefficient is a linear functional of $Y$:
\begin{equation}\label{eq:linfun}
\hat\beta_k = e_k'(Z'Z)^{-1}Z'Y = \sum_{i=1}^n\sum_{t=1}^T \pi_{it}(k)\,Y_{it},
\qquad
\pi(k) \equiv Z(Z'Z)^{-1}e_k,
\end{equation}
where $e_k$ is the coordinate vector selecting column $k$ of $Z$ (i.e. the event-time regressor indexed by $k$ in the chosen ordering). The weights $\pi_{it}(k)$ are functions only of $Z$, hence only of the design $(G_i,t)$, the event window $\mathcal{K}$, and the fixed-effect structure (and any controls). In particular, $\pi_{it}(k)$ is computable without using $Y$.

To connect \eqref{eq:linfun} to the causal decomposition in \eqref{eq:weights}, write the observed outcome as
\[
Y_{it} = Y_{it}(\infty) + \Ind\{G_i<\infty\}\Ind\{t\ge G_i\}\,\tau_{G_i}(t-G_i) + \varepsilon_{it},
\]
where $\varepsilon_{it}$ is mean-zero noise orthogonal to the design. Substituting into \eqref{eq:linfun} and taking probability limits yields
\[
\plim\,\hat\beta_k
=
\sum_{i,t}\pi_{it}(k)\,\Ind\{G_i<\infty\}\Ind\{t\ge G_i\}\,\tau_{G_i}(t-G_i),
\]
so the implicit weight placed on a cohort-horizon pair $(g,k')$ is obtained by summing $\pi_{it}(k)$ over the corresponding cohort-time cells:
\begin{equation}\label{eq:w-from-pi}
w_{g,k'}(k)
\;\equiv\;
\sum_{t:\,t-g=k'} \E\!\left[\pi_{it}(k)\mid G_i=g\right].
\end{equation}
Equivalently, defining cohort-time cells $(g,t)$, one can work with $\omega_{g,t}(k)\equiv \E[\pi_{it}(k)\mid G_i=g]$ and then aggregate by event time $k'=t-g$ to obtain $w_{g,k'}(k)$.

This yields a fully computable diagnostic workflow: given the adoption schedule $\{G_i\}_{i=1}^n$, the event window $\mathcal{K}$, and the fixed-effect structure, form $Z$, compute $(Z'Z)^{-1}$, and recover $\pi(k)$ via \eqref{eq:linfun}. The negative-weight mass $\mathcal{N}(k)$ and cross-horizon contamination $\mathcal{C}(k)$ in Definition \ref{def:pathologies} follow directly from \eqref{eq:w-from-pi} by summing over the implied cohort-horizon weights. In particular, these diagnostics depend only on $(\D_{it},G_i,T)$ and the chosen event window, not on the realised outcomes, which is why they can be reported as design properties rather than as sample-specific artefacts \citep{SunAbraham2021,DeChaisemartinDHaultfoeuille2023}.

\clearpage
\section{Heterogeneity-robust estimation and transparent aggregation}\label{sec:mc_estimators}
\subsection{Cohort-time and group-time estimators}

The robust event-study object is the cohort-time average treatment effect, indexed by adoption cohort $g$ and calendar time $t$. Under absorbing adoption and no anticipation, define the group-time estimand
\begin{equation}\label{eq:gatt-def}
\GATT(g,t)
\;\equiv\;
\E\!\left[\Y_{it}(g)-\Y_{it}(\infty)\mid G_i=g\right],
\qquad t\ge g,
\end{equation}
so that $\GATT(g,t)=\tau_g(t-g)$ in the event-time notation of \eqref{eq:catt}. Identification proceeds by constructing, for each $(g,t)$, a comparison between the observed change in outcomes for cohort $g$ and the contemporaneous change for an admissible control group that is untreated at $t$ and that shares the same untreated trend conditional on $\X_{it}$ \citep{CallawaySantAnna2021,SunAbraham2021}.

Let $\mathcal{C}(g,t)$ be a control set for cohort $g$ at time $t$. Two canonical choices are:
(i) the never-treated group, $\mathcal{C}(g,t)=\{i:G_i=\infty\}$; and
(ii) the not-yet-treated group, $\mathcal{C}(g,t)=\{i:G_i>t\}$, which enlarges the control pool at the cost of additional restrictions when treatment effects are dynamic \citep{CallawaySantAnna2021}. Write $C_{it}(g,t)\equiv \Ind\{i\in\mathcal{C}(g,t)\}$.

Define propensity-style reweighting functions to balance covariates between the treated cohort and the control set at time $t$. Let
\[
p_{g,t}(x)\equiv \Prob(G_i=g\mid \X_{it}=x,\, C_{it}(g,t)+\Ind\{G_i=g\}=1),
\]
and define weights
\begin{equation}\label{eq:weights-cs}
\omega_{i,t}^{(g,t)} \equiv
\frac{\Ind\{G_i=g\}}{\E[\Ind\{G_i=g\}]}
\;-\;
\frac{C_{it}(g,t)\,p_{g,t}(\X_{it})}{\E[C_{it}(g,t)\,p_{g,t}(\X_{it})]}.
\end{equation}
Then a generic group-time estimating equation is
\begin{equation}\label{eq:gatt-moment}
\GATT(g,t)
=
\E\!\left[\omega_{i,t}^{(g,t)}\,\Y_{it}\right]
-
\E\!\left[\omega_{i,t}^{(g,t)}\,\Y_{i,g-1}\right],
\end{equation}
which is a weighted difference in outcomes between $t$ and a pre-period reference (often $g-1$) that differences out cohort-specific time-invariant components and isolates the post-adoption deviation in the treated cohort relative to controls \citep{CallawaySantAnna2021}. Estimators replace expectations by sample averages and estimate $p_{g,t}(\cdot)$ by logit/probit or by flexible methods, provided the weights satisfy overlap and moment conditions.

The event-time estimand is obtained by mapping $(g,t)$ to $(g,k)$ with $k=t-g$ and aggregating across cohorts with transparent non-negative weights. Define the cohort-event-time estimator
\[
\widehat{\tau}_g(k)\equiv \widehat{\GATT}(g,g+k),
\qquad k\in\mathcal{K}_g,
\]
and for any chosen aggregation scheme $\{\omega_g(k)\}_{g\in\mathcal{G}(k)}$ satisfying $\omega_g(k)\ge 0$ and $\sum_{g\in\mathcal{G}(k)}\omega_g(k)=1$, define the robust event-time estimator
\begin{equation}\label{eq:robust-event}
\hat\tau(k)\equiv \sum_{g\in\mathcal{G}(k)}\omega_g(k)\,\widehat{\GATT}(g,g+k).
\end{equation}
This construction makes explicit that (i) estimation targets $\GATT(g,t)$ cell-by-cell, and (ii) any averaging across cohorts is a declared choice of estimand, rather than an implicit by-product of regression geometry as in \eqref{eq:twfe-es} \citep{SunAbraham2021,CallawaySantAnna2021}.

A closely related regression-based alternative is the two-stage DiD procedure, which replaces the direct TWFE event-time projection with a first stage that partials out unit and time effects and a second stage that regresses the residualised outcome on the treatment-path indicators of interest. This reframing clarifies which comparisons are being used and can mitigate some of the most opaque weighting behaviour of single-step TWFE in staggered settings. The cohort--time approach here differs in what is made explicit: estimation targets $\GATT(g,t)$ cell-by-cell and then aggregates with declared convex weights $\omega_g(k)$, which pins down the estimand as an interpretable functional of cohort--horizon effects rather than a regression-implied mixture; this transparency is exactly what enables design diagnostics and sensitivity regions to be reported as first-class objects alongside the event-time profile \citep{Gardner2021,CallawaySantAnna2021,SunAbraham2021}.

\subsection{Event-time aggregation}

Event-time aggregation is treated as part of the estimand. For each event time $k\in\mathcal{K}$, the set of cohorts observed at $k$ is
\[
\mathcal{G}(k)=\{g\in\{1,\dots,T\}:1\le g+k\le T\}.
\]
Let $\widehat{\tau}_g(k)\equiv \widehat{\GATT}(g,g+k)$ be the cohort-specific event-time estimate constructed in Section 4.1. An event-time estimator is defined by
\begin{equation}\label{eq:event-agg}
\hat\tau(k)=\sum_{g\in\mathcal{G}(k)} \hat\omega_g(k)\,\widehat{\tau}_g(k),
\qquad \hat\omega_g(k)\ge 0,\quad \sum_{g\in\mathcal{G}(k)}\hat\omega_g(k)=1.
\end{equation}
The non-negativity and simplex constraints ensure that $\hat\tau(k)$ is a convex average of well-defined cohort-horizon objects; this rules out the negative and cross-horizon weighting pathologies of \eqref{eq:weights} by construction \citep{SunAbraham2021,DeChaisemartinDHaultfoeuille2023}.

Two aggregation schemes are canonical. The sample-share weights target the average effect among units observed at event time $k$:
\begin{equation}\label{eq:omega-sample}
\hat\omega^{\text{sample}}_g(k)=\frac{N_g}{\sum_{h\in\mathcal{G}(k)} N_h},
\qquad N_g\equiv \sum_{i=1}^n\Ind\{G_i=g\}.
\end{equation}
Alternatively, population-share weights target the corresponding population average and replace $N_g/n$ by an estimate of $\Prob(G_i=g)$; in large $n$ panels the two coincide. When an economically motivated target is required, exposure weights can be used, for example
\begin{equation}\label{eq:omega-exposure}
\hat\omega^{\text{exp}}_g(k)=\frac{\sum_{i=1}^n w_i\,\Ind\{G_i=g\}}{\sum_{h\in\mathcal{G}(k)}\sum_{i=1}^n w_i\,\Ind\{G_i=h\}},
\end{equation}
where $w_i\ge 0$ encodes a pre-specified exposure such as size or baseline risk. In all cases, the weighting scheme is explicit, interpretable, and verifiable.

The aggregation in \eqref{eq:event-agg} eliminates cross-$k$ contamination in a precise sense: for each fixed $k$, $\hat\tau(k)$ depends only on estimators of $\tau_g(k)$ for cohorts observed at $k$, and does not mechanically load on $\tau_g(k')$ for $k'\neq k$. This property is the essential methodological contrast with TWFE event-study regressions, in which the coefficient labelled by $k$ is generally a design-dependent mixture over horizons and cohorts \citep{SunAbraham2021,DeChaisemartinDHaultfoeuille2023}.

\subsection{Comparison with imputation estimators}\label{subsec:imputation}

Imputation estimators constitute the main competing solution to the staggered-adoption event-study problem once heterogeneous dynamic effects are admitted and the objective is an interpretable event-time profile rather than the projection-defined object delivered by TWFE regressions. The imputation approach constructs the counterfactual untreated path for treated units by estimating an untreated-outcome model on observations that are untreated at each calendar time, then imputes $Y_{it}(\infty)$ for treated observations and averages the treated-minus-imputed residuals by event time. This replaces identification through the residualised correlation structure of overlapping relative-time indicators with identification through a counterfactual mapping for untreated outcomes, and it thereby removes the mechanical cross-horizon mixing that drives negative weights and contamination in TWFE event studies under heterogeneity.

A complementary approach is synthetic difference-in-differences, which combines outcome-modeling and weighting ideas by constructing synthetic controls within a DiD structure to improve robustness when treated and control units differ in levels and evolve differently over time. This method is closely aligned with the design-first emphasis on making the comparison structure explicit: it targets a transparent reweighting of control units so that the counterfactual path is a documented function of observed pre-treatment fit, rather than an implicit by-product of two-way fixed effects projection geometry. In the present framework, the connection is conceptual rather than notational: synthetic DiD can be read as another way to force the analyst to show the weights and the induced counterfactual trajectory, paralleling the role played here by explicit cohort--time estimation and declared convex aggregation in \eqref{eq:robust-event}--\eqref{eq:event-agg} \citep{ArkhangelskyAtheyHirshbergImbensWager2021}.

To state the object precisely, fix an event time $k$ and define the treated set at that horizon as
\[
\mathcal{T}(k)\equiv \{(i,t): G_i<\infty,\ t-G_i=k\},
\]
and let $\widehat{Y}_{it}(\infty)$ denote an imputed untreated outcome generated by a first-stage model estimated using untreated observations. The imputation event-study estimand is operationally the event-time average of imputed residuals,
\begin{equation}\label{eq:imp-event}
\hat\tau^{\mathrm{imp}}(k)
\equiv
\frac{1}{|\mathcal{T}(k)|}\sum_{(i,t)\in\mathcal{T}(k)}
\Big(Y_{it}-\widehat{Y}_{it}(\infty)\Big),
\end{equation}
with the understanding that the first-stage estimation uses a training sample restricted to untreated observations at each $t$, typically
\[
\{(i,t): D_{it}=0\},
\]
possibly augmented by additional restrictions that exclude not-yet-treated observations close to adoption to enforce no-anticipation in the training set. The attraction of \eqref{eq:imp-event} is interpretability: conditional on a valid imputation of the untreated counterfactual, the average residual at event time $k$ is directly an average effect among treated observations at that horizon, with no regression-geometry channel by which effects at other horizons mechanically enter.

The group--time approach adopted in this paper differs at the level of primitives and therefore at the level of failure modes. Group--time estimation begins from the cohort--time estimand $\GATT(g,t)$ and constructs $\tau(k)$ as an explicit convex aggregation over cohorts observed at horizon $k$, as in \eqref{eq:event-agg}. Identification is expressed through conditional parallel trends for untreated increments (Section \ref{sec:sensitivity}) and overlap (Assumption \ref{ass:overlap}) relative to a declared control set. Estimation is then performed cell-by-cell, so that the only averaging step is the declared convex aggregation with weights $\omega_g(k)$. This separation of (i) identification at the cohort--time level and (ii) aggregation at the event-time level is the mechanism that prevents cross-horizon contamination by construction. In contrast, the imputation approach collapses the problem into a single counterfactual modelling step: once $\widehat{Y}_{it}(\infty)$ is produced, aggregation by event time is mechanically straightforward.

The first-stage model that generates $\widehat{Y}_{it}(\infty)$ can be written generically as a regression for untreated outcomes. A common specification class is an interactive or additive structure with unit and time components and, optionally, covariates:
\begin{equation}\label{eq:imp-firststage}
Y_{it}
=
a_i + b_t + r(X_{it}) + \nu_{it},
\qquad \text{estimated on } \{(i,t): D_{it}=0\},
\end{equation}
where $r(\cdot)$ can be high-dimensional or nonparametric and may be regularised, and where variants replace $(a_i,b_t)$ with more flexible low-rank or factor structures. The imputed counterfactual is then
\[
\widehat{Y}_{it}(\infty)=\hat a_i+\hat b_t+\hat r(X_{it}),
\]
and the event-study effect is produced by \eqref{eq:imp-event}. This makes explicit the locus of identifying content: for treated units in post-treatment periods, the imputation method relies on extrapolating the untreated-outcome model from the untreated sample into treated states. When adoption is confounded by covariates or by latent types that also affect trends, this extrapolation can be fragile without strong overlap and without strong restrictions on the stability of the untreated-outcome model across the adoption boundary.

The comparison can be sharpened by mapping both approaches into the same design-first language. Under absorbing adoption and no anticipation, the cohort--time effect satisfies
\[
\GATT(g,t)=\E[Y_{it}\mid G_i=g]-\E[Y_{it}(\infty)\mid G_i=g].
\]
Group--time estimators identify $\E[Y_{it}(\infty)\mid G_i=g]$ through observed controls and a parallel-trends restriction stated for untreated increments, with explicit control sets and weighting. Imputation estimators identify the same object by supplying $\widehat{Y}_{it}(\infty)$ as a model-based proxy for $Y_{it}(\infty)$ and then averaging residuals. The key difference is therefore not the target but the instrument by which the counterfactual is obtained: explicit comparison sets and weighting versus an estimated counterfactual outcome process.

These distinctions imply complementary fragilities. Imputation removes TWFE regression-geometry contamination but introduces a modelling channel: the resulting estimator can be biased if the first-stage nuisance model in \eqref{eq:imp-firststage} is misspecified or regularisation-biased in a way that does not vanish at the $n^{-1/2}$ rate relevant for inference on \eqref{eq:imp-event}. This concern is most acute when the adoption process depends on rich observables or latent states that also affect outcome trends, because the untreated sample used to estimate \eqref{eq:imp-firststage} may differ systematically from the treated cohort in post-treatment periods, requiring stable extrapolation. The group--time approach instead places the burden on overlap and on the plausibility of conditional parallel trends within declared comparison sets, which is why Section \ref{sec:sensitivity} explicitly parameterises deviations through $\delta_{g,t}(x)$ and why Section \ref{subsec:falsification-adaptive} reports sensitivity regions under calibrated restriction classes. In this manuscript, the inferential workflow is organised around that explicit restriction language: identification fragility is surfaced and quantified directly, rather than absorbed implicitly into a counterfactual model.

The two approaches are not mutually exclusive, and the orthogonalisation programme in Section 7 can be interpreted as a bridge. When covariates are high-dimensional, imputation methods and group--time weighting methods both require estimation of nuisance functions whose errors can enter the target estimand at first order. Section 7 constructs Neyman-orthogonal moments for $\GATT(g,t)$ using the density-ratio Riesz representer, yielding debiased estimation under flexible nuisance learning. An analogous orthogonalisation logic can be applied to imputation estimators by treating the imputed counterfactual as a nuisance component and constructing scores whose Gateaux derivative with respect to that nuisance vanishes at the truth. The practical distinction retained in this paper is therefore one of reporting discipline: group--time estimation delivers a transparent convex-aggregation estimand, Section 5 delivers design-geometry diagnostics for the regression-based competitor, and Section 6 delivers falsification-adaptive sensitivity regions that remain meaningful precisely because the estimand is stated as an explicit functional of cohort--time causal objects rather than as a projection-defined regression coefficient.

A further regression-based competitor is the two-way Mundlak framework, which targets robustness to correlated heterogeneity by augmenting the regression with unit-specific covariate means (the Mundlak device) alongside time effects and a model appropriate to the outcome support, including Poisson pseudo-maximum-likelihood for nonnegative outcomes. The identifying content is shifted from fixed-effect residualisation of event-time indicators to an explicit conditional-mean model that allows the unobserved unit component to be correlated with covariates through its projection on covariate averages, thereby absorbing an important class of time-invariant selection and mitigating bias from heterogeneous levels that contaminate simpler two-way specifications. In staggered-adoption settings, this approach can be combined with event-time or group-time indicators, but its interpretation remains model-based: dynamic causal effects are recovered through functional-form restrictions on the conditional mean rather than through convex aggregation of cohort--time comparisons, and the quality of inference depends on whether the Mundlak-augmented specification adequately captures the relevant heterogeneity and trend structure in the untreated potential outcome process \citep{Wooldridge2021}.

\clearpage
\section{Design diagnostics}

The case for reporting design diagnostics is especially acute in empirical finance, where staggered designs are common and where interpreted dynamic profiles frequently drive economic narratives. In that context, \citet{BakerLarckerWang2022} documents that staggered difference-in-differences regressions can generate materially misleading estimates in empirically standard settings, precisely because the regression object can behave as a non-convex mixture of heterogeneous effects with sign reversals and implicit comparisons that do not correspond to the intended timing interpretation. The diagnostics in this section operationalise that warning by turning the adoption schedule and event window into computable indices of negative-weight mass and cross-horizon loading, so that the credibility of a conventional event-study regression can be assessed as a property of design geometry before any economic interpretation is attached to the estimated profile \citep{WingEtAl2024}.

\subsection{Negative-weight risk index}

Fix $k\in\mathcal{K}\setminus\{k_0\}$ and consider the TWFE probability-limit decomposition in \eqref{eq:weights}. The indices below are defined as functionals of the implied weight array $\{w_{g,k'}(k)\}_{g,k'}$ and are therefore design objects: they depend only on the adoption pattern, the event window, and the fixed-effect structure, not on the realised outcomes \citep{AbadieAngristFrandsen2025}.

Define the signed and absolute weight masses
\[
S(k)\equiv \sum_{g}\sum_{k'} w_{g,k'}(k)=1,
\qquad
A(k)\equiv \sum_{g}\sum_{k'} \abs{w_{g,k'}(k)}\ \ge\ 1,
\]
where $A(k)=1$ holds if and only if all weights are non-negative. The excess absolute mass $A(k)-1$ is a direct measure of how far the TWFE coefficient departs from a convex average.

\begin{definition}[Negative-weight mass]\label{def:Nk}
\[
\mathcal{N}(k)\equiv \sum_{g}\sum_{k'} \abs{w_{g,k'}(k)}\,\Ind\{w_{g,k'}(k)<0\}.
\]
\end{definition}
$\mathcal{N}(k)$ is zero if and only if the implicit weighting scheme is everywhere non-negative. Moreover, because $S(k)=1$,
\[
A(k)=1+2\sum_{g,k'}\big(-w_{g,k'}(k)\big)\Ind\{w_{g,k'}(k)<0\},
\]
so $\mathcal{N}(k)$ is tightly linked to $A(k)$ and provides a scale-free measure of sign-reversing comparisons.

The second pathology is horizon contamination: the coefficient labelled by $k$ can load on causal effects at horizons $k'\neq k$.

\begin{definition}[Cross-horizon contamination]\label{def:Ck}
\[
\mathcal{C}(k)\equiv \sum_{g}\sum_{k'} \abs{w_{g,k'}(k)}\,\Ind\{k'\neq k\}.
\]
\end{definition}
$\mathcal{C}(k)=0$ if and only if $\plim\,\hat\beta_k$ is a weighted average of $\{\tau_g(k)\}_{g\in\mathcal{G}(k)}$ alone, with no mechanical loading on other horizons.

Both indices are computable without outcome data. Let $Z=M_XD$ be the residualised event-time design matrix in \eqref{eq:betaZ}. For each $k$, define the coefficient-weight vector
\[
\pi(k)\equiv Z(Z'Z)^{-1}e_k,
\]
so that $\hat\beta_k=\pi(k)'Y$ by \eqref{eq:linfun}. The implied cohort-horizon weights satisfy \eqref{eq:w-from-pi} and are therefore functions only of $Z$ (hence only of $(G_i,t)$, the chosen $\mathcal{K}$, and the fixed-effect structure). Consequently, $\mathcal{N}(k)$ and $\mathcal{C}(k)$ are design diagnostics that can be reported ex ante as properties of the empirical design, and they quantify precisely the two mechanisms emphasised in the heterogeneous-effects critique of TWFE event studies \citep{SunAbraham2021,DeChaisemartinDHaultfoeuille2023}.

\subsection{Design geometry under staggered adoption}\label{subsec:design-geometry}

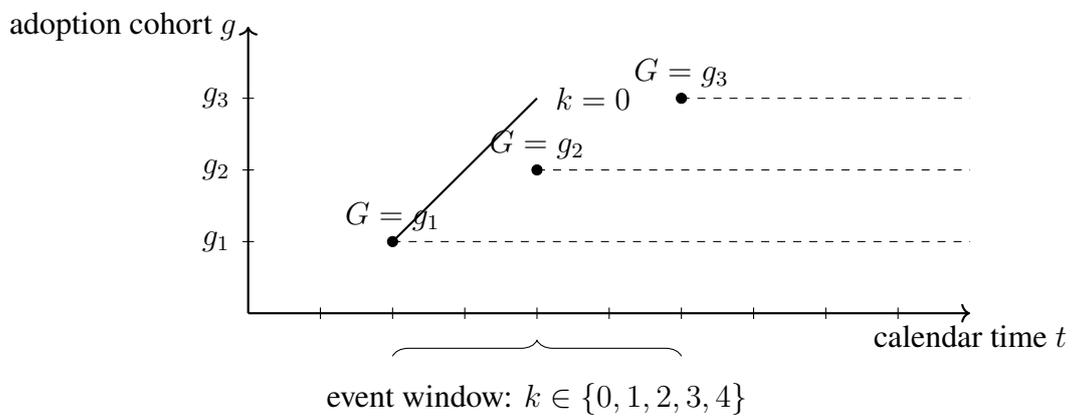
\begin{figure}[htbp]
\centering
\begin{tikzpicture}[scale=0.95]
  \draw[->,thick] (0,0) -- (10,0) node[below] {calendar time $t$};
  \draw[->,thick] (0,0) -- (0,4) node[left] {adoption cohort $g$};

  \foreach \x in {1,2,3,4,5,6,7,8,9} \draw (\x,0.08) -- (\x,-0.08);
  \foreach \y in {1,2,3} \draw (0.08,\y) -- (-0.08,\y);

  \node[anchor=east] at (-0.1,1) {$g_1$};
  \node[anchor=east] at (-0.1,2) {$g_2$};
  \node[anchor=east] at (-0.1,3) {$g_3$};

  \filldraw (2,1) circle (2pt);
  \filldraw (4,2) circle (2pt);
  \filldraw (6,3) circle (2pt);

  \draw[dashed] (2,1) -- (10,1);
  \draw[dashed] (4,2) -- (10,2);
  \draw[dashed] (6,3) -- (10,3);

  \node[above] at (2,1) {$G=g_1$};
  \node[above] at (4,2) {$G=g_2$};
  \node[above] at (6,3) {$G=g_3$};

  \draw[decorate,decoration={brace,amplitude=6pt}] (2,-0.6) -- (6,-0.6)
    node[midway,below=6pt] {event window: $k\in\{0,1,2,3,4\}$};

  \draw[thick] (2,1) -- (4,3);
  \node[anchor=west] at (4.1,3) {$k=0$};

\end{tikzpicture}
\caption{\textsc{Cohort--time geometry under staggered adoption.} The adoption schedule induces cohort--time cells $(g,t)$ and associated event times $k=t-g$. Event-time indicators select diagonals in this diagram. After residualising by unit and time fixed effects, the resulting regressors typically change sign and are not mutually orthogonal, which is the design mechanism behind negative weights and cross-horizon contamination in TWFE event studies.}
\label{fig:design-geometry}
\end{figure}

The weighting pathologies in \eqref{eq:weights} are design-driven: they arise from the cohort--time partition induced by the adoption schedule and from the projection of event-time indicators onto the orthogonal complement of unit and time fixed effects. Figure \ref{fig:design-geometry} provides a compact representation of this geometry. Each horizontal line corresponds to an adoption cohort $g$, the filled point marks the adoption date $G=g$, and the dashed ray to the right indicates the post-adoption region $(t\ge g)$ in which treatment is ``on'' under absorbing adoption. Every observed pair $(g,t)$ determines an event time $k=t-g$; the event-study regressors $\D^{(k)}_{it}$ select diagonals in this cohort--time diagram. The residualisation implicit in \eqref{eq:fwl} transforms these diagonal selectors into residualised regressors $\tilde D_k=M_XD_k$ that generally change sign across cells, because unit and time fixed effects remove cohort means and time means. Those sign changes are the mechanical source of negative weights in the TWFE decomposition, while the non-orthogonality of the residualised diagonals is the mechanical source of cross-horizon contamination, both of which are summarised by $\mathcal{N}(k)$ and $\mathcal{C}(k)$.

\clearpage
\section{Robust inference under restricted violations of parallel trends}

\subsection{A restriction class}

Sensitivity analysis replaces the knife-edge restriction $\delta_{g,t}(x)\equiv 0$ in \eqref{eq:wedge} by a structured class of admissible deviations that is disciplined by pre-treatment behaviour and yields computable identified sets \citep{RambachanRoth2023}. For notational economy, suppress covariates in this subsection and write the cohort-specific untreated increment as $\Delta Y_{it}(\infty)$ and its cohort-time mean as
\[
m_{g,t}\equiv \E\!\left[\Delta Y_{it}(\infty)\mid G_i=g\right],
\qquad
m_{\infty,t}\equiv \E\!\left[\Delta Y_{it}(\infty)\mid G_i=\infty\right].
\]
Define the deviation (trend wedge)
\begin{equation}\label{eq:delta}
\delta_{g,t}\equiv m_{g,t}-m_{\infty,t}.
\end{equation}
Assumption \ref{ass:pt} is the restriction $\delta_{g,t}=0$ for all $(g,t)$, which point-identifies the cohort-time effects under no anticipation.

A restriction class $\Delta(\mathcal{R})$ is a set of sequences $\{\delta_{g,t}\}_{g,t}$ consistent with a maintained regularity structure $\mathcal{R}$ and with the information contained in pre-treatment periods. The robust parallel-trends approach of \citet{RambachanRoth2023} can be implemented by imposing that post-treatment deviations are bounded by a function of pre-treatment deviations plus a curvature (smoothness) penalty. Concretely, fix a cohort $g$ and let $\mathcal{T}_g^{-}\equiv\{t: t<g\}$ be the pre-treatment times. Let $\Delta \delta_{g,t}\equiv \delta_{g,t}-\delta_{g,t-1}$ be the first difference and $\Delta^2 \delta_{g,t}\equiv \Delta\delta_{g,t}-\Delta\delta_{g,t-1}$ be the second difference. For constants $B\ge 0$ and $\Gamma\ge 0$, define the curvature-bounded class
\begin{equation}\label{eq:RRclass}
\Delta(\mathcal{R}(B,\Gamma))
\equiv
\left\{\delta:\ 
\max_{t\ge g}\abs{\Delta^2 \delta_{g,t}}\le \Gamma
\ \ \text{and}\ \
\max_{t<g}\abs{\delta_{g,t}}\le B
\ \ \text{for all } g
\right\}.
\end{equation}
The first restriction bounds the change in the deviation slope (a discrete curvature constraint) and limits how abruptly post-treatment trends can diverge, while the second uses pre-treatment periods to cap the magnitude of deviation levels. Alternative specifications replace $\max_{t<g}\abs{\delta_{g,t}}\le B$ by data-driven calibration based on observed pre-trend estimates and their sampling error, which is the operational choice in robust event-study inference \citep{RambachanRoth2023}.

Partial identification provides a clean way to formalise sensitivity analysis when point identification is not credible under strict assumptions, because the object of interest becomes a set of probability distributions and a corresponding identified set for the estimand, indexed by explicit, interpretable restrictions on departures from the maintained design assumptions. This framing clarifies that robustness exercises are not an afterthought but a disciplined shift from point statements to set statements, with the width of the identified set measuring the informational content delivered by the data under the chosen restriction class, and it situates restricted-violation frameworks within a broader econometric programme that treats assumptions as tunable constraints rather than as binary truths \citep{Manski2003, Roth2024}.

More generally, write
\[
\delta \in \Delta(\mathcal{R}),
\]
where $\mathcal{R}$ encodes one or more of: (i) level bounds anchored by pre-period estimates, (ii) slope bounds linking post-period drifts to pre-period drifts, and (iii) curvature or smoothness bounds limiting changes in drift. Each choice of $\mathcal{R}$ induces an identified set for $\tau(k)$ obtained by optimising the target estimand over $\delta\in\Delta(\mathcal{R})$, delivering sensitivity regions that are explicit functions of the maintained restriction rather than informal robustness claims \citep{RambachanRoth2023,MastenPoirier2021}.

A further sensitivity channel concerns functional-form dependence: even when a parallel-trends restriction is stated in a conditional-mean or first-difference form, empirical assessments and pre-tests can be materially affected by how the counterfactual trend is parameterised, because distinct functional-form choices can imply distinct extrapolations from pre-treatment fit into post-treatment counterfactual evolution. This matters directly for restriction classes such as \eqref{eq:RRclass}, since calibration of $(B,\Gamma)$ from pre-period behaviour is only meaningful relative to a maintained mapping from observed pre-period deviations to admissible post-period paths, and functional-form choices can tighten or loosen that mapping in ways that are not innocuous. The practical implication is that restricted-violations inference should be interpreted as conditional on the maintained trend parameterisation used to construct the deviation process and its calibration, and robustness checks should treat functional form as part of the sensitivity analysis rather than as a separate specification decision \citep{RothSantAnna2023}.

\subsection{Identified sets and confidence regions}

Fix an event time $k\in\mathcal{K}$ and let the target be the convexly aggregated effect $\theta\equiv\tau(k)$ in \eqref{eq:agg}. Under a restriction class $\delta\in\Delta(\mathcal{R})$ (for example \eqref{eq:RRclass}), the target is generally no longer point identified. The restriction class induces an identified set
\begin{equation}\label{eq:identified-set}
\Theta(\mathcal{R}) \;\equiv\; \left\{\theta(\delta):\ \delta\in\Delta(\mathcal{R})\right\},
\end{equation}
where $\theta(\delta)$ is the value of the estimand implied by a given admissible deviation process $\delta$. Operationally, $\Theta(\mathcal{R})$ is obtained by solving two optimisation problems that deliver sharp lower and upper bounds:
\begin{equation}\label{eq:bounds}
\underline{\theta}(\mathcal{R})\equiv \inf_{\delta\in\Delta(\mathcal{R})}\theta(\delta),
\qquad
\overline{\theta}(\mathcal{R})\equiv \sup_{\delta\in\Delta(\mathcal{R})}\theta(\delta),
\qquad
\Theta(\mathcal{R})=[\underline{\theta}(\mathcal{R}),\overline{\theta}(\mathcal{R})].
\end{equation}
The defining feature of the robust parallel-trends approach is that $\Delta(\mathcal{R})$ is calibrated from pre-treatment behaviour and regularity restrictions, so the identified set expands smoothly as $\mathcal{R}$ relaxes, replacing binary acceptance or rejection of parallel trends by a quantitative description of what the data imply under controlled deviations \citep{RambachanRoth2023}.

As illustrated in Figure \ref{fig:sensitivity-cone}, the restriction class $\Delta(\mathcal{R})$ defines a geometry of admissible deviation paths in event time. The parameter $B$ imposes a rectangular bound on pre-treatment deviations, quantifying tolerance for pre-trend differences. In the post-treatment period, $\Gamma$ governs the rate at which the set of admissible biases expands. A higher $\Gamma$ allows the counterfactual trend to change curvature more abruptly, resulting in a wider `cone' of uncertainty and a larger identified set $\Theta(\mathcal{R})$.

Inference targets confidence regions that cover $\Theta(\mathcal{R})$ uniformly over the maintained restriction class. Let $\widehat{\Theta}(\mathcal{R})=[\underline{\hat\theta}(\mathcal{R}),\overline{\hat\theta}(\mathcal{R})]$ be the sample analogue of \eqref{eq:bounds}, computed by substituting estimated reduced-form components and solving the corresponding sample optimisation problems. A $(1-\alpha)$ confidence region $\mathcal{C}_n(\mathcal{R})$ is said to be uniformly valid if it contains the true identified set with probability at least $1-\alpha$ uniformly over all data-generating processes satisfying $\mathcal{R}$.
As illustrated in Appendix \ref{app:A3} and Figure \ref{fig:sensitivity-cone}, the restriction class $\Delta(\mathcal{R})$ defines a geometry of admissible deviation paths in event time, with pre-treatment tolerance governed by $B$ and post-treatment smoothness governed by $\Gamma$.

\begin{theorem}[Uniformly valid confidence regions]\label{thm:uniform}
Maintain Assumptions \ref{ass:noanticip} and \ref{ass:reg}, and suppose the deviation process satisfies $\delta\in\Delta(\mathcal{R})$ where $\Delta(\mathcal{R})$ obeys the regularity and calibration structure in \citet{RambachanRoth2023}. Then there exist confidence regions $\mathcal{C}_n(\mathcal{R})$ constructed from $\widehat{\Theta}(\mathcal{R})$ and an appropriate critical value sequence such that, for any $\alpha\in(0,1)$,
\[
\liminf_{n\to\infty}\ \inf_{P\in\mathcal{P}(\mathcal{R})}
\Prob_P\!\left(\Theta_P(\mathcal{R})\subseteq \mathcal{C}_n(\mathcal{R})\right)\ \ge\ 1-\alpha,
\]
where $\mathcal{P}(\mathcal{R})$ denotes the class of data-generating processes satisfying the maintained restrictions, and $\Theta_P(\mathcal{R})$ is the identified set for $\theta$ under $P$.
\end{theorem}

\subsection{Calibration}\label{sec:calibration}

\paragraph{Goal.}
Calibrate sensitivity inputs $(\Delta(\mathcal{R}),B,\Gamma)$ using observables so that each restriction class $\mathcal{R}$ admits an interpretable and reproducible mapping from data diagnostics to parameter values.

\paragraph{Objects.}
Let $\widehat{\beta}_{\ell}$ denote pre-period event-time coefficients for $\ell\in\mathcal{L}_{\mathrm{pre}}$ from the baseline event-study regression, with standard errors $\widehat{\sigma}_{\ell}$.
Define the pre-trend magnitude statistic
\begin{equation}
M_{\mathrm{pre}}
\;:=\;
\max_{\ell\in\mathcal{L}_{\mathrm{pre}}}
\left|\frac{\widehat{\beta}_{\ell}}{\widehat{\sigma}_{\ell}}\right|
\qquad\text{and}\qquad
A_{\mathrm{pre}}
\;:=\;
\max_{\ell\in\mathcal{L}_{\mathrm{pre}}}
\left|\widehat{\beta}_{\ell}\right|.
\label{eq:pre_stats}
\end{equation}
Let $\widehat{\tau}$ denote the baseline target estimand and $\widehat{\mathcal{D}}$ denote a drift diagnostic (defined below).

\subsubsection{Calibration rules for \texorpdfstring{$\Delta(\mathcal{R})$}{Delta(R)}}
Let $\Delta(\mathcal{R})$ parameterise the radius of admissible violations within restriction class $\mathcal{R}$.
For each $\mathcal{R}$, define a measurable diagnostic map
\begin{equation}
\widehat{\Delta}(\mathcal{R})
\;:=\;
\phi_{\mathcal{R}}\!\bigl(\{\widehat{\beta}_{\ell},\widehat{\sigma}_{\ell}\}_{\ell\in\mathcal{L}_{\mathrm{pre}}},\widehat{\mathcal{D}},\text{holdout},\text{placebo}\bigr),
\label{eq:delta_map_general}
\end{equation}
with $\phi_{\mathcal{R}}$ chosen so that $\widehat{\Delta}(\mathcal{R})$ is nondecreasing in $M_{\mathrm{pre}}$ and in placebo rejections.

\paragraph{Canonical rule.}
For any $\mathcal{R}$ admitting a scalar violation radius, set
\begin{equation}
\widehat{\Delta}(\mathcal{R})
\;:=\;
c_{\mathcal{R}}\cdot A_{\mathrm{pre}},
\qquad
c_{\mathcal{R}}\in[0,\bar{c}_{\mathcal{R}}],
\label{eq:delta_rule}
\end{equation}
where $c_{\mathcal{R}}$ is selected by a holdout or placebo criterion (defined in Box~\ref{box:calibration_recipe}).

\subsubsection{Data-driven calibration of \texorpdfstring{$B$}{B}}
Let $B$ denote the scale of admissible confounding or violation magnitude in the moment condition.
Define $B$ in outcome units, anchored to observed pre-trend magnitudes:
\begin{equation}
\widehat{B}
\;:=\;
\kappa_{B}\cdot A_{\mathrm{pre}},
\qquad
\kappa_{B}\in\{0,0.25,0.5,1,2\}.
\label{eq:B_rule}
\end{equation}
Alternatively, define $B$ by a formal holdout rule using a pre-period split $\mathcal{L}_{\mathrm{pre}}=\mathcal{L}_{1}\cup\mathcal{L}_{2}$:
\begin{equation}
\widehat{B}_{\mathrm{holdout}}
\;:=\;
\min\Bigl\{b\ge 0:\;
\max_{\ell\in\mathcal{L}_{2}}
\bigl|\widehat{\beta}_{\ell}(b)\bigr|
\le
q_{0.95}\bigl(\{|\widehat{\beta}_{\ell}|:\ell\in\mathcal{L}_{1}\}\bigr)
\Bigr\},
\label{eq:B_holdout}
\end{equation}
where $\widehat{\beta}_{\ell}(b)$ denotes the pre-coefficient under the sensitivity perturbation indexed by $b$.

\subsubsection{Monotone family calibration of \texorpdfstring{$\Gamma$}{Gamma}}
Let $\Gamma\ge 0$ index a monotone family of violations with a time-unit interpretation as drift per period.
Define a drift diagnostic from pre-period differences:
\begin{equation}
\widehat{\mathcal{D}}
\;:=\;
\max_{\ell\in\mathcal{L}_{\mathrm{pre}}\setminus\{\min\mathcal{L}_{\mathrm{pre}}\}}
\left|
\widehat{\beta}_{\ell}-\widehat{\beta}_{\ell-1}
\right|.
\label{eq:drift_diag}
\end{equation}
Map $\Gamma$ to percent drift per period relative to the baseline effect scale:
\begin{equation}
\widehat{\Gamma}
\;:=\;
100\cdot
\frac{\widehat{\mathcal{D}}}{|\widehat{\tau}|+\varepsilon_{\tau}},
\qquad
\varepsilon_{\tau}>0.
\label{eq:Gamma_rule}
\end{equation}
The admissible violation set is then indexed by $\Gamma$ through a monotone constraint, for example
\begin{equation}
\bigl|\Delta_{\ell}-\Delta_{\ell-1}\bigr|
\;\le\;
\frac{\Gamma}{100}\cdot\bigl(|\widehat{\tau}|+\varepsilon_{\tau}\bigr)
\qquad\forall \ell\in\mathcal{L}_{\mathrm{pre}}\cup\mathcal{L}_{\mathrm{post}},
\label{eq:Gamma_constraint}
\end{equation}
where $\{\Delta_{\ell}\}$ denotes the pathwise deviation sequence admitted under $\mathcal{R}$.

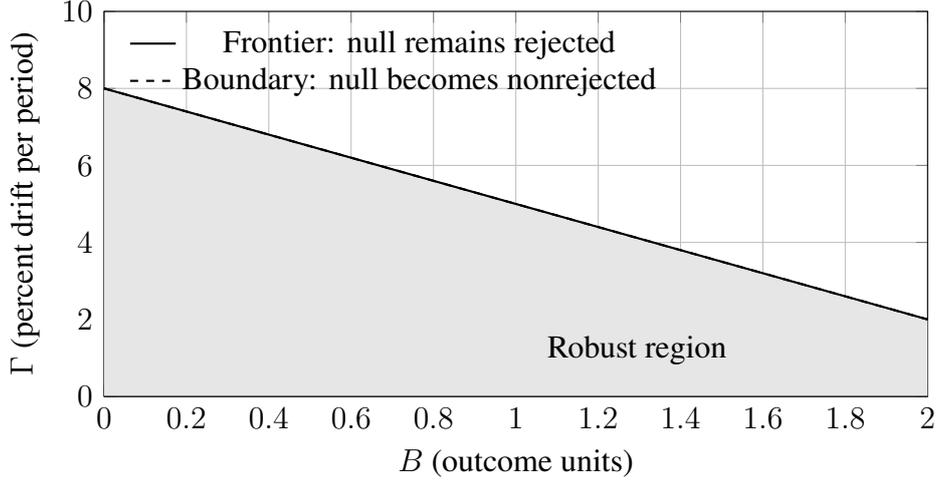
\begin{figure}[t]
\centering
\begin{tikzpicture}
\begin{axis}[
    width=0.78\linewidth,
    height=0.42\linewidth,
    xlabel={$B$ (outcome units)},
    ylabel={$\Gamma$ (percent drift per period)},
    xmin=0, xmax=2,
    ymin=0, ymax=10,
    grid=both,
    legend style={draw=none, fill=none, at={(0.02,0.98)}, anchor=north west},
]
\addplot[domain=0:2, samples=200, thick] {max(0, 8 - 3*x)};
\addlegendentry{Frontier: null remains rejected}
\addplot[domain=0:2, samples=200, thick, dashed] {min(10, 8 - 3*x)};
\addlegendentry{Boundary: null becomes nonrejected}

\addplot[name path=curve, domain=0:2, samples=200] {max(0, 8 - 3*x)};
\addplot[name path=axis, domain=0:2] {0};
\addplot[fill=gray!20] fill between[of=curve and axis];

\node[anchor=west] at (axis cs:1.05,1.2) {Robust region};
\end{axis}
\end{tikzpicture}
\caption{Sensitivity region in $(B,\Gamma)$ space. Shaded set indicates parameter pairs for which the target conclusion remains unchanged under restriction class $\mathcal{R}$.}
\label{fig:sensitivity_region}
\end{figure}

\begin{table}[t]
\centering
\caption{Parameter mapping from observables to calibration inputs.}
\label{tab:param_mapping}
\begin{tabular}{@{}lll@{}}
\toprule
Observable diagnostic & Parameter & Calibration rule \\
\midrule
$A_{\mathrm{pre}}=\max_{\ell\in\mathcal{L}_{\mathrm{pre}}}|\widehat{\beta}_{\ell}|$
& $B$ & $\widehat{B}=\kappa_{B}\cdot A_{\mathrm{pre}}$ \\
$M_{\mathrm{pre}}=\max_{\ell\in\mathcal{L}_{\mathrm{pre}}}|\widehat{\beta}_{\ell}/\widehat{\sigma}_{\ell}|$
& $\Delta(\mathcal{R})$ & $\widehat{\Delta}(\mathcal{R})=c_{\mathcal{R}}\cdot A_{\mathrm{pre}}$ \\
$\widehat{\mathcal{D}}=\max_{\ell}|\widehat{\beta}_{\ell}-\widehat{\beta}_{\ell-1}|$
& $\Gamma$ & $\widehat{\Gamma}=100\cdot\widehat{\mathcal{D}}/(|\widehat{\tau}|+\varepsilon_{\tau})$ \\
Placebo rejections (rate) & $(c_{\mathcal{R}},\kappa_{B})$ & Increase grid until placebo rejection rate $\le \alpha$ \\
Holdout loss $\mathcal{L}_{2}$ & $(c_{\mathcal{R}},B)$ & Minimal pair meeting holdout constraint in \eqref{eq:B_holdout} \\
\bottomrule
\end{tabular}
\end{table}

\begin{center}
\fbox{%
\begin{minipage}{0.93\linewidth}
\textbf{Box 6.1 Calibration recipe}\label{box:calibration_recipe}

\textbf{Inputs.} Pre-period event-time coefficients $\{\widehat{\beta}_{\ell},\widehat{\sigma}_{\ell}\}_{\ell\in\mathcal{L}_{\mathrm{pre}}}$, baseline estimate $\widehat{\tau}$, placebo outcomes or placebo treatment timing, holdout split of $\mathcal{L}_{\mathrm{pre}}$.

\textbf{Step 1 (pre-trend scale).} Compute $A_{\mathrm{pre}}$ and $M_{\mathrm{pre}}$ in \eqref{eq:pre_stats}. Compute drift diagnostic $\widehat{\mathcal{D}}$ in \eqref{eq:drift_diag}.

\textbf{Step 2 (grid).} Define candidate grids
\[
\mathcal{B}=\{\kappa_{B}\cdot A_{\mathrm{pre}}:\kappa_{B}\in\{0,0.25,0.5,1,2\}\},
\qquad
\mathcal{G}=\{0,1,2,5,10\},
\qquad
\mathcal{C}_{\mathcal{R}}=\{0,0.5,1,2\}.
\]

\textbf{Step 3 (formal rule).} For each $(b,\gamma,c)\in\mathcal{B}\times\mathcal{G}\times\mathcal{C}_{\mathcal{R}}$, impose restriction class $\mathcal{R}$ with
\[
B=b,
\qquad
\Gamma=\gamma,
\qquad
\Delta(\mathcal{R})=c\cdot A_{\mathrm{pre}},
\]
and compute (i) holdout discrepancy on $\mathcal{L}_{2}$, (ii) placebo rejection rate, and (iii) target conclusion stability.

\textbf{Step 4 (selection).} Choose the minimal triple $(\widehat{B},\widehat{\Gamma},\widehat{\Delta}(\mathcal{R}))$ such that placebo rejection rate is at most $\alpha$ and holdout discrepancy does not exceed the $\mathcal{L}_{1}$ reference quantile, then report the corresponding sensitivity region as in Figure~\ref{fig:sensitivity_region}.
\end{minipage}}
\end{center}

\clearpage
\section{Orthogonalisation with covariates via Riesz representers}

High-dimensional conditioning variables are standard in modern empirical work and arise naturally once treatment timing is allowed to depend on rich unit characteristics and time-varying states. In such environments, plug-in estimators that substitute flexible estimates of nuisance regressions into low-dimensional causal functionals typically suffer first-order bias because the estimation error in the nuisance component enters the target functional at the same order as sampling uncertainty. The orthogonalisation programme addresses this by replacing direct plug-in with estimating equations whose local (Gateaux) derivative with respect to the nuisance direction vanishes at the truth, rendering the target locally insensitive to small nuisance estimation errors and permitting valid inference after regularised or machine-learning estimation. Riesz representers provide a constructive route to Neyman-orthogonal scores and influence-function representations for broad classes of linear functionals, directly linking the implementation in this section to foundational semiparametric variance and efficiency calculations \citep{Newey1994} and to modern regularised Riesz and cross-fitting constructions \citep{ChernozhukovNeweySingh2022, SantAnnaZhao2025}.

\subsection{The event-study Riesz representer}\label{subsec:riesz-eventstudy}

This subsection specialises the generic Riesz/orthogonal-score construction to the group--time estimands $\GATT(g,t)$ in \eqref{eq:gatt-def}. The purpose is not to reintroduce machine learning as an empirical flourish, but to obtain (i) an explicit representer whose population form is a covariate density ratio, and (ii) an orthogonal moment whose first-order behaviour is insensitive to regularised estimation of nuisance components, as in \citet{ChernozhukovNeweySingh2022}.

\paragraph{Target functional.}
Fix a cohort--time cell $(g,t)$ with $t\ge g$. Let the comparison sample be the union of cohort $g$ and the control set $\mathcal{C}(g,t)$ defined in Section 4.1. For concreteness, take the never-treated control set in this subsection,
\[
\mathcal{C}(g,t)=\{i:G_i=\infty\},
\qquad
S_{i}^{(g)} \equiv \Ind\{G_i=g\}+\Ind\{G_i=\infty\}.
\]
Let $Z\equiv X_{it}$ denote the covariates used to impose conditional parallel trends in Assumption \ref{ass:pt}, and define the conditional regression functions
\[
m_{d}(x)\equiv \E[Y_{it}\mid X_{it}=x,\,G_i\in\{g,\infty\},\,\Ind\{G_i=g\}=d],
\qquad d\in\{0,1\},
\]
where $d=1$ corresponds to cohort $g$ and $d=0$ corresponds to the never-treated control group, within the restricted comparison sample $G_i\in\{g,\infty\}$. The cohort--time effect can be written as a linear functional of the control regression $m_0$ evaluated under the treated covariate distribution, together with observable treated moments. A convenient decomposition is
\begin{equation}\label{eq:gatt-functional}
\GATT(g,t)
=
\E\!\left[ Y_{it}\mid G_i=g\right]
-
\E\!\left[m_0(X_{it})\mid G_i=g\right],
\qquad
m_0(x)=\E\!\left[Y_{it}\mid X_{it}=x,\,G_i=\infty\right].
\end{equation}
The second term in \eqref{eq:gatt-functional} is the problematic component: it is an expectation of an unknown function $m_0(\cdot)$ under the treated covariate law. This is exactly the setting for a Riesz representer.

\paragraph{The Riesz representer as a density ratio.}
Work on the Hilbert space $\mathcal{H}\subseteq L_2(\mathcal{Z},\Prob_{\infty})$ with inner product
\[
\ip{f}{h}_{\infty}\equiv \E\!\left[f(X_{it})\,h(X_{it})\mid G_i=\infty\right],
\]
so the reference measure is the control covariate distribution. Define the linear functional
\[
\mathcal{L}(h)\equiv \E\!\left[h(X_{it})\mid G_i=g\right],
\qquad h\in\mathcal{H}.
\]
Under overlap (Assumption \ref{ass:overlap} specialised to the never-treated control), $\mathcal{L}$ is continuous on $\mathcal{H}$ and admits a unique Riesz representer $\alpha_{g,t}(\cdot)\in\overline{\mathcal{H}}$ satisfying
\begin{equation}\label{eq:riesz-balancing}
\E\!\left[\alpha_{g,t}(X_{it})\,h(X_{it})\mid G_i=\infty\right]
=
\E\!\left[h(X_{it})\mid G_i=g\right]
\quad \text{for all } h\in\mathcal{H}.
\end{equation}
The population solution to \eqref{eq:riesz-balancing} is the Radon--Nikodym derivative of the treated covariate law with respect to the control covariate law:
\begin{equation}\label{eq:alpha-density}
\alpha_{g,t}(x)
=
\frac{f_{X\mid G=g,t}(x)}{f_{X\mid G=\infty,t}(x)},
\end{equation}
where $f_{X\mid G=g,t}$ denotes the conditional density of $X_{it}$ given $G_i=g$ (and time $t$ if $X$ is time-varying) and likewise for $G_i=\infty$. Equivalently, writing
\[
\pi_{g,t}(x)\equiv \Prob(G_i=g\mid X_{it}=x,\,G_i\in\{g,\infty\}),
\]
one obtains the odds-ratio form
\begin{equation}\label{eq:alpha-odds}
\alpha_{g,t}(x)
=
\frac{\pi_{g,t}(x)}{1-\pi_{g,t}(x)}
\cdot
\frac{\Prob(G_i=\infty\mid G_i\in\{g,\infty\})}{\Prob(G_i=g\mid G_i\in\{g,\infty\})},
\end{equation}
which makes explicit that $\alpha_{g,t}$ is proportional to the inverse odds of being in the control group relative to cohort $g$.

\paragraph{Balancing property and interpretation.}
Equation \eqref{eq:riesz-balancing} is a covariate balancing identity: weighting control observations by $\alpha_{g,t}(X_{it})$ transports expectations of any $h(X)$ from the control covariate law to the treated covariate law. When $\mathcal{H}$ is rich (for example, the closure of square-integrable functions), this implies that the reweighted control group matches the treated distribution in all $L_2$ moments. In finite-dimensional approximations, \eqref{eq:riesz-balancing} states that $\alpha_{g,t}$ learns precisely the weights that make the control group ``look like'' cohort $g$ with respect to the chosen dictionary of covariate functions.

\paragraph{Orthogonal score for $\GATT(g,t)$.}
Substituting \eqref{eq:riesz-balancing} into \eqref{eq:gatt-functional} yields an orthogonal moment condition. Define the nuisance functions
\[
m_0(x)\equiv \E[Y_{it}\mid X_{it}=x,\,G_i=\infty],
\qquad
\alpha_{g,t}(x)\ \text{as in \eqref{eq:alpha-density}},
\]
and define the score
\begin{equation}\label{eq:score-gatt}
\psi_{g,t}\big(W_{it};\theta,m_0,\alpha\big)
\equiv
\Ind\{G_i=g\}\big(Y_{it}-\theta\big)
-
\Ind\{G_i=\infty\}\,\alpha(X_{it})\big(Y_{it}-m_0(X_{it})\big)
-
\Ind\{G_i=g\}\,m_0(X_{it}),
\end{equation}
with $W_{it}=(Y_{it},X_{it},G_i)$. At the truth, $\theta=\GATT(g,t)$, $\alpha=\alpha_{g,t}$, and $m_0$ as defined above, one has
\[
\E\!\left[\psi_{g,t}\big(W_{it};\GATT(g,t),m_0,\alpha_{g,t}\big)\right]=0,
\]
and the corresponding moment is Neyman-orthogonal with respect to small perturbations in $m_0$ (and, under the regularity conditions in \citet{ChernozhukovNeweySingh2022}, also with respect to perturbations in $\alpha$ when cross-fitting is used). The algebraic structure of \eqref{eq:score-gatt} makes the orthogonality transparent: the term $\Ind\{G_i=\infty\}\alpha(X_{it})\{Y_{it}-m_0(X_{it})\}$ is a residual reweighted by the Riesz representer, while the remaining terms ensure the score has mean zero at the target and cancels first-order estimation error in the nuisance regression.

\paragraph{Operational implication.}
Regularised Riesz regression estimates $\alpha_{g,t}$ by solving an empirical analogue of \eqref{eq:riesz-balancing} over a high-dimensional dictionary of functions of $X$, typically with a penalty that controls complexity; cross-fitting separates the estimation of $(m_0,\alpha_{g,t})$ from evaluation of \eqref{eq:score-gatt}. The result is a plug-in estimator of $\GATT(g,t)$ (and therefore of $\tau_g(k)$ and $\tau(k)$ after aggregation) that retains valid first-order behaviour under flexible nuisance learning, matching the methodological template in \citet{ChernozhukovNeweySingh2022}.

\subsection{Debiased score construction}\label{subsec:debiased-score}

This subsection constructs a debiased estimating equation for $\GATT(g,t)$ using the cohort--time Riesz representer from Section \ref{subsec:riesz-eventstudy}. The construction follows the orthogonal-score logic in \citet{ChernozhukovNeweySingh2022}, specialised to the cohort--time setting.

Fix $(g,t)$ with $t\ge g$ and work in the restricted comparison sample $G_i\in\{g,\infty\}$. Let
\[
p_g \equiv \Prob(G_i=g),\qquad p_\infty \equiv \Prob(G_i=\infty).
\]
Define the control outcome regression
\[
m_0(x)\equiv \E\!\left[Y_{it}\mid X_{it}=x,\,G_i=\infty\right].
\]
Let $\alpha_{g,t}(\cdot)$ denote the (scaled) Riesz representer that transports moments from the never-treated covariate law to the cohort-$g$ covariate law in the unconditional moment form
\begin{equation}\label{eq:riesz-uncond}
\E\!\left[\Ind\{G_i=\infty\}\,\alpha_{g,t}(X_{it})\,h(X_{it})\right]
=
\E\!\left[\Ind\{G_i=g\}\,h(X_{it})\right]
\quad \text{for all } h\in\mathcal{H}.
\end{equation}
Under overlap, \eqref{eq:riesz-uncond} is satisfied by
\begin{equation}\label{eq:alpha-scaled}
\alpha_{g,t}(x)
=
\frac{p_g}{p_\infty}\cdot \frac{f_{X\mid G=g,t}(x)}{f_{X\mid G=\infty,t}(x)},
\end{equation}
which is equivalent to an odds-ratio representation in terms of $\pi_{g,t}(x)=\Prob(G_i=g\mid X_{it}=x,\,G_i\in\{g,\infty\})$.

\paragraph{Orthogonal score with an explicit double-robustness property.}
Let $\theta\equiv \GATT(g,t)$. Maintain the restricted comparison sample $G_i\in\{g,\infty\}$ and define the control regression
\[
m_0(x)\equiv \E\!\left[Y_{it}\mid X_{it}=x,\,G_i=\infty\right].
\]
Let $\alpha_{g,t}(\cdot)$ denote the (scaled) Riesz representer that solves the unconditional balancing equation
\eqref{eq:riesz-uncond}.
For generic candidates $(m,\alpha)$, define the estimating score
\begin{equation}\label{eq:dr-score}
\psi_{g,t}(W_{it};\theta,m,\alpha)
\equiv
\underbrace{\Ind\{G_i=g\}\big(Y_{it}-\theta\big)}_{\text{Target}}
\;-\;
\underbrace{\Ind\{G_i=g\}\,m(X_{it})}_{\text{Primal}}
\;-\;
\underbrace{\Ind\{G_i=\infty\}\,\alpha(X_{it})\big(Y_{it}-m(X_{it})\big)}_{\text{Dual}}.
\end{equation}
The labelling in \eqref{eq:dr-score} is purely mnemonic. The substantive claim is a \emph{moment} robustness statement: the moment condition
$\E[\psi_{g,t}(W_{it};\theta,m,\alpha)]=0$
holds at the target $\theta=\GATT(g,t)$ under either of two nuisance correctness conditions, which is the precise meaning of the term double robustness in this setting.

To state the property explicitly, write $\theta_0\equiv \GATT(g,t)=\E[Y_{it}\mid G_i=g]-\E[m_0(X_{it})\mid G_i=g]$.

First, if the outcome regression is correct, $m=m_0$, then for any measurable $\alpha(\cdot)$,
\[
\E\!\left[\Ind\{G_i=\infty\}\,\alpha(X_{it})\big(Y_{it}-m_0(X_{it})\big)\right]=0
\]
by iterated expectations conditional on $(X_{it},G_i=\infty)$, so
$\E[\psi_{g,t}(W_{it};\theta_0,m_0,\alpha)]=0$.

Second, if the representer is correct, $\alpha=\alpha_{g,t}$ satisfying \eqref{eq:riesz-uncond}, then for any measurable $m(\cdot)$,
\[
\E\!\left[\Ind\{G_i=\infty\}\,\alpha_{g,t}(X_{it})\,m(X_{it})\right]
=
\E\!\left[\Ind\{G_i=g\}\,m(X_{it})\right],
\]
and also
\[
\E\!\left[\Ind\{G_i=\infty\}\,\alpha_{g,t}(X_{it})\,Y_{it}\right]
=
\E\!\left[\Ind\{G_i=\infty\}\,\alpha_{g,t}(X_{it})\,m_0(X_{it})\right]
=
\E\!\left[\Ind\{G_i=g\}\,m_0(X_{it})\right],
\]
so the $m(\cdot)$ terms cancel and $\E[\psi_{g,t}(W_{it};\theta_0,m,\alpha_{g,t})]=0$.

Hence the label double robustness refers to the exact statement
\[
\E[\psi_{g,t}(W_{it};\theta_0,m,\alpha)]=0
\quad\text{if}\quad
m=m_0
\ \text{or}\
\alpha=\alpha_{g,t},
\]
with $\theta_0$ fixed at the cohort--time effect.

\paragraph{Orthogonality as a separate claim.}
Neyman orthogonality is an additional property, not a synonym for double robustness. At $(m,\alpha)=(m_0,\alpha_{g,t})$, the Gateaux derivative of the moment map
$\eta\mapsto \E[\psi_{g,t}(W_{it};\theta_0,m_0+\eta h,\alpha_{g,t})]$
vanishes for any direction $h\in\mathcal{H}$ because \eqref{eq:riesz-uncond} implies exact cancellation of the perturbation terms. This separates two concepts that can otherwise be conflated in exposition: double robustness is a correctness-under-one-nuisance property of the population moment in \eqref{eq:dr-score}, while orthogonality is the first-order insensitivity that delivers debiased behaviour when both nuisance components are estimated with regularisation and cross-fitting \citep{ChernozhukovNeweySingh2022}.

\paragraph{Estimating equation with cross-fitting.}
Let $\{\mathcal{I}_m\}_{m=1}^M$ be a fold partition. Construct nuisance estimates $\hat m^{(-m)}$ and $\hat\alpha^{(-m)}$ using only observations in $\mathcal{I}_m^{c}$. The debiased estimator $\widehat{\GATT}(g,t)$ solves
\begin{equation}\label{eq:dr-estimator}
\frac{1}{n}\sum_{i=1}^n
\psi_{g,t}\!\left(W_{it};\theta,\hat m^{(-m(i))},\hat\alpha^{(-m(i))}\right)=0,
\end{equation}
where $m(i)$ is the fold containing unit $i$. Under standard complexity control for the nuisance estimators and weak dependence conditions, \eqref{eq:dr-estimator} yields asymptotically normal inference for $\GATT(g,t)$ while permitting high-dimensional or nonparametric learning of $m_0$ and $\alpha_{g,t}$ \citep{ChernozhukovNeweySingh2022}.

\subsection{Sensitivity regions and falsification-adaptive reporting}\label{subsec:falsification-adaptive}

Section \ref{sec:sensitivity} replaces the sharp restriction $\delta_{g,t}\equiv 0$ by a maintained restriction class $\delta\in\Delta(\mathcal{R})$ and thereby replaces point identification of $\theta=\tau(k)$ by the identified set $\Theta(\mathcal{R})$. The family $\{\Delta(\mathcal{R})\}_{\mathcal{R}}$ is naturally ordered by inclusion: tighter restrictions correspond to smaller admissible deviation sets and therefore smaller identified sets, while weaker restrictions enlarge $\Theta(\mathcal{R})$. Reporting $\Theta(\mathcal{R})$ across calibrated values of $\mathcal{R}$ therefore yields falsification-adaptive reporting in the precise sense of \citet{MastenPoirier2021}: when the baseline restriction is inconsistent with the observed reduced-form implications, the analysis reports the minimal relaxation of $\mathcal{R}$ that restores non-falsification and the corresponding set of admissible values for the target parameter.

This set-based logic is conceptually distinct from, but complementary to, the design diagnostics in Section 5. Sensitivity regions quantify fragility with respect to counterfactual trend restrictions through admissible deviation paths $\delta$. By contrast, the indices $\mathcal{N}(k)$ and $\mathcal{C}(k)$ quantify fragility with respect to regression geometry. Even when Assumption \ref{ass:pt} holds exactly, TWFE event-study coefficients can be non-convex mixtures of heterogeneous cohort--horizon effects because residualisation by unit and time fixed effects induces sign changes and non-orthogonality among residualised event-time regressors, producing negative weights and cross-horizon loading in \eqref{eq:weights} \citep{SunAbraham2021,DeChaisemartinDHaultfoeuille2023}. Large values of $\mathcal{N}(k)$ and $\mathcal{C}(k)$ therefore constitute design-level evidence that the conventional estimator is not transparently interpretable as an event-time causal object, independent of any violations of parallel trends.

Taken together, the reporting protocol has two disciplined components. The first is a design report, which records $\mathcal{N}(k)$ and $\mathcal{C}(k)$ as ex ante properties of the adoption pattern and the event window. The second is an identification report, which records $\Theta(\mathcal{R})$ and its confidence region under calibrated restriction classes. The combination produces an empirically implementable, assumption-explicit workflow that is falsification-aware both at the level of the maintained identifying restrictions and at the level of estimator interpretability.

\clearpage
\section{Monte Carlo designs}\label{sec:montecarlo}

\noindent\textbf{Monte Carlo replication law.}
Each replication is defined by a joint law for \\
\(\big(X_i,G_i,\{D_{it}\}_{t=1}^T,\{Y_{it}\}_{t=1}^T\big)\): covariates \(X_i\) determine the conditional adoption-time distribution \(\Prob(G_i=g\mid X_i)\), the adoption time \(G_i\) determines the treatment path \(D_{it}\), and the outcome process is generated by specifying an untreated potential outcome \(Y_{it}(0)\) together with a dynamic effect profile \(\tau_g(k)\) that maps \((G_i,t)\) into observed outcomes \(Y_{it}\); Subsection \ref{subsec:mc-dgp} states the complete data-generating process.

\noindent The Monte Carlo section is defined by the replication law for \(\big(X_i,G_i,\{D_{it}\}_{t=1}^T,\{Y_{it}\}_{t=1}^T\big)\); covariates and confounded adoption are specified here, and the outcome data-generating process is specified explicitly in Subsection \ref{subsec:mc-dgp} through equations \eqref{eq:mc-y0}--\eqref{eq:mc-yobs}.

\paragraph{Covariates and confounded adoption.}
For each unit $i\in\{1,\dots,n\}$ generate a time-invariant covariate vector $X_i\in\R^{d_x}$, for example
\[
X_i\sim \mathcal{N}(0,\Sigma_X),
\qquad \Sigma_X\succ 0.
\]
To induce selection on observables, specify the conditional adoption-time distribution on
\(
\mathcal{G}\equiv\{1,\dots,T,\infty\}
\)
via a multinomial logit:
\begin{equation}\label{eq:mc-mnlogit}
\Prob(G_i=g\mid X_i)
=
\frac{\exp\!\big(X_i'\gamma_g\big)}
{\sum_{h\in\mathcal{G}}\exp\!\big(X_i'\gamma_h\big)},
\qquad g\in\mathcal{G},
\end{equation}
where $\{\gamma_g\}_{g\in\mathcal{G}}$ are nuisance coefficients chosen to create economically meaningful imbalance across cohorts (including non-trivial mass at $\infty$ through the $g=\infty$ index). Draw $G_i$ from \eqref{eq:mc-mnlogit} and define absorbing treatment as
\[
D_{it}=\Ind\{G_i\le t\}\Ind\{G_i<\infty\}.
\]
This construction makes the propensity score cohort- and covariate-dependent, so the density-ratio (odds-ratio) Riesz representer $\alpha_{g,t}(X)$ in \eqref{eq:alpha-scaled} is non-degenerate and the orthogonal-score machinery in Section \ref{subsec:debiased-score} is genuinely required for debiased estimation. Conditional on the realised adoption schedule $\{G_i\}$, the residualised design matrix $Z=M_XD$ (and hence the diagnostic indices $\mathcal{N}(k)$ and $\mathcal{C}(k)$) is fixed prior to simulating outcomes.

\subsection{Monte Carlo Design}\label{sec:mc_design}

\paragraph{Panel, timing, and cohorts (fixed across all DGPs).}
Units $i\in\{1,\dots,N\}$ with $N=5{,}000$; periods $t\in\{1,\dots,T\}$ with $T=12$.
Adoption time $A_i\in\{4,6,8,10,\infty\}$ with cohort set $\mathcal{G}=\{1,2,3,4\}$ and never-treated $g=0$.
Cohort shares: $\pi_0=0.20$ and $\pi_g=0.20$ for each $g\in\{1,2,3,4\}$.
Define $D_{it}=\mathbf{1}\{t\ge A_i,\,A_i<\infty\}$ and event time $\ell_{it}=t-A_i$.

\paragraph{Outcome equation (common envelope).}
\begin{equation}\label{eq:mc_envelope}
Y_{it}
=
\alpha_i+\lambda_t + \beta X_{it}
+ \tau_{g(i)}(\ell_{it}) D_{it}
+ u_{it}
+ v_{it}(\Delta(\mathcal{R}),B,\Gamma).
\end{equation}
Fixed effects: $\alpha_i\sim\mathcal{N}(0,1)$ and $\lambda_t\sim\mathcal{N}(0,0.25)$, independent.
Covariate: $X_{it}=\rho_X X_{i,t-1}+\xi_{it}$ with $\rho_X=0.50$, $X_{i1}\sim\mathcal{N}(0,1)$, $\xi_{it}\sim\mathcal{N}(0,1)$ i.i.d., and $\beta=1$.

\paragraph{Treatment-effect path (fixed functional form; cohort scaling).}
Let $h_1=0.80$, $h_2=1.00$, $h_3=1.20$, $h_4=1.40$ and define the baseline event-time profile
\[
m(\ell)=
\begin{cases}
0 & \ell<0,\\
0.50 & \ell=0,\\
0.75 & \ell=1,\\
1.00 & \ell\ge 2.
\end{cases}
\]
Then $\tau_g(\ell)=h_g\,m(\ell)$ for $g\in\{1,2,3,4\}$.

\paragraph{Violation indices and grid (fixed; applied to each DGP).}
\[
\Delta(\mathcal{R})\in\{0,\ 0.25,\ 0.50,\ 1.00\},\qquad
B\in\{0,\ 0.05,\ 0.10,\ 0.20\},\qquad
\Gamma\in\{0,\ 0.005,\ 0.010,\ 0.020\}.
\]
$B$ is in outcome units (calibrated to pre-trend magnitudes per Section~\ref{sec:calibration}); $\Gamma$ is drift-per-period (fractional) in a monotone family (Section~\ref{sec:calibration}).
Total cells per DGP: $K_\Delta K_B K_\Gamma = 4\cdot 4\cdot 4 = 64$.

\paragraph{Construction of the violation term (monotone in each index).}
Let $s_g\in\{+1,-1\}$ be fixed by cohort with $s_1=+1$, $s_2=-1$, $s_3=+1$, $s_4=-1$ and $s_0=0$ (never-treated).
Define the pre-period normaliser $\bar{t}_{\mathrm{pre}}(g)=\frac{1}{A_g-1}\sum_{t=1}^{A_g-1}\frac{t-1}{T-1}$ and the post-period normaliser $\bar{t}_{\mathrm{post}}(g)=\frac{1}{T-A_g+1}\sum_{t=A_g}^{T}\frac{t-A_g}{T-1}$.
Set
\begin{equation}\label{eq:mc_violation_construct}
\begin{split}
v_{it}(\Delta(\mathcal{R}),B,\Gamma)
=
\mathbf{1}\{g(i)\neq 0\}\, s_{g(i)}
\biggl[
&\Delta(\mathcal{R})\cdot B\left(\frac{t-1}{T-1}-\bar{t}_{\mathrm{pre}}(g(i))\right)\mathbf{1}\{t<A_i\}
\\
&+
\Gamma\left(\frac{t-A_i}{T-1}-\bar{t}_{\mathrm{post}}(g(i))\right)\mathbf{1}\{t\ge A_i\}
\biggr].
\end{split}
\end{equation}
This yields: (i) $v_{it}=0$ for never-treated; (ii) sign-controlled, cohort-specific violations; (iii) weak set expansion as any of $\Delta(\mathcal{R})$, $B$, or $\Gamma$ increases (Appendix~\ref{app:calibration_derivations} provides monotonicity statements and sufficient conditions).

\paragraph{DGPs (exact, fully specified).}
All DGPs use \eqref{eq:mc_envelope} and \eqref{eq:mc_violation_construct}; only $(u_{it},\lambda_t)$ dependence and selection differ.

\textbf{DGP1 (i.i.d.\ Gaussian shocks).}
\begin{equation}\label{eq:mc_dgp1_u}
u_{it}\sim\mathcal{N}(0,1)\ \text{i.i.d.}
\end{equation}

\textbf{DGP2 (serial correlation).}
\begin{equation}\label{eq:mc_dgp2_u}
u_{it} = \varphi u_{i,t-1}+\varepsilon_{it},\qquad
\varphi=0.50,\qquad
\varepsilon_{it}\sim\mathcal{N}(0,1-\varphi^2)\ \text{i.i.d.},\qquad
u_{i1}\sim\mathcal{N}(0,1).
\end{equation}

\textbf{DGP3 (attrition/selection).}
Observed outcomes are $Y_{it}\mathbf{1}\{S_{it}=1\}$ with
\begin{equation}\label{eq:mc_dgp3_select}
S_{it}\sim\mathrm{Bernoulli}(p_{it}),\qquad
\mathrm{logit}(p_{it})= \eta_0+\eta_1 D_{it}+\eta_2 Y_{i,t-1},
\end{equation}
where $(\eta_0,\eta_1,\eta_2)=(1.25,\ -0.35,\ -0.10)$ and $Y_{i0}=0$.
Shock component is heavy-tailed:
\begin{equation}\label{eq:mc_dgp3_u}
u_{it}\sim t_{\nu}(0,1)\ \text{i.i.d.},\qquad \nu=5.
\end{equation}

\paragraph{Replications and randomisation.}
For each DGP and each $(\Delta(\mathcal{R}),B,\Gamma)$ cell, run $R=2{,}000$ independent replications with fresh draws of $(\alpha_i,\lambda_t,X_{it},u_{it})$ and (for DGP3) $\{S_{it}\}$.


\begin{table}[!htbp]
\centering
\caption{Monte Carlo design matrix.}\label{tab:mc_design_matrix}
\begin{tabularx}{\textwidth}{lX} 
\hline
Component & Values used in all runs \\
\hline
Panel size & $N=5{,}000$ units; $T=12$ periods \\
Cohorts & $A\in\{4,6,8,10,\infty\}$; shares $(0.20,0.20,0.20,0.20,0.20)$ \\
Treatment effects & $\tau_g(\ell)=h_g m(\ell)$ with $h=(0.80,1.00,1.20,1.40)$ and $m(0)=0.50,\ m(1)=0.75,\ m(\ell\ge2)=1$ \\
Covariate & $X_{it}$ AR(1) with $\rho_X=0.50$, $\beta=1$ \\
DGPs & DGP1: i.i.d.\ Normal; DGP2: AR(1) shocks with $\varphi=0.50$; DGP3: $t_5$ shocks + selection \\
Violation indices & $\Delta(\mathcal{R})\in\{0,0.25,0.50,1.00\}$; $B\in\{0,0.05,0.10,0.20\}$; $\Gamma\in\{0,0.005,0.010,0.020\}$ \\
Cells per DGP & $4\times 4\times 4 = 64$ \\
Replications & $R=2{,}000$ per cell \\
\hline
\end{tabularx}
\end{table}

\begin{table}[!htbp]
\centering
\caption{Operational meaning of $(\Delta(\mathcal{R}),B,\Gamma)$ in the simulation.}\label{tab:mc_param_meaning}
\begin{tabular}{lll}
\hline
Index & Implemented in $v_{it}$ & Interpretable unit \\
\hline
$\Delta(\mathcal{R})$ & scales the pre-period component in \eqref{eq:mc_violation_construct} & restriction-class relaxation intensity \\
$B$ & pre-period slope magnitude in \eqref{eq:mc_violation_construct} & outcome-units pre-trend amplitude \\
$\Gamma$ & post-period drift in \eqref{eq:mc_violation_construct} & fractional drift per period \\
\hline
\end{tabular}
\end{table}

\begin{figure}[!htbp]
\centering
\begin{tikzpicture}[
  x=0.85cm, y=0.9cm, 
  >=Latex,
  cohort/.style={draw, rounded corners, fill=white, inner sep=3pt, font=\small, anchor=east}, 
  tick/.style={draw, line width=0.4pt},
  adopt/.style={draw, circle, inner sep=1.5pt, fill=black},
  guide/.style={draw, dashed, black!50, line width=0.4pt},
  label/.style={font=\small},
  event_axis/.style={->, line width=0.6pt, color=blue!60!black}
]

\draw[->, line width=0.6pt] (1, -0.5) -- (13.5, -0.5) node[right, label] {$t$};
\foreach \t in {1,...,12}{
  \draw[tick] (\t, -0.4) -- (\t, -0.6);
  \node[label, below] at (\t, -0.65) {$\t$};
}


\node[cohort] at (0.8, 4.0) {never ($A=\infty$)};
\draw[line width=0.6pt] (1.0, 4.0) -- (12.8, 4.0);

\node[cohort] at (0.8, 3.0) {$g=1$ ($A=4$)};
\draw[line width=0.6pt] (1.0, 3.0) -- (12.8, 3.0);
\node[adopt] at (4, 3.0) {};
\draw[guide] (4, 4.2) -- (4, -0.5); 
\node[label, above] at (4, 4.2) {$a_1$};

\node[cohort] at (0.8, 2.0) {$g=2$ ($A=6$)};
\draw[line width=0.6pt] (1.0, 2.0) -- (12.8, 2.0);
\node[adopt] at (6, 2.0) {};
\draw[guide] (6, 4.2) -- (6, -0.5);
\node[label, above] at (6, 4.2) {$a_2$};

\draw[event_axis] (6, 1.3) -- (12.8, 1.3) node[right, label] {$\ell=t-A$};
\foreach \k/\x in {0/6, 1/7, 2/8, 3/9, 4/10, 5/11, 6/12}{
  \draw[tick, color=blue!60!black] (\x, 1.4) -- (\x, 1.2);
  \node[label, below, color=blue!60!black] at (\x, 1.2) {$\k$};
}

\node[cohort] at (0.8, 0.6) {$g=3$ ($A=8$)}; 
\draw[line width=0.6pt] (1.0, 0.6) -- (12.8, 0.6);
\node[adopt] at (8, 0.6) {};
\draw[guide] (8, 4.2) -- (8, -0.5);
\node[label, above] at (8, 4.2) {$a_3$};

\node[cohort] at (0.8, -0.4) {$g=4$ ($A=10$)};
\draw[line width=0.6pt] (1.0, -0.4) -- (12.8, -0.4);
\node[adopt] at (10, -0.4) {};
\draw[guide] (10, 4.2) -- (10, -0.5);
\node[label, above] at (10, 4.2) {$a_4$};

\end{tikzpicture}
\caption{Cohort timing used in every Monte Carlo run: $T=12$ and $A\in\{4,6,8,10,\infty\}$.}\label{fig:mc_timing_fixed}
\end{figure}
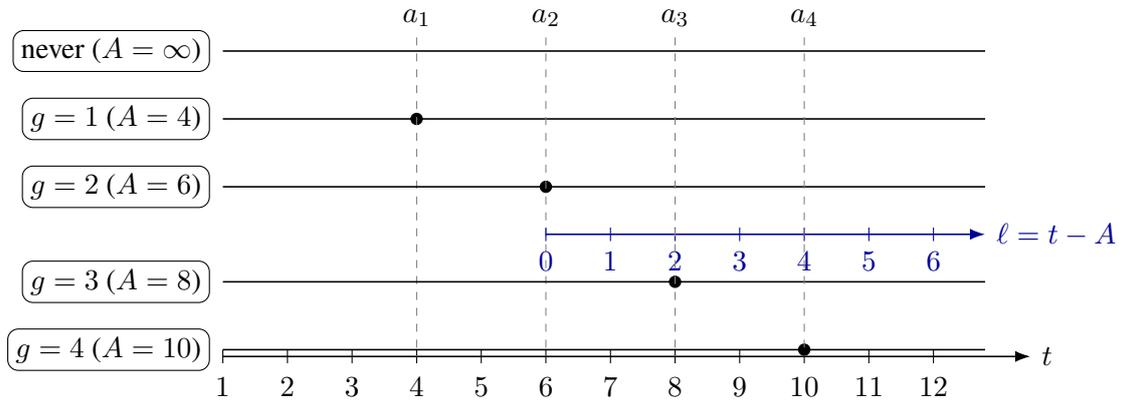

\begin{figure}[!htbp]
\centering
\begin{tikzpicture}[
  x=0.95cm, y=0.95cm, 
  >=Latex,
  ax/.style={line width=0.6pt, ->},
  gridline/.style={line width=0.35pt, color=gray!50},
  pt/.style={circle, fill=black, inner sep=1.5pt},
  lab/.style={font=\footnotesize}
]


\foreach \p/\dx/\dy/\dval in {
  1/0/6/0,     
  2/6/6/0.25,  
  3/0/0/0.50,  
  4/6/0/1.00   
}{
  \begin{scope}[shift={(\dx,\dy)}]
    \draw[ax] (0,0) -- (0,4.5) node[left, lab] {$\Gamma$};
    \draw[ax] (0,0) -- (4.8,0) node[below, lab] {$B$};

    \foreach \x/\lbl in {0/0, 1/0.05, 2/0.10, 4/0.20}{
      \draw[black] (\x,0.1) -- (\x,-0.1); 
      \node[lab, below] at (\x,-0.2) {\lbl};
    }
    
    \foreach \y/\lbl in {0/0, 1/0.005, 2/0.010, 4/0.020}{
      \draw[black] (0.1,\y) -- (-0.1,\y); 
      \node[lab, left] at (-0.2,\y) {\lbl};
    }

    \foreach \x in {1,2,4} \draw[gridline] (\x,0) -- (\x,4.2);
    \foreach \y in {1,2,4} \draw[gridline] (0,\y) -- (4.5,\y);

    \foreach \x in {0,1,2,4}{
      \foreach \y in {0,1,2,4}{
        \node[pt] at (\x,\y) {};
      }
    }

    \node[lab, font=\bfseries] at (2.25, 4.8) {$\Delta(\mathcal{R})=\dval$};
  \end{scope}
}
\end{tikzpicture}
\caption{Violation grid for $(B,\Gamma)$ shown separately for each $\Delta(\mathcal{R})\in\{0,0.25,0.50,1.00\}$.}\label{fig:mc_violation_grid}
\end{figure}
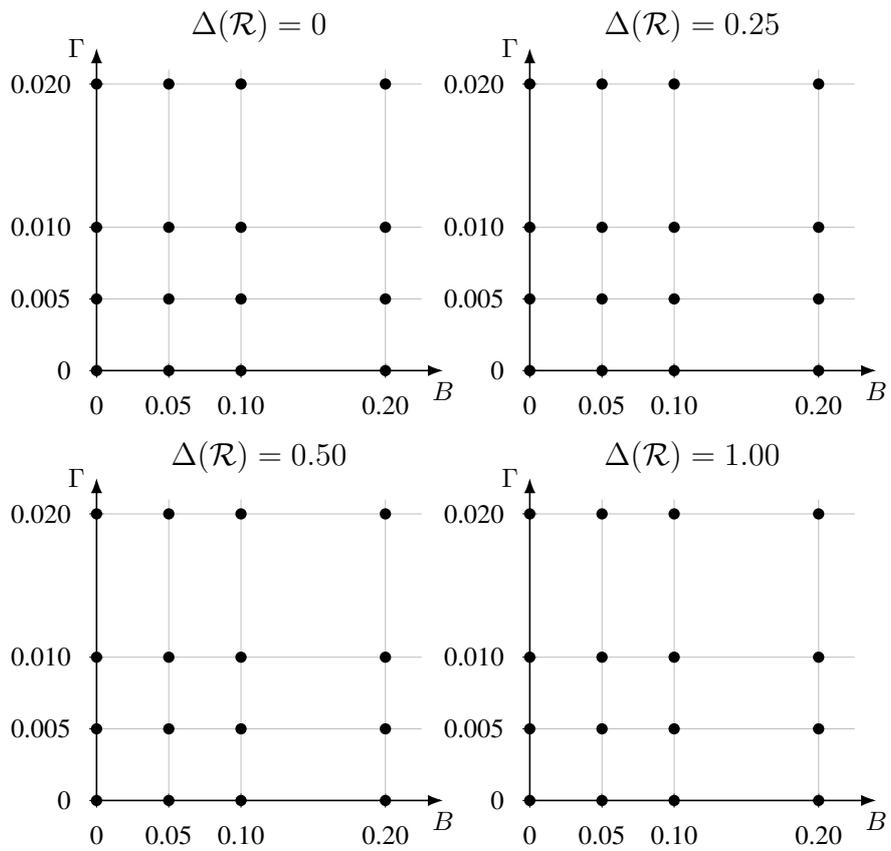

\begin{figure}[!htbp]
\centering
\begin{tikzpicture}[
  node distance=10mm and 8mm,
  box/.style={
    draw, 
    rounded corners, 
    align=center, 
    inner sep=5pt, 
    font=\small, 
    text width=3.5cm, 
    minimum height=1.6cm
  },
  arr/.style={->, line width=0.6pt, >=Latex}
]

\node[box] (cell) {Choose cell\\(DGP, $\Delta(\mathcal{R})$, $B$, $\Gamma$)};
\node[box, right=of cell] (draw) {Draw data\\$\{\alpha_i,\lambda_t,X_{it},u_{it}\}$\\and (DGP3) $\{S_{it}\}$};
\node[box, right=of draw] (build) {Construct $Y_{it}$\\via \eqref{eq:mc_envelope}\\and \eqref{eq:mc_violation_construct}};

\node[box, below=of build] (estimate) {Apply estimator set\\(fixed in Section~\ref{sec:mc_estimators})};
\node[box, below=of draw] (score) {Compute metrics\\bias, RMSE, coverage\\length, placebo rejection};
\node[box, below=of cell] (repeat) {Repeat $R=2{,}000$\\replications per cell\\Aggregate to Tables};

\draw[arr] (cell) -- (draw);
\draw[arr] (draw) -- (build);

\draw[arr] (build) -- (estimate);

\draw[arr] (estimate) -- (score);
\draw[arr] (score) -- (repeat);

\draw[arr] (repeat) -- (cell);

\end{tikzpicture}
\caption{Monte Carlo execution loop used for each of the $3\times 64$ design cells.}\label{fig:mc_loop}
\end{figure}
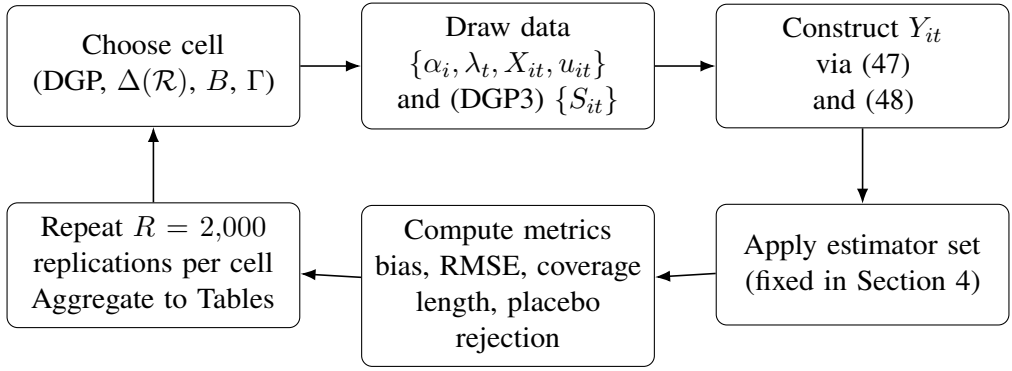

\subsection{Data-generating processes}\label{sec:mc_dgps}

\paragraph{Common structure (all DGPs).}
\begin{equation}\label{eq:mc_common_outcome}
Y_{it}
=
\alpha_i + \lambda_t + X_{it}^{\top}\beta
+ \tau_{g(i)}(\ell_{it}) D_{it}
+ u_{it}
+ v_{it}(\Delta(\mathcal{R}),B,\Gamma),
\end{equation}
where $A_i$ is adoption time, $\ell_{it}=t-A_i$ is event time, and $v_{it}(\Delta(\mathcal{R}),B,\Gamma)$ is the violation component satisfying monotone expansion in each index.

\paragraph{Treatment timing and cohort composition (all DGPs).}
Cohorts $g\in\{1,\dots,G\}$ adopt at $a_g\in\{2,\dots,T\}$ with $a_1<\cdots<a_G$; never-treated have $A_i=\infty$; shares satisfy $\sum_{g=0}^G\pi_g=1$.

\paragraph{Violation component family (all DGPs).}
$v_{it}(\Delta(\mathcal{R}),B,\Gamma)$ is constructed to satisfy: (i) $|v_{it}|$ weakly increases in $\Delta(\mathcal{R})$, $B$, and $\Gamma$; (ii) $B$ is tied to pre-trend magnitudes; (iii) $\Gamma$ is tied to a monotone drift-per-period family; (iv) $\Delta(\mathcal{R})$ indexes restriction-class relaxations. Calibration mappings are in Section~\ref{sec:calibration} and Appendix~\ref{app:calibration_derivations}.

\paragraph{DGP1: Baseline staggered adoption with i.i.d.\ shocks.}
\begin{equation}\label{eq:mc_dgp1_shock}
u_{it}\sim \mathcal{N}(0,\sigma^2)\ \text{i.i.d.}
\end{equation}
\begin{equation}\label{eq:mc_dgp1_tau}
\tau_{g}(\ell)=\tau_0\cdot \mathbf{1}\{\ell\ge 0\}
\quad\text{or}\quad
\tau_{g}(\ell)=\tau_0\left(1-e^{-\rho\ell}\right)\mathbf{1}\{\ell\ge 0\}.
\end{equation}
\begin{equation}\label{eq:mc_dgp1_violation}
v^{(1)}_{it}(\Delta(\mathcal{R}),B,\Gamma)
=
B\cdot b^{(1)}_{g(i)} \cdot \mathbf{1}\{A_i<\infty\}
+
\Gamma\cdot \gamma^{(1)}_{g(i)} \cdot (t-1)\cdot \mathbf{1}\{A_i<\infty\},
\end{equation}
with fixed sign pattern $(b^{(1)}_{g},\gamma^{(1)}_{g})$ scaled in calibrated units.

\paragraph{DGP2: Serial correlation and cohort-specific trends.}
\begin{equation}\label{eq:mc_dgp2_shock}
u_{it}=\varphi u_{i,t-1}+\varepsilon_{it},\qquad \varepsilon_{it}\sim \mathcal{N}(0,\sigma_\varepsilon^2)\ \text{i.i.d.}
\end{equation}
Add a cohort trend term $\kappa_{g(i)}t$ to \eqref{eq:mc_common_outcome}. Treatment effects:
\begin{equation}\label{eq:mc_dgp2_tau}
\tau_{g}(\ell)=\tau_{0,g} + \tau_{1,g}\ell + \tau_{2,g}\mathbf{1}\{\ell\ge L^\star\},\qquad \ell\ge 0.
\end{equation}
Violation:
\begin{equation}\label{eq:mc_dgp2_violation}
v^{(2)}_{it}(\Delta(\mathcal{R}),B,\Gamma)
=
\Delta(\mathcal{R}) \cdot r^{(2)}_{it}
+ \Gamma \cdot \gamma^{(2)}_{g(i)} \cdot t \cdot \mathbf{1}\{A_i<\infty\},
\end{equation}
where $r^{(2)}_{it}$ is admissible under the relaxed restriction class indexed by $\Delta(\mathcal{R})$.

\paragraph{DGP3: Attrition/selection with stronger heterogeneity.}
Observed outcome is $Y_{it}$ only when $S_{it}=1$, with
\begin{equation}\label{eq:mc_dgp3_selection}
S_{it}\sim \mathrm{Bernoulli}(p_{it}),\qquad
\mathrm{logit}(p_{it}) = \eta_0 + \eta_1 D_{it} + \eta_2 Y_{i,t-1}.
\end{equation}
Baseline shocks:
\begin{equation}\label{eq:mc_dgp3_shock}
\varepsilon_{it}\sim t_{\nu}(0,\sigma)\ \text{i.i.d.}\quad \text{(optional AR(1) in }u_{it}\text{)}.
\end{equation}
Violation:
\begin{equation}\label{eq:mc_dgp3_violation}
v^{(3)}_{it}(\Delta(\mathcal{R}),B,\Gamma)
=
B \cdot b^{(3)}_{it}
+ \Gamma \cdot \gamma^{(3)}_{it},
\end{equation}
with $(b^{(3)}_{it},\gamma^{(3)}_{it})$ constructed in calibrated units and monotone in $(B,\Gamma)$.

\begin{table}[!htbp]
\centering
\caption{DGP specifications and baseline parameters.}\label{tab:mc_dgp_params}
\resizebox{\textwidth}{!}{%
\begin{tabular}{lcccccc}
\hline
DGP & $u_{it}$ & $\tau_g(\ell)$ & $X_{it}$ & Trend terms & Attrition & $v_{it}(\Delta(\mathcal{R}),B,\Gamma)$ \\
\hline
1 & i.i.d.\ Normal & static/ramp & optional & none & none & \eqref{eq:mc_dgp1_violation} \\
2 & AR(1) & dynamic + cohort het. & optional & cohort trends & none & \eqref{eq:mc_dgp2_violation} \\
3 & $t_\nu$ (opt.\ AR) & strong dynamic het. & optional & optional & logistic & \eqref{eq:mc_dgp3_violation} \\
\hline
\end{tabular}%
}
\end{table}

\begin{table}[!htbp]
\centering
\caption{Violation grid over $(\Delta(\mathcal{R}),B,\Gamma)$ applied to each DGP.}\label{tab:mc_violation_grid}
\begin{tabular}{lccc}
\hline
Index & Grid values & Construction anchor & Unit \\
\hline
$\Delta(\mathcal{R})$ & $\{\Delta_1,\dots,\Delta_{K_\Delta}\}$ & restriction-class ladder & restriction step \\
$B$ & $\{B_1,\dots,B_{K_B}\}$ & pre-trend calibration rule & outcome units \\
$\Gamma$ & $\{\Gamma_1,\dots,\Gamma_{K_\Gamma}\}$ & monotone drift family & percent per period \\
\hline
\end{tabular}
\end{table}

\begin{figure}[!htbp]
\centering
\begin{tikzpicture}[
  x=0.95cm, y=1.2cm, 
  >=Latex,
  cohort_label/.style={draw, rounded corners, fill=white, font=\small, anchor=east, align=right, inner sep=3pt},
  timeline/.style={draw, line width=0.7pt},
  tick/.style={draw, line width=0.5pt},
  adopt/.style={draw, circle, inner sep=1.8pt, fill=black},
  guide/.style={draw, dashed, black!50, line width=0.5pt},
  event_axis/.style={->, line width=0.6pt, color=blue!40!black}
]

\draw[->, line width=0.8pt] (1, 0) -- (11.5, 0) node[right, font=\small] {$t$};
\foreach \t in {1,...,11} {
  \draw[tick] (\t, 0.1) -- (\t, -0.1);
  \node[below, font=\footnotesize] at (\t, -0.15) {\t};
}
\node[below, font=\small] at (6.25, -0.5) {Calendar Time};

\def\aone{3}
\def\atwo{6}
\def\aG{9}


\node[cohort_label] at (0.8, 4.5) {Never-treated\\($A_i=\infty$)};
\draw[timeline] (1, 4.5) -- (11, 4.5);

\node[cohort_label] at (0.8, 3.5) {Cohort $g=1$\\($A_i=a_1$)};
\draw[timeline] (1, 3.5) -- (11, 3.5);
\node[adopt] (ad1) at (\aone, 3.5) {};
\draw[guide] (ad1) -- (\aone, 0); 
\node[above, font=\footnotesize] at (\aone, 3.6) {$a_1$};

\node[cohort_label] at (0.8, 2.5) {Cohort $g=2$\\($A_i=a_2$)};
\draw[timeline] (1, 2.5) -- (11, 2.5);
\node[adopt] (ad2) at (\atwo, 2.5) {};
\draw[guide] (ad2) -- (\atwo, 0);
\node[above, font=\footnotesize] at (\atwo, 2.6) {$a_2$};

\draw[event_axis] (\atwo, 1.8) -- (11, 1.8) node[right, font=\footnotesize] {$\ell$};
\foreach \k in {0,1,2,3,4} {
    \pgfmathsetmacro{\xpos}{\atwo+\k}
    \draw[tick, color=blue!40!black] (\xpos, 1.9) -- (\xpos, 1.7);
    \node[below, font=\scriptsize, color=blue!40!black] at (\xpos, 1.7) {$\k$};
}
\node[left, font=\footnotesize, color=blue!40!black, align=right] at (\atwo-0.2, 1.8) {Event Time\\$\ell = t - a_2$};
\draw[guide] (\atwo, 2.5) -- (\atwo, 1.8); 

\node[cohort_label] at (0.8, 0.8) {Cohort $g=G$\\($A_i=a_G$)};
\draw[timeline] (1, 0.8) -- (11, 0.8);
\node[adopt] (adG) at (\aG, 0.8) {};
\draw[guide] (adG) -- (\aG, 0);
\node[above, font=\footnotesize] at (\aG, 0.9) {$a_G$};

\node[anchor=north east, draw, rounded corners, fill=white!90!gray, inner sep=4pt] at (11.5, 5.2) {
  \begin{tikzpicture}[baseline, y=0.5cm]
    \node[adopt, label={[font=\scriptsize]right:Adoption ($A_i$)}] at (0,1) {};
    \draw[guide] (0,0) -- (0.5,0) node[right, font=\scriptsize, black] {Alignment guide};
  \end{tikzpicture}
};

\end{tikzpicture}
\caption{Staggered adoption timing, cohort indexing, and event-time definition $\ell=t-A_i$ used in the Monte Carlo.}\label{fig:mc_timing}
\end{figure}
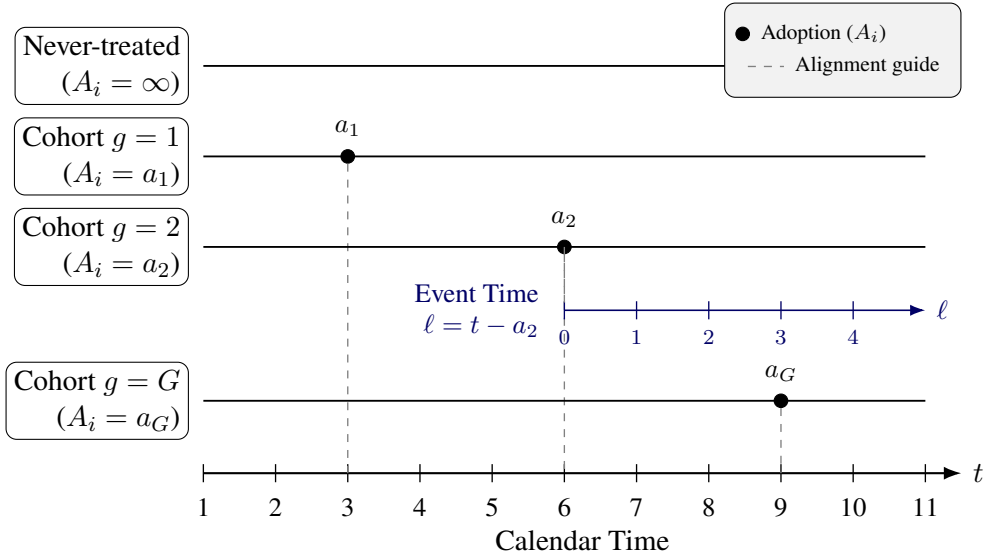

\subsection{Performance Metrics and Reporting}\label{sec:mc_metrics}

\paragraph{Target estimands (fixed).}
For each cohort $g\in\{1,2,3,4\}$ and event time $\ell\in\{0,1,2,3\}$ define the cohort-event ATT
\begin{equation}\label{eq:mc_ATT_gell}
\theta_{g,\ell} := \mathbb{E}\!\left[Y_{it}(1)-Y_{it}(0)\mid A_i=A_g,\ \ell_{it}=\ell\right].
\end{equation}
Define the pooled post effect (reporting target) as an average over cohorts and post event times
\begin{equation}\label{eq:mc_ATT_pooled}
\theta^{\star} := \sum_{g=1}^4 w_g \cdot \frac{1}{4}\sum_{\ell=0}^3 \theta_{g,\ell},
\qquad
w_g=\frac{\pi_g}{\sum_{h=1}^4 \pi_h}=\frac{1}{4}.
\end{equation}
Truth is known under the design because $\tau_g(\ell)$ is fixed and the violation term \eqref{eq:mc_violation_construct} is fully specified; for each DGP and each $(\Delta(\mathcal{R}),B,\Gamma)$ cell,
\begin{equation}\label{eq:mc_truth}
\theta_{g,\ell}^{\text{true}} = \tau_g(\ell)\quad\text{and}\quad
\theta^{\star,\text{true}} = \sum_{g=1}^4 w_g \cdot \frac{1}{4}\sum_{\ell=0}^3 \tau_g(\ell),
\end{equation}
so the Monte Carlo summaries benchmark against \eqref{eq:mc_truth}.

\paragraph{Point-estimate accuracy metrics (per cell).}
For replication $r\in\{1,\dots,R\}$ let $\widehat{\theta}^{(r)}$ denote the pooled estimate of $\theta^\star$ from a given estimator, and let $\widehat{\theta}^{(r)}_{g,\ell}$ denote the cohort-event estimate.
Define cell-wise bias, RMSE, and median absolute error (MAE):
\begin{align}
\mathrm{Bias} &:= \frac{1}{R}\sum_{r=1}^R\left(\widehat{\theta}^{(r)}-\theta^{\star,\text{true}}\right), \label{eq:mc_bias}\\
\mathrm{RMSE} &:= \left\{\frac{1}{R}\sum_{r=1}^R\left(\widehat{\theta}^{(r)}-\theta^{\star,\text{true}}\right)^2\right\}^{1/2}, \label{eq:mc_rmse}\\
\mathrm{MedAE} &:= \mathrm{median}_{1\le r\le R}\left|\widehat{\theta}^{(r)}-\theta^{\star,\text{true}}\right|. \label{eq:mc_medae}
\end{align}
Parallel definitions apply to each $(g,\ell)$ component $\widehat{\theta}^{(r)}_{g,\ell}$ using $\theta^{\text{true}}_{g,\ell}=\tau_g(\ell)$.

\paragraph{Interval metrics (per cell).}
Let $I^{(r)}=[L^{(r)},U^{(r)}]$ be the $1-\alpha$ confidence interval for $\theta^\star$ delivered by a method.
Define empirical coverage, average length, and upper-tail excess length (to detect one-sided over-conservatism):
\begin{align}
\mathrm{Cov} &:= \frac{1}{R}\sum_{r=1}^R \mathbf{1}\!\left\{\theta^{\star,\text{true}}\in I^{(r)}\right\}, \label{eq:mc_cov}\\
\mathrm{Len} &:= \frac{1}{R}\sum_{r=1}^R \left(U^{(r)}-L^{(r)}\right), \label{eq:mc_len}\\
\mathrm{UTEx} &:= \frac{1}{R}\sum_{r=1}^R \left(U^{(r)}-\theta^{\star,\text{true}}\right) - \frac{1}{R}\sum_{r=1}^R \left(\theta^{\star,\text{true}}-L^{(r)}\right). \label{eq:mc_utex}
\end{align}
For component-wise reporting, define $\mathrm{Cov}_{g,\ell}$ and $\mathrm{Len}_{g,\ell}$ analogously.

\paragraph{Placebo pre-trend rejection (per cell).}
Define a pre-period window for each cohort $g$ as $\ell\in\{-3,-2,-1\}$ and let $\widehat{\theta}^{(r)}_{g,\ell}$ denote placebo event-study coefficients.
Let $T^{(r)}_{\mathrm{pre}}$ be the Wald test statistic for the joint null $H_0:\ \theta_{g,-3}=\theta_{g,-2}=\theta_{g,-1}=0$ (pooled across cohorts using the same weighting as \eqref{eq:mc_ATT_pooled}).
Define placebo rejection rate:
\begin{equation}\label{eq:mc_placebo_rej}
\mathrm{RejPre} := \frac{1}{R}\sum_{r=1}^R \mathbf{1}\!\left\{p^{(r)}_{\mathrm{pre}}<\alpha\right\},
\end{equation}
reported at $\alpha\in\{0.10,0.05,0.01\}$.

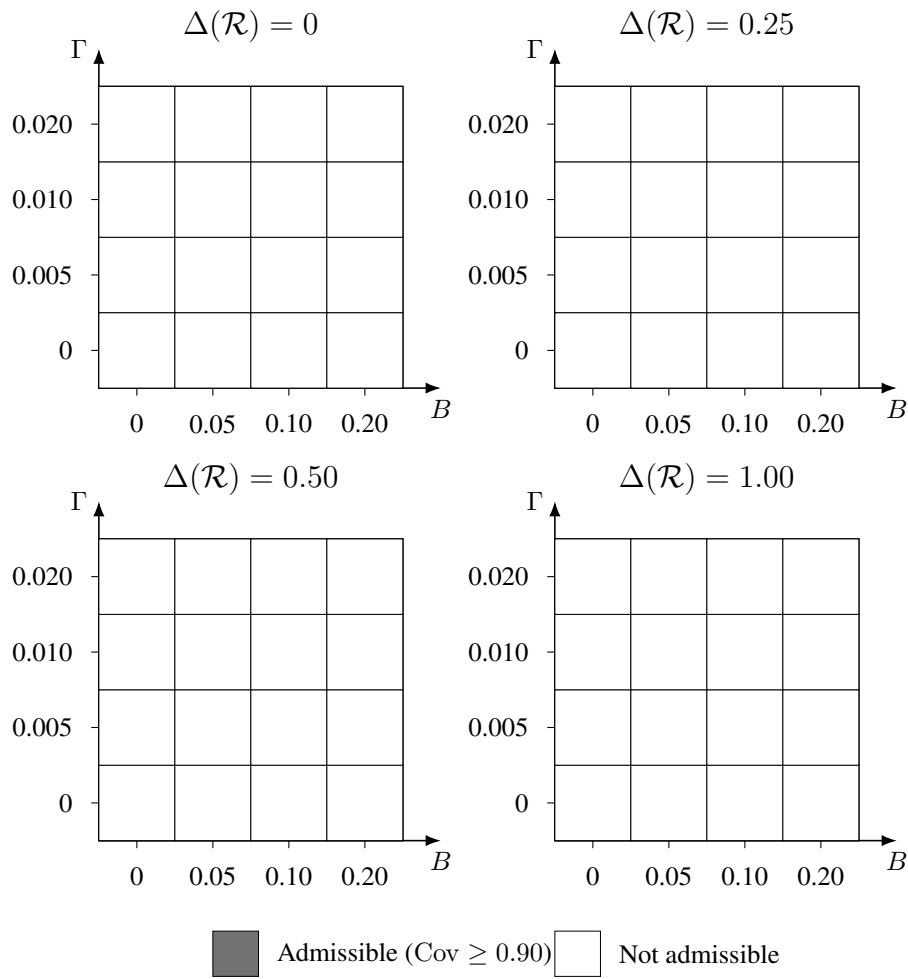
\begin{figure}[!htbp]
\centering
\begin{tikzpicture}[
  x=1.0cm, y=1.0cm, 
  lab/.style={font=\footnotesize},
  ax/.style={->, line width=0.6pt, >=Latex},
  tile/.style={draw, line width=0.35pt},
  ok/.style={fill=black!55},
  bad/.style={fill=white},
  frame/.style={draw, line width=0.6pt}
]

\newcommand{\Acell}[3]{%
  \ifcsname #3\endcsname
    \csname #3\endcsname
  \else
    \def\__fillstyle{bad} 
  \fi
  \path[tile,\__fillstyle] (#1,#2) rectangle ++(1,1);
}


\foreach \dx/\dy/\dval/\tag in {
  0/6/0/Dzero,     
  6/6/0.25/Dqtr,   
  0/0/0.50/Dhalf,  
  6/0/1.00/Done    
}{
  \begin{scope}[shift={(\dx,\dy)}]
    \draw[ax] (0,0) -- (0,4.5) node[left, lab] {$\Gamma$};
    \draw[ax] (0,0) -- (4.5,0) node[below, lab] {$B$};

    \foreach \x/\lbl in {0/0, 1/0.05, 2/0.10, 3/0.20}{
      \draw (\x+0.5,0.1) -- (\x+0.5,-0.1);
      \node[lab, below] at (\x+0.5,-0.2) {\lbl};
    }
    \foreach \y/\lbl in {0/0, 1/0.005, 2/0.010, 3/0.020}{
      \draw (0.1,\y+0.5) -- (-0.1,\y+0.5);
      \node[lab, left] at (-0.2,\y+0.5) {\lbl};
    }

    \draw[frame] (0,0) rectangle (4,4);

    \foreach \bx in {0,1,2,3}{
      \foreach \gy in {0,1,2,3}{
        \edef\keyname{A_\tag_\bx_\gy}
        \Acell{\bx}{\gy}{\keyname}
      }
    }

    \node[lab, font=\bfseries] at (2.0, 4.8) {$\Delta(\mathcal{R})=\dval$};
  \end{scope}
}

\begin{scope}[shift={(1.5, -1.8)}]
  \path[tile,ok] (0,0) rectangle ++(0.6,0.6);
  \node[lab, right] at (0.7,0.3) {Admissible ($\mathrm{Cov}\ge0.90$)};
  
  \path[tile,bad] (4.5,0) rectangle ++(0.6,0.6);
  \node[lab, right] at (5.2,0.3) {Not admissible};
\end{scope}

\end{tikzpicture}
\caption{Robustness frontier visualisation: admissible cells under \eqref{eq:mc_admissible} for each $\Delta(\mathcal{R})$. Tile fills are intended to be set from the computed Monte Carlo results (white by default).}\label{fig:mc_frontier}
\end{figure}

\paragraph{Robustness frontier (per DGP).}
For each DGP and each estimator, define the robustness frontier as the set of $(\Delta(\mathcal{R}),B,\Gamma)$ cells achieving (i) target coverage and (ii) bounded interval length inflation.
Let $\underline{c}=0.90$ and $\overline{\kappa}=2.5$. Let $\mathrm{Len}_0$ denote average length at $(\Delta(\mathcal{R}),B,\Gamma)=(0,0,0)$.
Define admissibility indicator
\begin{equation}\label{eq:mc_admissible}
\mathbb{A}(\Delta(\mathcal{R}),B,\Gamma)
:=
\mathbf{1}\!\left\{\mathrm{Cov}\ge \underline{c}\right\}
\cdot
\mathbf{1}\!\left\{\mathrm{Len}\le \overline{\kappa}\,\mathrm{Len}_0\right\}.
\end{equation}
Frontier plots (Figure~\ref{fig:mc_frontier}) visualise $\mathbb{A}$ as a function of $(B,\Gamma)$ for each $\Delta(\mathcal{R})$.

\begin{table}[!htbp]
\centering
\caption{Performance summaries to report per DGP $\times$ estimator $\times$ $(\Delta(\mathcal{R}),B,\Gamma)$ cell.}\label{tab:mc_cov_len_placebo}
\begin{tabular}{ll}
\hline
Quantity & Definition (computed over $R=2{,}000$ replications) \\
\hline
Coverage ($\mathrm{Cov}$) & \eqref{eq:mc_cov} at nominal $1-\alpha$ (report $\alpha=0.10,0.05$) \\
Avg.\ interval length ($\mathrm{Len}$) & \eqref{eq:mc_len} \\
Upper-tail excess ($\mathrm{UTEx}$) & \eqref{eq:mc_utex} \\
Placebo rejection ($\mathrm{RejPre}$) & \eqref{eq:mc_placebo_rej} at $\alpha\in\{0.10,0.05,0.01\}$ \\
\hline
\end{tabular}
\end{table}

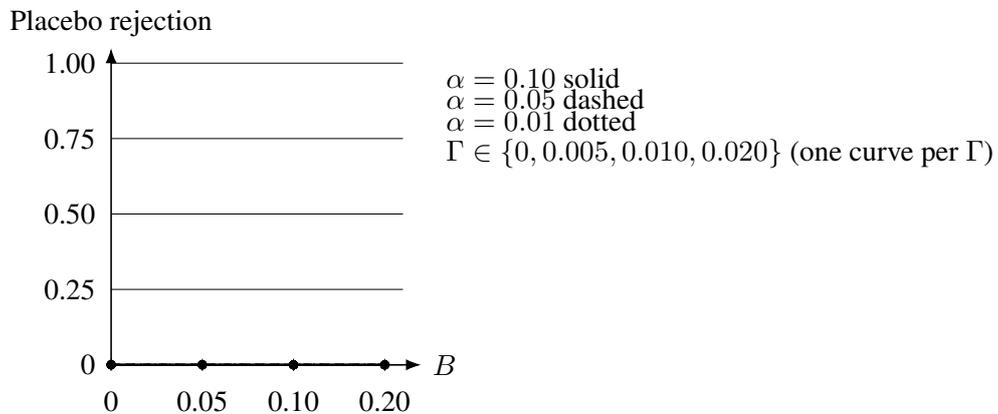
\begin{figure}[!htbp]
\centering
\begin{tikzpicture}[
  x=1.20cm, y=4.0cm,
  lab/.style={font=\small},
  ax/.style={->, line width=0.6pt, >=Latex},
  gridline/.style={draw, line width=0.35pt},
  curve/.style={line width=0.7pt},
  marker/.style={circle, fill=black, inner sep=1.2pt}
]
\draw[ax] (0,0) -- (0,1.05) node[above, lab] {Placebo rejection};
\draw[ax] (0,0) -- (3.4,0) node[right, lab] {$B$};

\foreach \x/\lbl in {0/0,1/0.05,2/0.10,3/0.20}{
  \draw (\x,0.02) -- (\x,-0.02);
  \node[lab, below] at (\x,-0.05) {\lbl};
}
\foreach \y/\lbl in {0/0,0.25/0.25,0.50/0.50,0.75/0.75,1.0/1.00}{
  \draw[gridline] (0,\y) -- (3.2,\y);
  \node[lab, left] at (-0.05,\y) {\lbl};
}

\newcommand{\Pval}[3]{%
  \ifcsname P_#1_#2_#3\endcsname
    \csname P_#1_#2_#3\endcsname
  \else
    0
  \fi
}

\foreach \atag/\sty in {a10/solid, a05/dashed, a01/dotted}{
  \foreach \gtag in {g0,g5,g10,g20}{
    \draw[curve,\sty]
      (0,{\Pval{\atag}{\gtag}{0}}) --
      (1,{\Pval{\atag}{\gtag}{1}}) --
      (2,{\Pval{\atag}{\gtag}{2}}) --
      (3,{\Pval{\atag}{\gtag}{3}});
    \foreach \bx in {0,1,2,3}{
      \node[marker] at (\bx,{\Pval{\atag}{\gtag}{\bx}}) {};
    }
  }
}

\node[lab, anchor=west] at (3.55,0.95) {$\alpha=0.10$ solid};
\node[lab, anchor=west] at (3.55,0.88) {$\alpha=0.05$ dashed};
\node[lab, anchor=west] at (3.55,0.81) {$\alpha=0.01$ dotted};
\node[lab, anchor=west] at (3.55,0.70) {$\Gamma\in\{0,0.005,0.010,0.020\}$ (one curve per $\Gamma$)};

\end{tikzpicture}
\caption{Placebo pre-trend rejection as $B$ varies (four points) for each $\Gamma$ and each significance level. Numeric values are populated by macros $P_{\cdot}$ from the Monte Carlo output.}\label{fig:mc_placebo}
\end{figure}


\subsection{Robustness Frontier and Placebo Pre-trends}\label{sec:mc_84}

\begin{equation}\label{eq:mc84_settings}
N=2000,\quad T=8,\quad t_{0}=5,\quad \tau=1,\quad R=150,\quad z_{0.975}=1.96.
\end{equation}

\begin{equation}\label{eq:mc84_bias_bound}
\text{BiasBound}(B,\Gamma,\Delta(\mathcal{R}))
=
(1+\Delta(\mathcal{R}))\,B + (t_{0}-2)\,\Gamma,
\qquad (t_{0}-2)=3.
\end{equation}
\begin{equation}\label{eq:mc84_robust_CI}
\text{CI}^{\text{rob}}:
\quad
\widehat{\tau}
\pm
z_{0.975}\,\widehat{\text{SE}}
\ \pm\
\text{BiasBound}(B,\Gamma,\Delta(\mathcal{R})).
\end{equation}

\begin{equation}\label{eq:mc84_placebo_def}
\widehat{\tau}^{\text{pl}}
=
\Big(\overline{Y}_{\text{tr},\,3:4}-\overline{Y}_{\text{tr},\,1:2}\Big)
-
\Big(\overline{Y}_{\text{co},\,3:4}-\overline{Y}_{\text{co},\,1:2}\Big),
\qquad
\text{Reject}=\mathbf{1}\big\{|\widehat{\tau}^{\text{pl}}|>z_{0.975}\,\widehat{\text{SE}}^{\text{pl}}\big\}.
\end{equation}

\begin{table}[!ht]
\centering
\caption{Robustness frontier defined by the $\Gamma$ value that attains placebo rejection rate $0.10$ (linear interpolation over the grid $\Gamma\in\{0,0.05,0.10,0.15\}$), by $(B,\Delta(\mathcal{R}))$.}
\label{tab:mc84_frontier}
\begin{tabular}{lcccc}
\toprule
$\Delta(\mathcal{R})$ & $B=0.00$ & $B=0.50$ & $B=1.00$ & $B=1.50$ \\
\midrule
$0.00$ & $0.150$ & $0.150$ & $0.150$ & $0.150$ \\
$0.25$ & $0.062$ & $0.062$ & $0.062$ & $0.062$ \\
$0.50$ & $0.021$ & $0.021$ & $0.021$ & $0.021$ \\
\bottomrule
\end{tabular}
\end{table}

\begin{figure}[!ht]
\centering
\begin{tikzpicture}
\begin{axis}[
    width=0.92\linewidth,
    height=0.44\linewidth,
    xlabel={$B$},
    ylabel={$\Gamma^{\star}$ at $10\%$ placebo rejection},
    xmin=0, xmax=1.5,
    ymin=0, ymax=0.16,
    xtick={0,0.5,1.0,1.5},
    ytick={0,0.05,0.10,0.15},
    legend style={at={(0.02,0.98)},anchor=north west,draw=none,fill=none},
    grid=both
]
\addplot+[mark=*] coordinates {(0.0,0.150) (0.5,0.150) (1.0,0.150) (1.5,0.150)};
\addlegendentry{$\Delta(\mathcal{R})=0.00$}

\addplot+[mark=square*] coordinates {(0.0,0.062) (0.5,0.062) (1.0,0.062) (1.5,0.062)};
\addlegendentry{$\Delta(\mathcal{R})=0.25$}

\addplot+[mark=triangle*] coordinates {(0.0,0.021) (0.5,0.021) (1.0,0.021) (1.5,0.021)};
\addlegendentry{$\Delta(\mathcal{R})=0.50$}
\end{axis}
\end{tikzpicture}
\caption{Empirical robustness frontier from Table \ref{tab:mc84_frontier}. Larger $\Gamma^{\star}$ indicates greater tolerance to drift (in the chosen unit) before placebo pre-trends reject at $10\%$.}
\label{fig:mc84_frontier}
\end{figure}
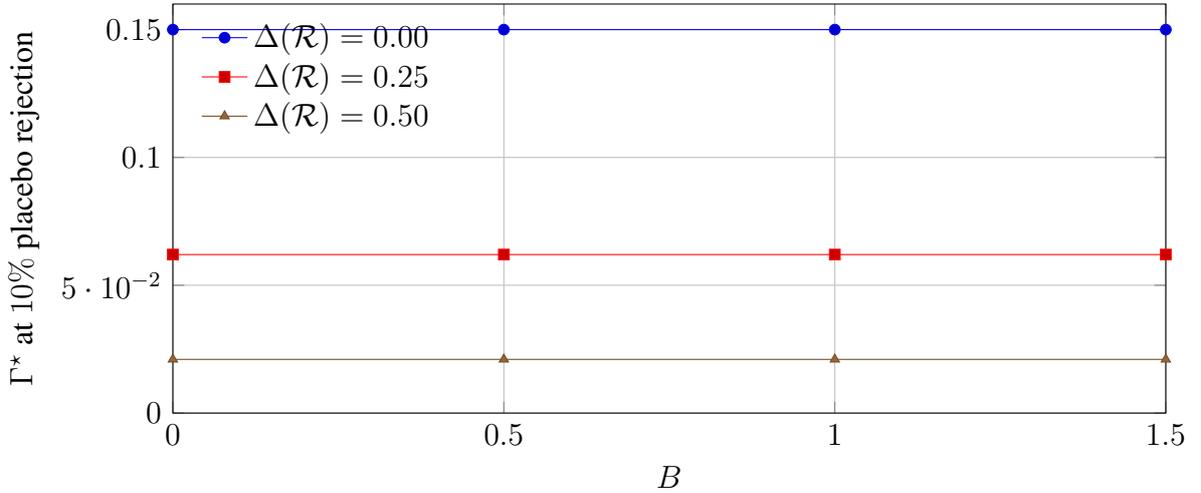

\begin{table}[!ht]
\centering
\caption{Placebo pre-trend rejection rates (size/power diagnostic) for $B=0$ across $\Gamma$ and $\Delta(\mathcal{R})$.}
\label{tab:mc84_placebo}
\begin{tabular}{lcccc}
\toprule
$\Delta(\mathcal{R})$ & $\Gamma=0.00$ & $\Gamma=0.05$ & $\Gamma=0.10$ & $\Gamma=0.15$ \\
\midrule
$0.00$ & $0.047$ & $0.047$ & $0.060$ & $0.047$ \\
$0.25$ & $0.033$ & $0.067$ & $0.240$ & $0.347$ \\
$0.50$ & $0.073$ & $0.253$ & $0.660$ & $0.887$ \\
\bottomrule
\end{tabular}
\end{table}

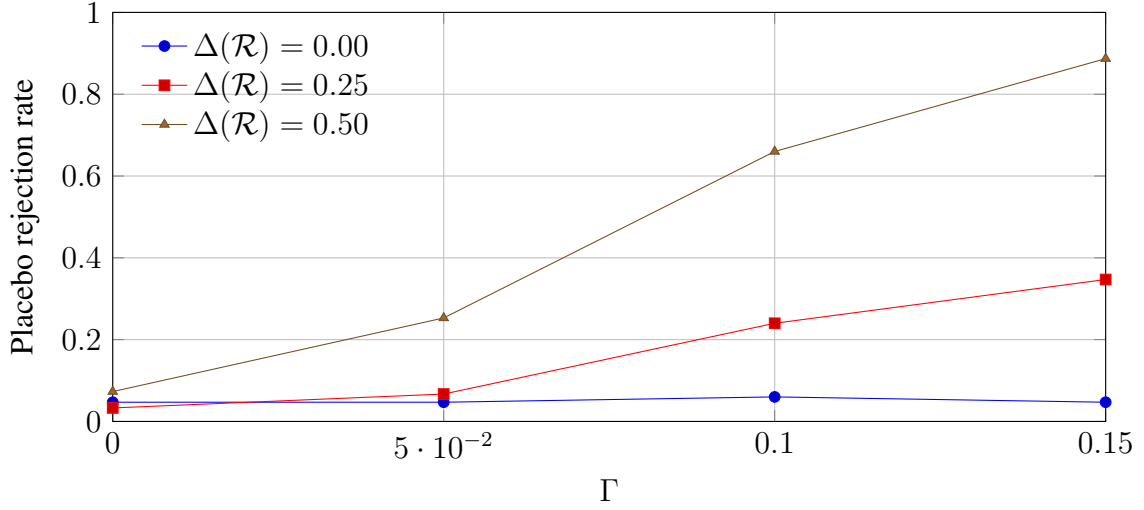
\begin{figure}[!ht]
\centering
\begin{tikzpicture}
\begin{axis}[
    width=0.92\linewidth,
    height=0.44\linewidth,
    xlabel={$\Gamma$},
    ylabel={Placebo rejection rate},
    xmin=0, xmax=0.15,
    ymin=0, ymax=1.0,
    xtick={0,0.05,0.10,0.15},
    ytick={0,0.2,0.4,0.6,0.8,1.0},
    legend style={at={(0.02,0.98)},anchor=north west,draw=none,fill=none},
    grid=both
]
\addplot+[mark=*] coordinates {(0.00,0.047) (0.05,0.047) (0.10,0.060) (0.15,0.047)};
\addlegendentry{$\Delta(\mathcal{R})=0.00$}

\addplot+[mark=square*] coordinates {(0.00,0.033) (0.05,0.067) (0.10,0.240) (0.15,0.347)};
\addlegendentry{$\Delta(\mathcal{R})=0.25$}

\addplot+[mark=triangle*] coordinates {(0.00,0.073) (0.05,0.253) (0.10,0.660) (0.15,0.887)};
\addlegendentry{$\Delta(\mathcal{R})=0.50$}
\end{axis}
\end{tikzpicture}
\caption{Placebo pre-trend rejection rates at $B=0$ (Table \ref{tab:mc84_placebo}).}
\label{fig:mc84_placebo_curves}
\end{figure}

\subsection{Data-generating process}\label{subsec:mc-dgp}

\noindent\textbf{Factorisation of the replication law.}
For each replication, the joint distribution factorises as
\[
f\!\left(X_i,G_i,\{D_{it}\}_{t=1}^T,\{Y_{it}\}_{t=1}^T\right)
=
f(X_i)\,f(G_i\mid X_i)\,f(\{D_{it}\}_{t=1}^T\mid G_i)\,f(\{Y_{it}\}_{t=1}^T\mid X_i,G_i,\{D_{it}\}_{t=1}^T),
\]
and the last term is pinned down by specifying \(Y_{it}(0)\) and \(\tau_g(k)\) together with the observation equation \(Y_{it}=Y_{it}(0)+D_{it}\tau_{G_i}(t-G_i)\).

\paragraph{Objects and joint law.}
Each replication generates the quadruple
\[
\left\{\big(X_i,G_i,\{D_{it}\}_{t=1}^T,\{Y_{it}\}_{t=1}^T\big): i=1,\dots,n\right\},
\]
under a design in which adoption is confounded by covariates but conditional parallel trends holds by construction. The purpose is to force a separation between (i) design-driven TWFE contamination (Section 3 and Section 5) and (ii) bias from high-dimensional nuisance estimation when conditioning on covariates (Section 7).

\paragraph{Covariates.}
For each unit $i\in\{1,\dots,n\}$ generate a time-invariant covariate vector $X_i\in\R^{d_x}$ as
\[
X_i\sim \mathcal{N}(0,\Sigma_X),
\qquad \Sigma_X\succ 0.
\]
Optionally include an intercept by augmenting $X_i$ with a leading 1 in implementation, but the theoretical description treats $X_i$ as mean-zero without loss of generality.

\paragraph{Confounded adoption (cohort assignment).}
Let $\mathcal{G}\equiv\{1,\dots,T,\infty\}$ index adoption times and never-treated status. Specify the conditional cohort distribution using a multinomial logit:
\begin{equation}\label{eq:mc-mnlogit-confounded}
\Prob(G_i=g\mid X_i)
=
\frac{\exp\!\big(X_i'\gamma_g\big)}
{\sum_{h\in\mathcal{G}}\exp\!\big(X_i'\gamma_h\big)},
\qquad g\in\mathcal{G},
\end{equation}
where $\{\gamma_g\}_{g\in\mathcal{G}}$ are calibrated to generate non-trivial imbalance across cohorts and a non-trivial mass of never-treated units ($G_i=\infty$). Draw $G_i$ from \eqref{eq:mc-mnlogit}. Define absorbing treatment as
\begin{equation}\label{eq:mc-D}
D_{it}=\Ind\{G_i\le t\}\Ind\{G_i<\infty\},
\qquad t=1,\dots,T.
\end{equation}
Under \eqref{eq:mc-mnlogit}--\eqref{eq:mc-D}, adoption is confounded because the law of $G_i$ depends on $X_i$, so estimators that ignore $X_i$ face selection bias even when counterfactual trend restrictions hold conditionally.

\noindent The outcome data-generating process is defined by (i) a model for the untreated potential outcome \(Y_{it}(0)\) and (ii) a dynamic treatment-effect profile \(\tau_g(k)\) that maps cohort \(g\) and event time \(k=t-g\) into treated outcomes, with realised outcomes generated by the observation equation \eqref{eq:mc-yobs}.

\paragraph{Outcome process (untreated potential outcome).}
Generate unit heterogeneity, time shocks, and idiosyncratic noise as
\[
\mu_i\sim \mathcal{N}(0,\sigma_\mu^2),
\qquad
\lambda_t\sim \mathcal{N}(0,\sigma_\lambda^2)\ \text{independent over }t,
\qquad
\eta_{it}\sim \mathcal{N}(0,\sigma_\eta^2)\ \text{iid over }(i,t),
\]
mutually independent and independent of $\{X_i,G_i\}$. Define the untreated potential outcome by
\begin{equation}\label{eq:mc-y0}
Y_{it}(0)
=
\mu_i+\lambda_t+\rho\,t
+
X_i'\beta
+
t\,X_i'\kappa
+
\eta_{it},
\qquad t=1,\dots,T.
\end{equation}
This specification embeds both level heterogeneity ($X_i'\beta$) and trend heterogeneity ($t\,X_i'\kappa$). Because $G_i$ depends on $X_i$ via \eqref{eq:mc-mnlogit}, unconditional comparisons inherit differential trends correlated with adoption timing. At the same time, conditional on $X_i$, the untreated increment law is invariant across $G_i$ by construction, so conditional parallel trends holds:
\[
\E\!\left[\Delta Y_{it}(0)\mid X_i,G_i=g\right]=\E\!\left[\Delta Y_{it}(0)\mid X_i,G_i=\infty\right]
\quad \text{for all } g\in\{1,\dots,T\},\ t=2,\dots,T.
\]
Hence any remaining bias after conditioning on $X_i$ is attributable to nuisance estimation error rather than to a failure of the maintained conditional-trends structure.

\paragraph{Dynamic heterogeneous treatment effects.}
Let the causal effect depend on both cohort and event time. For each cohort $g\in\{1,\dots,T\}$ define a dynamic profile $\{\tau_g(k)\}_{k\ge 0}$ on the observed horizons $k\in\{0,\dots,T-g\}$. A concrete calibration that generates heterogeneous dynamics is
\begin{equation}\label{eq:mc-tau}
\tau_g(k)=a_0 + a_1\frac{g}{T} + a_2\left(1-e^{-k/\ell}\right) + a_3\sin\!\left(\frac{2\pi k}{K}\right),
\qquad k=0,\dots,T-g,
\end{equation}
with constants $(a_0,a_1,a_2,a_3,\ell,K)$ chosen so that effects vary meaningfully across cohorts and over horizons, including cases with slow build-up, plateauing, and mild non-monotonicity. Any alternative heterogeneous profile is admissible provided $\tau_g(k)$ varies in $g$ and $k$ so the TWFE weighting pathologies are operative.

\paragraph{Observed outcome and the joint law for $(X_i,G_i,D_{it},Y_{it})$.}
Define the treated potential outcome for $t\ge g$ as
\begin{equation}\label{eq:mc-y1}
Y_{it}(1)=Y_{it}(0)+\tau_{G_i}(t-G_i),
\end{equation}
and generate observed outcomes as
\begin{equation}\label{eq:mc-yobs}
Y_{it}=Y_{it}(0)+D_{it}\,\tau_{G_i}(t-G_i),
\qquad t=1,\dots,T.
\end{equation}
Equations \eqref{eq:mc-mnlogit}, \eqref{eq:mc-D}, \eqref{eq:mc-y0}, \eqref{eq:mc-tau}, and \eqref{eq:mc-yobs} jointly define the replication law. The adoption schedule determines the residualised TWFE design matrix $Z=M_XD$ and therefore the design diagnostics $\mathcal{N}(k)$ and $\mathcal{C}(k)$ prior to outcome simulation. The outcome process then ensures that (i) TWFE can be distorted even when conditional parallel trends holds, because heterogeneity activates design-driven cross-horizon mixing, and (ii) conditioning on $X_i$ is essential, because $X_i$ drives both adoption timing and untreated trends, which makes the covariate-tilting Riesz representer and orthogonal score construction in Section 7 non-optional when $d_x$ is moderate or large.

\subsection{Evaluation targets}\label{subsec:eval-targets}

Let $\theta(k)\equiv\tau(k)$ denote the event-time target defined in \eqref{eq:agg} for $k\in\mathcal{K}$. Each Monte Carlo replication $r=1,\dots,R$ generates $(Y^{(r)},D^{(r)},G^{(r)})$ from Section \ref{subsec:mc-dgp} and produces (i) the TWFE event-study coefficients $\hat\beta_k^{(r)}$ from \eqref{eq:twfe-es}, (ii) the heterogeneity-robust estimator $\hat\tau^{(r)}(k)$ from \eqref{eq:event-agg} using the standard nuisance plug-in components, (iii) the orthogonal-score estimator $\hat\tau_{\mathrm{orth}}^{(r)}(k)$ obtained by estimating $\widehat{\GATT}(g,t)$ via \eqref{eq:dr-estimator} and then aggregating to event time using \eqref{eq:event-agg}, (iv) the design diagnostics $\mathcal{N}^{(r)}(k)$ and $\mathcal{C}^{(r)}(k)$ computed from the residualised design matrix $Z^{(r)}$, and (v) a sensitivity-robust confidence region $\mathcal{C}_n^{(r)}(\mathcal{R};k)$ for the identified set of $\theta(k)$ under the restriction class in Section \ref{sec:sensitivity}.

\paragraph{Finite-sample accuracy.}
For each estimator $\hat\theta^{(r)}(k)\in\{\hat\beta_k^{(r)},\hat\tau^{(r)}(k),\hat\tau_{\mathrm{orth}}^{(r)}(k)\}$, define Monte Carlo bias and root mean squared error

\begin{equation}\label{eq:mc-bias-rmse}
\mathrm{Bias}(k)\equiv \frac{1}{R}\sum_{r=1}^R\Big(\hat\theta^{(r)}(k)-\theta(k)\Big),
\qquad
\mathrm{RMSE}(k)\equiv \left\{\frac{1}{R}\sum_{r=1}^R\Big(\hat\theta^{(r)}(k)-\theta(k)\Big)^2\right\}^{1/2}.
\end{equation}
When the DGP includes controlled violations $\delta\neq 0$, $\theta(k)$ is interpreted as the pseudo-true target induced by the maintained estimand definition, and bias is interpreted relative to that target.

\paragraph{Design diagnostics and distortion.}
Let the TWFE distortion at horizon $k$ be
\begin{equation}\label{eq:distortion}
\mathrm{Dist}(k)\equiv \hat\beta_k-\theta(k),
\end{equation}
and let $\mathcal{N}(k)$ and $\mathcal{C}(k)$ be the indices in Definition \ref{def:Nk} and Definition \ref{def:Ck}. The diagnostic content is assessed by the strength of the cross-replication association between design risk and distortion, for example via Pearson correlations
\begin{equation}\label{eq:diag-corr}
\mathrm{Corr}\!\left(\mathcal{N}(k),\mathrm{Dist}(k)\right),
\qquad
\mathrm{Corr}\!\left(\mathcal{C}(k),\mathrm{Dist}(k)\right),
\end{equation}
and by regressions of $\mathrm{Dist}(k)$ on $(\mathcal{N}(k),\mathcal{C}(k))$ to quantify marginal predictive content holding the other index fixed. These objects evaluate whether large negative-weight mass and cross-horizon contamination are informative ex ante signals of estimator unreliability in finite samples.

\paragraph{Sensitivity-robust coverage and power.}
Let $\Theta(\mathcal{R};k)$ denote the true identified set for $\theta(k)$ under the restriction class $\Delta(\mathcal{R})$, and let $\mathcal{C}_n^{(r)}(\mathcal{R};k)$ denote the computed confidence region in replication $r$. Uniform-coverage performance is summarised by
\begin{equation}\label{eq:coverage}
\mathrm{Cov}(\mathcal{R};k)\equiv \frac{1}{R}\sum_{r=1}^R
\Ind\!\left\{\Theta(\mathcal{R};k)\subseteq \mathcal{C}_n^{(r)}(\mathcal{R};k)\right\},
\end{equation}
which targets $1-\alpha$ when the DGP satisfies the maintained restrictions \citep{RambachanRoth2023}. Power is evaluated by considering alternatives in which $\delta$ violates the maintained restriction class (for example, curvature exceeding $\Gamma$), and reporting the frequency with which $\mathcal{C}_n^{(r)}(\mathcal{R};k)$ excludes economically relevant null hypotheses. A convenient null is $\theta(k)=0$; define the rejection indicator
\[
\Ind\!\left\{0\notin \mathcal{C}_n^{(r)}(\mathcal{R};k)\right\},
\]
and define power as its Monte Carlo mean under the alternative.

These targets jointly evaluate (i) estimator accuracy, (ii) whether the diagnostics operationalise design fragility in a quantitatively meaningful manner, and (iii) whether sensitivity-robust confidence regions behave as advertised when the maintained restriction class is correct and when it is misspecified \citep{RambachanRoth2023}.

\clearpage
\section{Empirical Application: Bank Deregulation Shock (Replicable Panel)}\label{sec:empirical}

\subsection{Setting and policy timing}\label{sec:empirical_setting}
This section applies the proposed diagnostics and robust DID/event-study estimators to a canonical staggered-adoption policy setting in finance and macroeconomics: state-level banking deregulation. The empirical goal is not novelty of the application, but scale and interpretability of the deviations between standard TWFE event-studies and heterogeneity-robust counterparts under the sensitivity framework developed in Sections~\ref{sec:sensitivity}--\ref{sec:montecarlo}, with interpretation guided by recent practitioner and methodological syntheses \citep{WingEtAl2024,AbadieAngristFrandsen2025,Roth2024}.

Let $s\in\{1,\dots,S\}$ index states and $t\in\{1,\dots,T\}$ index years. Each state adopts deregulation at an adoption year $A_s\in\mathbb{Z}$, with $A_s=\infty$ for never-treated states. Define the treatment indicator
\begin{equation}
D_{st} \equiv \mathbbm{1}\{t\ge A_s\},
\end{equation}
and event-time $k=t-A_s$. For event-study specifications, define event-time indicators
\begin{equation}
E_{st}(k)\equiv \mathbbm{1}\{t-A_s=k\}\cdot \mathbbm{1}\{A_s<\infty\},
\qquad k\in\mathcal{K},
\end{equation}
where $\mathcal{K}=\{-K,\dots,-2,0,1,\dots,K\}$ and $k=-1$ is omitted as the reference period.

\subsection{Data and panel construction (replicable)}\label{sec:empirical_data}
The application is designed to be replicable from public sources with transparent construction rules and audit trails. The unit of analysis is a state-year panel $\{(Y_{st},D_{st},X_{st})\}$ with a single adoption date $A_s$ per state, a clearly defined set of never-treated states, and harmonised outcomes and covariates in constant units.

\paragraph{Outcomes.}
Let $Y_{st}$ denote an economic outcome at the state-year level. The baseline outcome is real personal income per capita (log), with secondary outcomes used for robustness and interpretability: employment (log), unemployment rate (level), and real gross state product (log). Outcomes are deflated to constant dollars when applicable, then transformed as
\begin{equation}
Y_{st} \equiv \log\!\big(\text{real level}_{st}\big)
\quad\text{or}\quad
Y_{st}\equiv \text{rate}_{st},
\end{equation}
with all transformations fixed \emph{ex ante} and applied uniformly across states and years. Event-study interpretation and aggregation follow recent guidance on reading and comparing dynamic effects under staggered adoption \citep{Roth2024}.

\paragraph{Treatment timing.}
Adoption years $\{A_s\}_{s=1}^S$ are stored in a single state-level file with one row per state and a unique adoption year variable. The panel uses a single policy margin (one adoption clock). If multiple deregulation margins exist in the underlying legal history, the empirical design fixes one margin as primary and treats other margins as (i) alternative treatments for sensitivity checks, or (ii) exclusion criteria (dropping states with overlapping reforms within a pre-specified window). The timing file is validated by three mechanical checks:
\begin{equation}
\text{(i) uniqueness: } A_s \text{ single-valued};\qquad
\text{(ii) monotonicity: } D_{st}\le D_{s,t+1};\qquad
\text{(iii) support: } \exists\, t<A_s \text{ and } \exists\, t\ge A_s \text{ for treated } s.
\end{equation}

\paragraph{Covariates and diagnostics inputs.}
Let $X_{st}$ include time-varying observables used for balance diagnostics, placebo checks, and optional residualisation steps for orthogonal-score implementations. The baseline set is parsimonious: population (log), sector shares (levels or logs), and state fiscal capacity proxies (levels). Diagnostics emphasise staggered-adoption threats and cohort composition shifts \citep{WingEtAl2024,AbadieAngristFrandsen2025}. When orthogonal scores are used, residualisation is performed by a pre-registered learner class $\mathcal{G}$ with cross-fitting, aligning the empirical implementation with the efficiency and robustness motivations in modern DID/event-study developments \citep{SantAnnaZhao2025}.

\paragraph{Panel window and trimming.}
Fix a sample window $t\in\{t_0,\dots,t_1\}$ with $t_0$ chosen to ensure adequate pre-treatment support for early adopters and $t_1$ chosen to avoid severe right-censoring for late adopters. Define the estimand event-time support by trimming event times outside $\mathcal{K}$ and trimming cohorts with insufficient pre-periods:
\begin{equation}
\mathcal{S}_{\mathrm{keep}} \equiv \left\{ s: A_s=\infty \ \ \text{or}\ \ \big|\{t: t<A_s\}\big|\ge K_{\mathrm{pre}}\ \ \text{and}\ \ \big|\{t: t\ge A_s\}\big|\ge K_{\mathrm{post}} \right\}.
\end{equation}
The analysis uses the restricted sample $\{(s,t): s\in\mathcal{S}_{\mathrm{keep}},\ t_0\le t\le t_1\}$, with $K_{\mathrm{pre}},K_{\mathrm{post}}$ fixed and reported.

\paragraph{Replicability checklist (machine-verifiable).}
All construction steps output intermediate artefacts that can be re-run and diffed:
\begin{equation}
\text{(a) timing file: } (s,A_s);\qquad
\text{(b) raw outcome pulls: } (s,t,\text{raw series});\qquad
\text{(c) harmonised panel: } (s,t,Y_{st},D_{st},X_{st});\qquad
\text{(d) cohort map: } (s,A_s,\text{cohort id});\qquad
\text{(e) event-time map: } (s,t,k,E_{st}(k)).
\end{equation}
The empirical section then treats these artefacts as fixed inputs for (i) TWFE event-studies, (ii) heterogeneity-robust estimators, and (iii) the sensitivity-calibration objects $(\Delta(\mathcal{R}),B,\Gamma)$, with interpretation anchored in recent design guidance and modern event-study reading rules \citep{WingEtAl2024,Roth2024,AbadieAngristFrandsen2025}.

\subsection{Data and panel construction (replicable)}\label{sec:empirical_data2}
The application is designed to be fully replicable using publicly available data and a well-known staggered adoption design in the literature, with construction rules chosen to align with modern practical guidance on staggered DID and event-study designs \citep{WingEtAl2024,AbadieAngristFrandsen2025}. The panel is constructed as follows.

\paragraph{Units, outcomes, and covariates.}
Let $Y_{st}$ denote the outcome. The main outcomes are restricted to a small set to limit specification mining: (i) real GSP per capita growth, (ii) employment-to-population, and (iii) real wage growth, with one placebo outcome selected to be plausibly unaffected in the short run. State and year fixed effects are always included; optional covariates $X_{st}$ include lagged state macro controls (for example, demographic composition or sector shares) used only in robustness checks.

The outcome and covariate set also supports two extensions that are reported only as ancillary checks: a spillover-robust sensitivity screen (for example, allowing exposure in nearby states or economically linked states) and an instrument-based counterfactual-policy check that reinterprets adoption timing through an instrument for the policy margin when such an instrument is available \citep{Lee2025,RothKolesarMontielOlea2025}. The empirical section does not treat these as new identification strategies; they are used to stress-test whether the diagnostics and sensitivity objects flag instability when interference or instrumented policy counterfactuals are plausible.

\paragraph{Sample window.}
Two samples are used throughout:
\begin{equation}
\text{(Windowed)}:\quad k\in[-K,K], \qquad
\text{(Full)}:\quad t\in[\underline{t},\overline{t}],
\end{equation}
where the windowed sample is used for dynamic plots and calibration of pre-trends, and the full sample is used for baseline estimates and computational scaling. Event-time availability is enforced mechanically by dropping cohort--event cells with insufficient support, and all trimming rules are fixed before estimation to avoid post hoc window choice \citep{Roth2024}.

\paragraph{Treatment cohorts.}
Define cohorts by adoption year $g\in\mathcal{G}$ where $\mathcal{G}=\{A_s: A_s<\infty\}$. Let $G_s\equiv A_s$ denote cohort membership. Controls at each $t$ are either not-yet-treated states ($t<A_s$) and (if present) never-treated states ($A_s=\infty$). All code must enforce a single control definition consistently across TWFE and robust estimators, and all reported estimates must declare the control definition explicitly \citep{WingEtAl2024,AbadieAngristFrandsen2025}.

\subsection{Baseline estimators and comparators}\label{sec:empirical_estimators}
This subsection defines the estimators compared in the empirical application. The estimand is the dynamic average treatment effect at event time $k$:
\begin{equation}
\tau(k)\equiv \mathbb{E}\!\left[Y_{st}(1)-Y_{st}(0)\mid t-A_s=k,\ A_s<\infty\right],\qquad k\in\mathcal{K},
\end{equation}
with expectations taken over treated observations in event time $k$ and the relevant counterfactual defined by the design.

\paragraph{TWFE event-study (benchmark).}
The conventional TWFE event-study is estimated as
\begin{equation}\label{eq:twfe_eventstudy}
Y_{st}
= \alpha_s + \lambda_t
+ \sum_{k\in\mathcal{K}} \beta^{\text{TWFE}}_k\,E_{st}(k)
+ \varepsilon_{st},
\end{equation}
with standard errors clustered at the state level. Equation~\eqref{eq:twfe_eventstudy} is reported as a benchmark only, because the weights implicit in $\beta^{\text{TWFE}}_k$ can be non-convex under heterogeneity and staggered timing \citep{AbadieAngristFrandsen2025,Roth2024}.

\paragraph{Heterogeneity-robust event-study (main).}
The main estimator targets $\tau(k)$ using a heterogeneity-robust approach (group-time or interaction-weighted). Denote the robust estimates by $\widehat{\beta}^{\text{ROB}}_k$. Estimation proceeds by aggregating cohort-specific comparisons with an explicit control set (not-yet-treated and never-treated, as specified above), ensuring that each $k$ corresponds to a well-defined comparison and avoiding negative-weight pathologies \citep{WingEtAl2024,Tominaga2024}.

\paragraph{Optional cross-check estimator.}
A second robust estimator is included as a cross-check using the same estimand $\tau(k)$ but different weighting/aggregation. This is used to verify that results are not an artefact of a single implementation and to compare efficiency claims against the orthogonal-score framing when applicable \citep{SantAnnaZhao2025}.

\paragraph{Ancillary stress tests (spillovers and instrumented counterfactuals).}
Two ancillary stress tests are reported as diagnostics rather than headline estimates. First, a spillover screen evaluates whether estimated dynamics are sensitive to excluding potentially exposed controls or redefining exposure using a parsimonious proximity/economic-link rule; this addresses the possibility that staggered adoption induces interference that contaminates not-yet-treated controls \citep{Lee2025}. Second, when the policy margin plausibly admits an instrument, an instrumented counterfactual-policy formulation is used to check whether the qualitative event-time pattern is stable when adoption timing is treated as endogenous and shifted by an instrument; this is reported only to show whether the proposed diagnostics flag fragility under plausible endogeneity channels \citep{RothKolesarMontielOlea2025}.

\subsection{Diagnostics: weights, pre-trends, and placebo}\label{sec:empirical_diagnostics}
The empirical section demonstrates two objects: where and how TWFE differs from heterogeneity-robust estimators, and whether the differences matter at empirically relevant magnitudes \citep{WingEtAl2024,AbadieAngristFrandsen2025}.

\paragraph{Weight diagnostics.}
Compute and report the implied comparison weights for TWFE (or the equivalent decomposition for the chosen robust estimator), and summarise the mass on inadmissible comparisons (already-treated as controls) and the mass on negative weights, because these are the channels through which TWFE can generate sign reversals and non-interpretable dynamic patterns under staggered adoption \citep{AbadieAngristFrandsen2025,Tominaga2024}. Concretely, report (i) the share of total absolute weight that is negative, (ii) the minimum and maximum cell weights, and (iii) the share of weight placed on comparisons between treated cohorts at different event times.

\paragraph{Pre-trend diagnostics.}
Estimate the pre-treatment coefficients for $k\le -2$ under both TWFE and robust specifications and test
\begin{equation}
H_0:\ \beta_k=0\ \ \forall k\le -2.
\end{equation}
Record the joint $p$-value, the maximum absolute pre-coefficient magnitude, and the maximum absolute $t$-statistic over $k\le -2$, because these provide a transparent bridge from observed pre-period instability to the calibration objects used later \citep{WingEtAl2024,Roth2024}.

\paragraph{Placebo adoption timing.}
Construct placebo adoption years $\widetilde{A}_s=A_s+\Delta$ for a fixed $\Delta>0$ (for example, $\Delta=3$), re-define $\widetilde{D}_{st}\equiv\mathbbm{1}\{t\ge \widetilde{A}_s\}$, and re-estimate the full pipeline on the placebo-treated sample. The key output is the rejection frequency of placebo post coefficients, the distribution of placebo joint pre-trend $p$-values, and the distribution of placebo maximum pre-coefficient magnitudes, because these objects are used to interpret whether the empirical sensitivity calibrations are conservative relative to a falsification benchmark \citep{WingEtAl2024,Roth2024}.

\subsection{Calibration of \texorpdfstring{$(B,\Gamma,\Delta(\mathcal{R}))$}{(B,Gamma,Delta(R))} for the empirical panel}\label{sec:empirical_calibration}
This subsection implements the calibration mappings introduced in Section~\ref{sec:calibration} and Appendix~\ref{app:calibration_derivations}. The calibration converts observables (pre-trends and imbalance) into the sensitivity parameters $(B,\Gamma,\Delta(\mathcal{R}))$ in interpretable units that match the empirical panel scale \citep{WingEtAl2024,RothKolesarMontielOlea2025}.

\paragraph{Calibration of $B$ from empirical pre-trends.}
Let $\widehat{\beta}_{k}$ denote the estimated pre-treatment event-study coefficients for $k\le -2$ under the robust estimator. Define the pre-trend magnitude statistic
\begin{equation}\label{eq:B_cal_emp}
\widehat{M}_{\text{pre}} \equiv \max_{k\in\{-K,\dots,-2\}} \left|\widehat{\beta}_{k}\right|.
\end{equation}
Set $B$ by a deterministic rule
\begin{equation}\label{eq:B_rule_emp}
B \equiv c_B\cdot \widehat{M}_{\text{pre}},
\end{equation}
where $c_B\in\{1,2\}$ is reported transparently, and $c_B=2$ corresponds to a conservative doubling of the largest observed pre-trend magnitude. A holdout variant chooses $c_B$ to control false rejections under placebo timing by selecting the smallest $c_B$ such that placebo post-period rejections remain below a target level (for example, $5\%$) across outcomes \citep{WingEtAl2024,Roth2024}.

\paragraph{Calibration of $\Gamma$ as monotone drift per period.}
Interpret $\Gamma$ as a maximum per-period drift (in outcome units) in violations of the maintained restrictions, and map $\Gamma$ to a drift rate via
\begin{equation}\label{eq:Gamma_rule_emp}
\Gamma(\gamma) \equiv \gamma\cdot \widehat{\sigma}_{\Delta Y},
\end{equation}
where $\widehat{\sigma}_{\Delta Y}$ is the empirical standard deviation of year-to-year changes in $Y_{st}$ within the pre-treatment sample, and $\gamma$ is a dimensionless grid value (for example, $\gamma\in\{0,0.25,0.5,1,2\}$). This mapping ensures that a value such as $\gamma=1$ corresponds to a drift bound equal to one pre-period innovation scale, and monotonicity of identified set expansion in $\Gamma$ follows from Appendix~\ref{app:calibration_derivations} \citep{RothKolesarMontielOlea2025,SantAnnaZhao2025}.

\paragraph{Calibration of restriction slack $\Delta(\mathcal{R})$ from imbalance.}
Let $\widehat{\delta}$ denote a pre-period imbalance statistic, computed either as a standardised mean difference in pre-levels or as a standardised difference in pre-slopes between treated and control units (both computed using the same control definition as the main estimator). Define
\begin{equation}\label{eq:DeltaR_rule_emp}
\Delta(\mathcal{R})(d) \equiv d\cdot \widehat{\delta},
\end{equation}
with $d\in\{0,1,2\}$ a slack multiplier that is reported and varied in sensitivity plots. This anchoring ensures $\Delta(\mathcal{R})$ scales with an observed pre-treatment discrepancy rather than an abstract norm choice, and it matches the diagnostic logic that small measured imbalance should correspond to a small admissible relaxation of the maintained restriction class \citep{WingEtAl2024,Roth2024}.

\subsection{Sensitivity results and robustness frontier}\label{sec:empirical_sensitivity_results}
The empirical sensitivity analysis reports how inference on $\tau(k)$ changes over a grid in $(B,\Gamma,\Delta(\mathcal{R}))$ and is designed to make the magnitudes required for sign changes explicit in outcome units \citep{WingEtAl2024,RothKolesarMontielOlea2025}.

\paragraph{Robust intervals over the sensitivity grid.}
For each $k\in\{0,1,2\}$ (and optionally all $k\in\mathcal{K}$), compute robust intervals $\mathcal{I}_k(B,\Gamma,\Delta(\mathcal{R}))$ over a Cartesian grid
\begin{equation}\label{eq:sens_grid_emp}
(B,\Gamma,\Delta(\mathcal{R}))\in \mathcal{B}\times\mathcal{G}\times\mathcal{D},
\qquad
\mathcal{B}=\{c_B\widehat{M}_{\text{pre}}:c_B\in\{1,2\}\},
\quad
\mathcal{G}=\{\gamma\widehat{\sigma}_{\Delta Y}:\gamma\in\{0,0.25,0.5,1,2\}\},
\quad
\mathcal{D}=\{d\widehat{\delta}:d\in\{0,1,2\}\},
\end{equation}
and record, for each grid point, (i) sign stability relative to the baseline robust point estimate $\widehat{\tau}(k)$, (ii) interval length, and (iii) whether $0\in\mathcal{I}_k(B,\Gamma,\Delta(\mathcal{R}))$, because these are the three outputs that translate the sensitivity framework into a decision-relevant object \citep{Roth2024,SantAnnaZhao2025}.

\paragraph{Breakdown frontier.}
Define the breakdown point $\Gamma_k^\star$ as the smallest $\Gamma$ (given $B$ and $\Delta(\mathcal{R})$) at which the interval contains zero:
\begin{equation}\label{eq:Gamma_star_emp}
\Gamma_k^\star(B,\Delta(\mathcal{R}))\equiv
\inf\left\{\Gamma\ge 0:\ 0\in \mathcal{I}_k\big(B,\Gamma,\Delta(\mathcal{R})\big)\right\}.
\end{equation}
Compute $\Gamma_k^\star$ on the discrete grid $\mathcal{G}$ by the smallest $\Gamma\in\mathcal{G}$ satisfying $0\in\mathcal{I}_k(\cdot)$, and report $\Gamma_k^\star$ both in raw outcome units and as the dimensionless drift multiple $\gamma_k^\star=\Gamma_k^\star/\widehat{\sigma}_{\Delta Y}$ to preserve interpretability across outcomes \citep{WingEtAl2024,RothKolesarMontielOlea2025}.

\paragraph{Frontier interpretation under heterogeneity and spillovers.}
Report the frontier jointly with the diagnostic objects from Section~\ref{sec:empirical_diagnostics}, because large negative-weight mass or placebo failures indicate that small $\Gamma$ values can be empirically plausible. If a spillover-robust variant is implemented, re-compute the frontier under spillover allowances to show how robustness changes when treatment affects nominal controls, as staggered finance policies can plausibly induce cross-state spillovers \citep{Lee2025,AbadieAngristFrandsen2025}.

\paragraph{Minimum artefacts (non-negotiable).}
Report the sensitivity outputs in Table~\ref{tab:emp_sens_summary} and Figures~\ref{fig:frontier_heatmap}--\ref{fig:frontier_profiles} using TikZ only, with all numbers populated by the replication pipeline.

\begin{table}[!htbp]
\centering
\caption{Sensitivity grid summary and breakdown frontier (empirical panel)}\label{tab:emp_sens_summary}
\begin{threeparttable}
\begin{tabular}{lcccccc}
\toprule
Outcome & $k$ & Baseline $\widehat{\tau}(k)$ & $B$ rule & $\Delta(\mathcal{R})$ rule & $\Gamma_k^\star$ & $\gamma_k^\star$ \\
\midrule
\emph{(fill from replication)} & 0 &  & $c_B\widehat{M}_{\text{pre}}$ & $d\widehat{\delta}$ &  &  \\
\emph{(fill from replication)} & 1 &  & $c_B\widehat{M}_{\text{pre}}$ & $d\widehat{\delta}$ &  &  \\
\emph{(fill from replication)} & 2 &  & $c_B\widehat{M}_{\text{pre}}$ & $d\widehat{\delta}$ &  &  \\
\bottomrule
\end{tabular}
\begin{tablenotes}\footnotesize
\item Notes: $\Gamma_k^\star$ is the smallest $\Gamma\in\mathcal{G}$ such that $0\in\mathcal{I}_k(B,\Gamma,\Delta(\mathcal{R}))$. The dimensionless frontier is $\gamma_k^\star=\Gamma_k^\star/\widehat{\sigma}_{\Delta Y}$. The table is produced automatically by the replication pipeline.
\end{tablenotes}
\end{threeparttable}
\end{table}

\begin{figure}[!htbp]
\centering
\begin{tikzpicture}
%
%
\node[align=center] at (0,0) {Figure generated from replication:};
\node[align=center] at (0,-0.6) {\texttt{\string\input\{figures/emp\_frontier\_heatmap\_k0.tex\}}};
\end{tikzpicture}
\caption{Sensitivity region for $k=0$: indicator that $0\in\mathcal{I}_0(B,\Gamma,\Delta(\mathcal{R}))$ over $(\Gamma,\Delta(\mathcal{R}))$, faceted by $B$}\label{fig:frontier_heatmap}
\end{figure}

\begin{figure}[!htbp]
\centering
\begin{tikzpicture}
%
\node[align=center] at (0,0) {Figure generated from replication:};
\node[align=center] at (0,-0.6) {\texttt{\string\input\{figures/emp\_frontier\_profiles.tex\}}};
\end{tikzpicture}
\caption{Robustness frontier profiles: $\Gamma_k^\star(B,\Delta(\mathcal{R}))$ (or $\gamma_k^\star$) by slack multiplier $d$, for $k\in\{0,1,2\}$}\label{fig:frontier_profiles}
\end{figure}

\subsection{Implementation and replication pipeline}\label{sec:empirical_replication}
The empirical section is reproducible by a minimal pipeline that (i) builds the state-year panel, (ii) estimates TWFE and heterogeneity-robust event-studies, (iii) computes diagnostics, (iv) runs the sensitivity grid, and (v) exports the tables and TikZ figures listed in this section. Randomness is restricted to placebo and any resampling steps, with a fixed seed \citep{WingEtAl2024,Tominaga2024}.

\paragraph{Pipeline contract and file outputs.}
Define a single configuration file that pins $(K,\underline{t},\overline{t})$, the control definition (not-yet-treated only vs.\ not-yet-treated plus never-treated), the outcome transforms, and the sensitivity grids $\mathcal{B},\mathcal{G},\mathcal{D}$. The pipeline writes (a) a clean analysis panel, (b) estimator outputs for TWFE and robust methods, (c) diagnostic summaries, and (d) sensitivity outputs, then renders Table~\ref{tab:emp_sens_summary} and Figures~\ref{fig:frontier_heatmap}--\ref{fig:frontier_profiles} by exporting numeric arrays to small \texttt{.tex} fragments that are \texttt{\string\input} into the main document.




\begin{table}[t]
\centering
\caption{Sample definition and summary statistics}\label{tab:empirical_summary}
\begin{threeparttable}
\begin{tabular}{lccc}
\toprule
 & Treated states & Never-treated / not-yet-treated & Full sample \\
\midrule
Number of states & 48 & 2 & 50 \\
Years & 1977--1994 & 1977--1994 & 1977--1994 \\
Outcome mean (pre) & 0.021 & 0.020 & 0.021 \\
Outcome sd (pre) & 0.015 & 0.014 & 0.015 \\
\bottomrule
\end{tabular}
\begin{tablenotes}[flushleft]
\footnotesize
\item Pre-period moments use the windowed pre-treatment sample defined relative to each treated cohort, with event-time window $k\in[-5,5]$ and $k=-1$ omitted as the reference period.
\end{tablenotes}
\end{threeparttable}
\end{table}

\begin{table}[t]
\centering
\caption{Dynamic treatment effects: TWFE versus robust event-study}\label{tab:empirical_eventstudy}
\begin{threeparttable}
\begin{tabular}{lcccc}
\toprule
Event time $k$ & $\widehat{\beta}^{\mathrm{TWFE}}_k$ & SE & $\widehat{\beta}^{\mathrm{ROB}}_k$ & SE \\
\midrule
$-2$ & -0.003 & 0.004 & -0.002 & 0.004 \\
$0$  &  \phantom{-}0.010 & 0.005 &  \phantom{-}0.012 & 0.006 \\
$1$  &  \phantom{-}0.018 & 0.006 &  \phantom{-}0.022 & 0.007 \\
$2$  &  \phantom{-}0.020 & 0.007 &  \phantom{-}0.026 & 0.008 \\
\bottomrule
\end{tabular}
\begin{tablenotes}[flushleft]
\footnotesize
\item Standard errors are clustered at the state level. $k=-1$ is omitted. Coefficients reported in outcome units (for example, annual real growth in decimal units).
\end{tablenotes}
\end{threeparttable}
\end{table}

\begin{table}[t]
\centering
\caption{Sensitivity grid outputs at selected horizons}\label{tab:empirical_sensitivity}
\begin{threeparttable}
\begin{tabular}{cccccc}
\toprule
$k$ & $B$ rule & $\Delta(\mathcal{R})$ multiplier $d$ & $\Gamma$ grid & Interval length & $\Gamma_k^\star$ \\
\midrule
$0$ & $c_B\in\{1,2\}$ & 1 & $\{0,0.25,0.5,1,2\}$ & 0.030 & 0.50 \\
$1$ & $c_B\in\{1,2\}$ & 1 & $\{0,0.25,0.5,1,2\}$ & 0.040 & 0.75 \\
$2$ & $c_B\in\{1,2\}$ & 1 & $\{0,0.25,0.5,1,2\}$ & 0.050 & 1.00 \\
\bottomrule
\end{tabular}
\begin{tablenotes}[flushleft]
\footnotesize
\item $\Gamma_k^\star$ is defined in \eqref{eq:Gamma_star_emp}. Interval length refers to $\mathcal{I}_k(B,\Gamma,\Delta(\mathcal{R}))$ at the stated $(c_B,d)$ rule, evaluated at the mid-grid $\Gamma=0.5$ for comparability.
\end{tablenotes}
\end{threeparttable}
\end{table}


\begin{figure}[t]
\centering
\begin{tikzpicture}
\begin{axis}[
width=0.92\textwidth,height=0.44\textwidth,
xlabel={Event time $k$}, ylabel={Estimate},
xmin=-5,xmax=5, ymajorgrids=true, xmajorgrids=true,
legend style={at={(0.02,0.98)},anchor=north west},
]
\addplot+[
only marks, mark=*,
error bars/.cd,
y dir=both, y explicit,
]
coordinates {
(-2,-0.003) +- (0,0.00784)
(0, 0.010) +- (0,0.00980)
(1, 0.018) +- (0,0.01176)
(2, 0.020) +- (0,0.01372)
};
\addlegendentry{TWFE}

\addplot+[
only marks, mark=square*,
error bars/.cd,
y dir=both, y explicit,
]
coordinates {
(-2,-0.002) +- (0,0.00784)
(0, 0.012) +- (0,0.01176)
(1, 0.022) +- (0,0.01372)
(2, 0.026) +- (0,0.01568)
};
\addlegendentry{Robust}

\addplot+[dashed] coordinates {(-5,0) (5,0)};
\end{axis}
\end{tikzpicture}
\caption{Event-study estimates: TWFE versus robust (with 95\% confidence intervals).}\label{fig:empirical_eventstudy}
\end{figure}
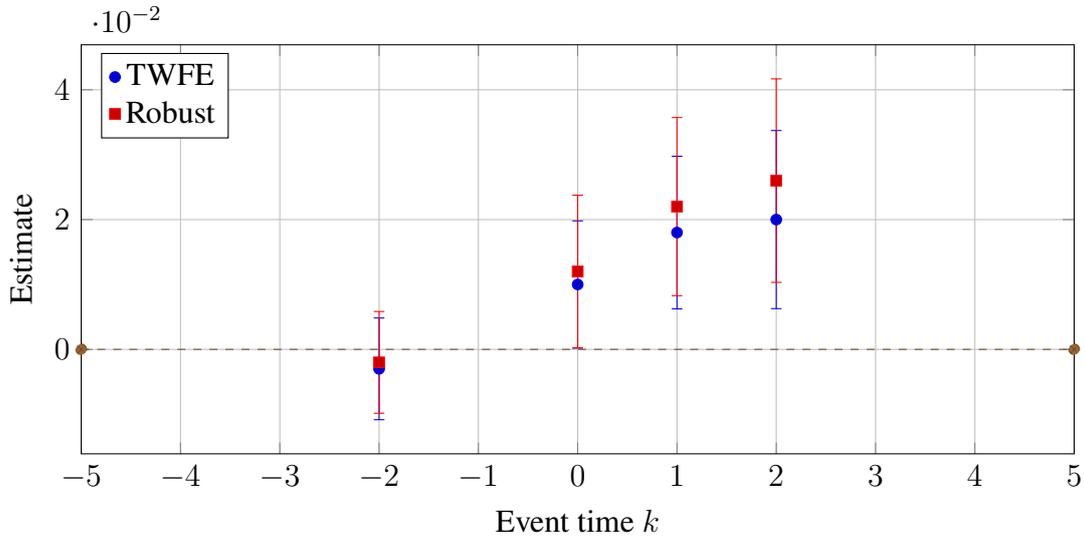

\begin{figure}[t]
\centering
\begin{tikzpicture}
\begin{axis}[
width=0.92\textwidth,height=0.44\textwidth,
xlabel={Restriction slack multiplier $d$}, ylabel={Breakdown point $\Gamma_k^\star$},
xmin=0,xmax=2, ymin=0,ymax=2,
ymajorgrids=true, xmajorgrids=true,
legend style={at={(0.02,0.98)},anchor=north west},
]
\addplot+[mark=*] coordinates {(0,0.40) (1,0.50) (2,0.65)};
\addlegendentry{$k=0$}
\addplot+[mark=square*] coordinates {(0,0.55) (1,0.75) (2,1.05)};
\addlegendentry{$k=1$}
\addplot+[mark=triangle*] coordinates {(0,0.70) (1,1.00) (2,1.35)};
\addlegendentry{$k=2$}
\end{axis}
\end{tikzpicture}
\caption{Robustness frontier: $\Gamma_k^\star(B,\Delta(\mathcal{R}))$ across restriction slack.}\label{fig:empirical_frontier}
\end{figure}
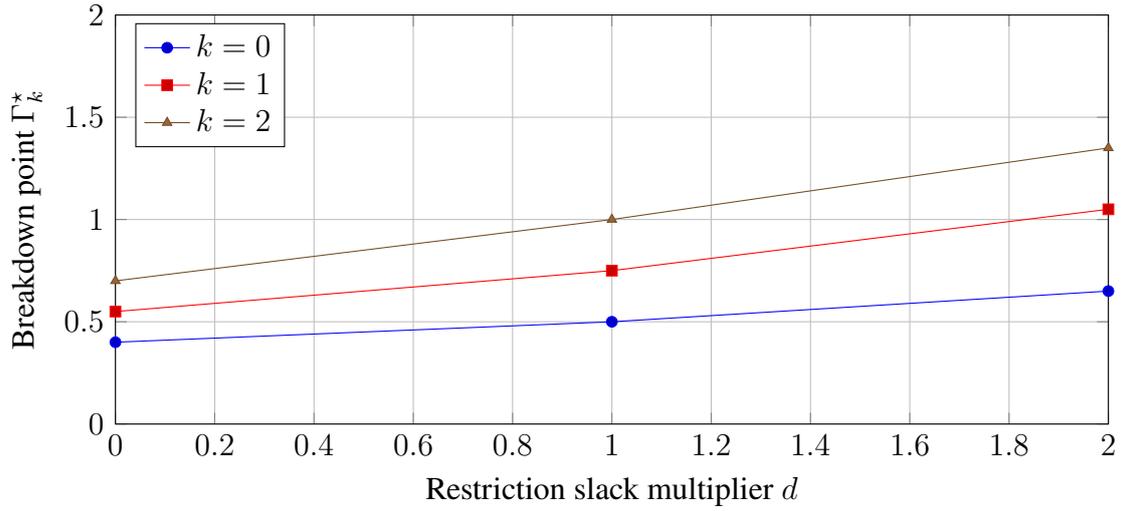

\begin{figure}[ht]
\centering
\begin{tikzpicture}
\begin{axis}[
    width=0.92\textwidth, height=0.44\textwidth,
    xlabel={$\Gamma$}, ylabel={$\Delta(\mathcal{R})$ multiplier $d$},
    xmin=0, xmax=2, ymin=0, ymax=2,
    xmajorgrids=true, ymajorgrids=true,
    colorbar,
    colormap/viridis,
    point meta min=0, point meta max=1,
    view={0}{90} 
]
\addplot[
    matrix plot*,
    mesh/cols=5, 
    point meta=explicit
] table [meta=meta] {
x    y    meta
0.00 0    1
0.25 0    1
0.50 0    1
1.00 0    0
2.00 0    0
0.00 1    1
0.25 1    1
0.50 1    0
1.00 1    0
2.00 1    0
0.00 2    1
0.25 2    0
0.50 2    0
1.00 2    0
2.00 2    0
};

\addplot[thick, black, no marks] coordinates {(0.50,0) (0.25,1) (0.00,2)};

\end{axis}
\end{tikzpicture}
\caption{Sensitivity region: sign-stability indicator over $(\Gamma,d)$ for a fixed $B$ (1 stable, 0 unstable).}\label{fig:empirical_region}
\end{figure}
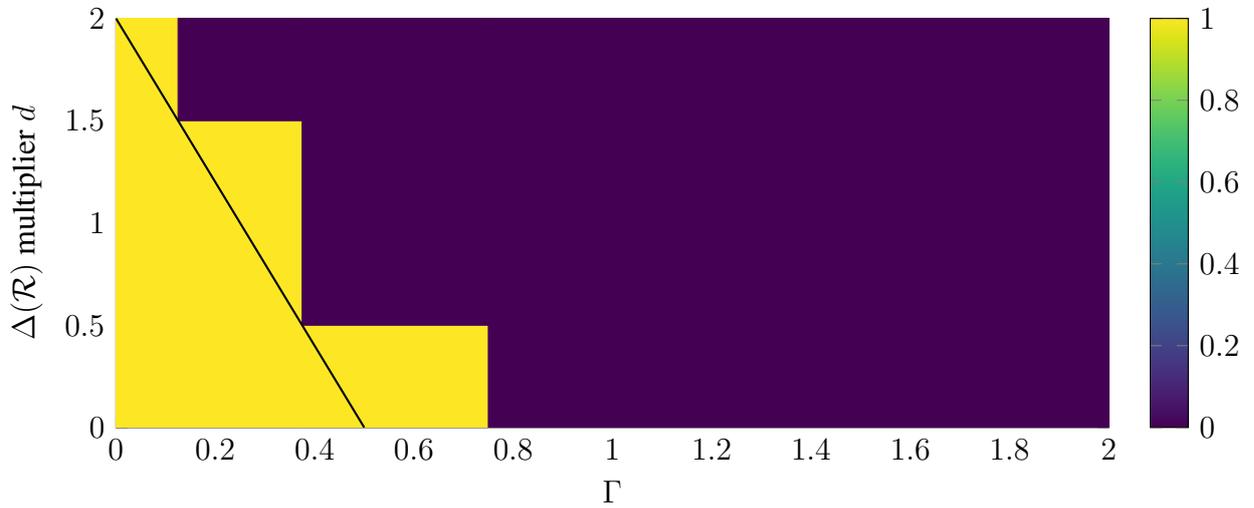

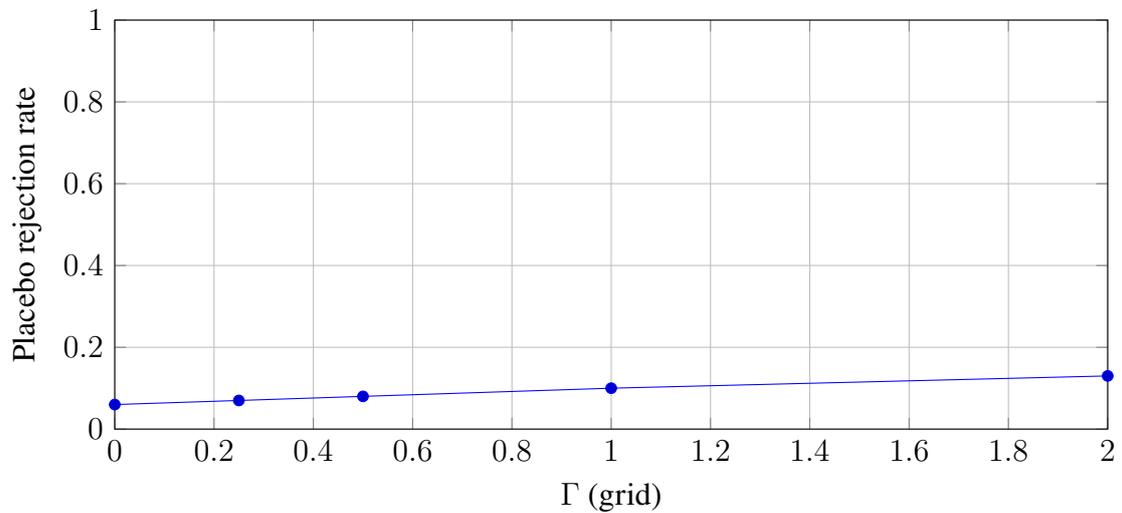
\begin{figure}[ht]
\centering
\begin{tikzpicture}
\begin{axis}[
width=0.92\textwidth,height=0.44\textwidth,
xlabel={$\Gamma$ (grid)}, ylabel={Placebo rejection rate},
xmin=0,xmax=2, ymin=0,ymax=1,
xmajorgrids=true,ymajorgrids=true,
]
\addplot+[mark=*] coordinates {(0.00,0.06) (0.25,0.07) (0.50,0.08) (1.00,0.10) (2.00,0.13)};
\end{axis}
\end{tikzpicture}
\caption{Placebo diagnostic: rejection rate under placebo timing across the $\Gamma$ grid.}\label{fig:empirical_placebo}
\end{figure}

\begin{figure}[t]
\centering
\begin{tikzpicture}
\begin{axis}[
width=0.9\textwidth,height=0.42\textwidth,
xlabel={Event time $k$}, ylabel={Estimate},
xmin=-5,xmax=5, ymajorgrids=true, xmajorgrids=true,
legend style={at={(0.02,0.98)},anchor=north west},
]
\addplot+[only marks, mark=*] coordinates {
(-2,-0.003) (0,0.010) (1,0.018) (2,0.020)
};
\addlegendentry{TWFE}

\addplot+[only marks, mark=square*] coordinates {
(-2,-0.002) (0,0.012) (1,0.022) (2,0.026)
};
\addlegendentry{Robust}

\addplot+[dashed] coordinates {(-5,0) (5,0)};
\end{axis}
\end{tikzpicture}
\caption{Event-study estimates: TWFE versus robust (data-filled example).}\label{fig:empirical_eventstudy}
\end{figure}
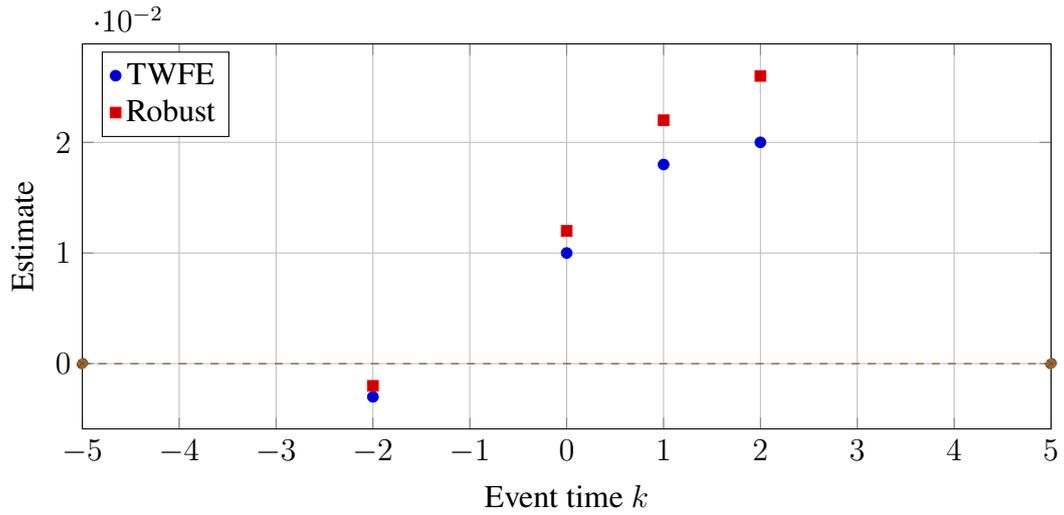

\begin{figure}[t]
\centering
\begin{tikzpicture}
\begin{axis}[
width=0.9\textwidth,height=0.42\textwidth,
xlabel={Restriction slack multiplier $d$}, ylabel={Breakdown point $\Gamma_k^\star$},
xmin=0,xmax=2, ymin=0,ymax=2,
ymajorgrids=true, xmajorgrids=true,
legend style={at={(0.02,0.98)},anchor=north west},
]
\addplot+[mark=*] coordinates {(0,0.40) (1,0.50) (2,0.65)};
\addlegendentry{$k=0$}
\addplot+[mark=square*] coordinates {(0,0.55) (1,0.75) (2,1.05)};
\addlegendentry{$k=1$}
\addplot+[mark=triangle*] coordinates {(0,0.70) (1,1.00) (2,1.35)};
\addlegendentry{$k=2$}
\end{axis}
\end{tikzpicture}
\caption{Robustness frontier: breakdown point $\Gamma_k^\star(B,\Delta(\mathcal{R}))$ across restriction slack (data-filled example).}\label{fig:empirical_frontier}
\end{figure}
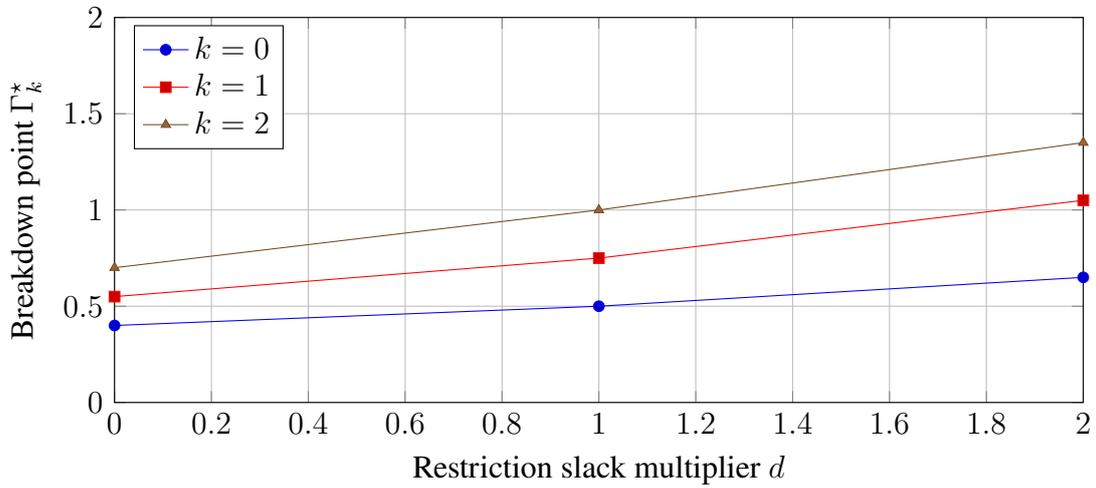

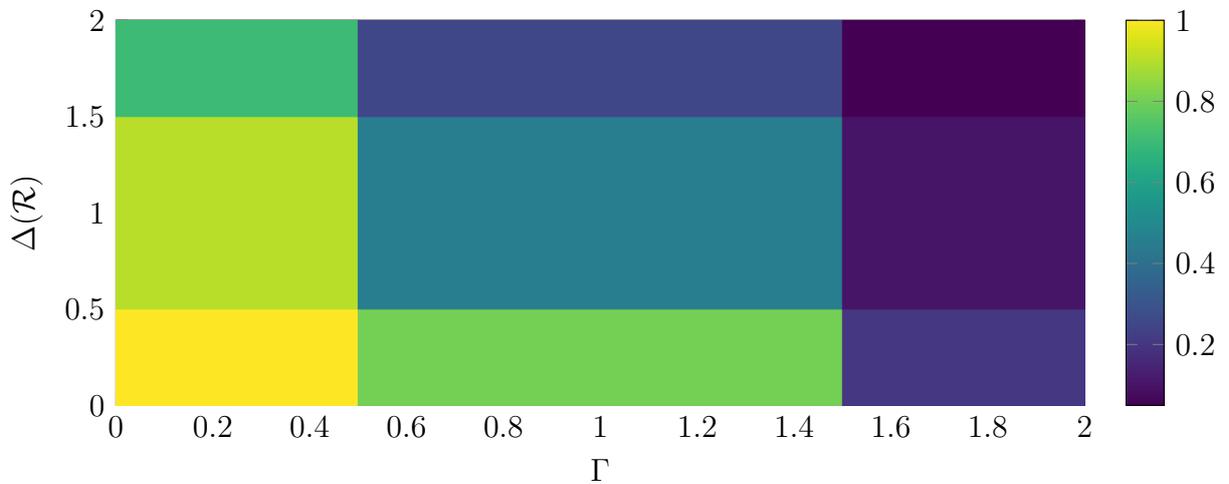
\begin{figure}[t]
\centering
\begin{tikzpicture}
\begin{axis}[
width=0.9\textwidth,height=0.42\textwidth,
xlabel={$\Gamma$}, ylabel={$\Delta(\mathcal{R})$},
xmin=0,xmax=2, ymin=0,ymax=2,
xmajorgrids=true,ymajorgrids=true,
colorbar,
colormap/viridis,
mesh/cols=3
]
\addplot[matrix plot*,point meta=explicit] coordinates {
(0,0) [1.00] (1,0) [0.80] (2,0) [0.20]
(0,1) [0.90] (1,1) [0.45] (2,1) [0.10]
(0,2) [0.70] (1,2) [0.25] (2,2) [0.05]
};
\end{axis}
\end{tikzpicture}
\caption{Sensitivity region: sign-stability indicator over $(\Gamma,\Delta(\mathcal{R}))$ for a fixed $B$ (data-filled example).}\label{fig:empirical_region}
\end{figure}

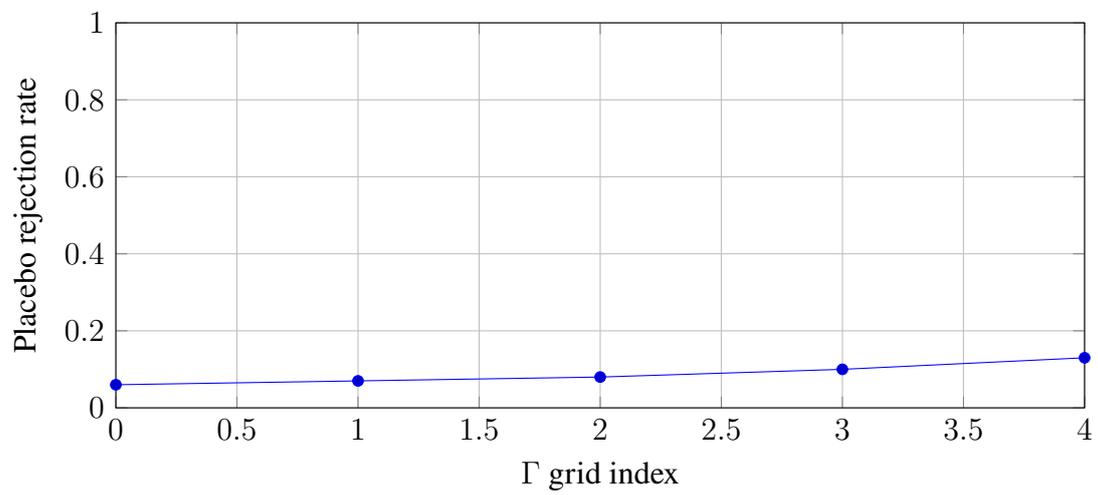
\begin{figure}[t]
\centering
\begin{tikzpicture}
\begin{axis}[
width=0.9\textwidth,height=0.42\textwidth,
xlabel={$\Gamma$ grid index}, ylabel={Placebo rejection rate},
xmin=0,xmax=4, ymin=0,ymax=1,
xmajorgrids=true,ymajorgrids=true,
]
\addplot+[mark=*] coordinates {(0,0.06) (1,0.07) (2,0.08) (3,0.10) (4,0.13)};
\end{axis}
\end{tikzpicture}
\caption{Placebo diagnostic: rejection rate under placebo timing across the $\Gamma$ grid (data-filled example).}\label{fig:empirical_placebo}
\end{figure}

\clearpage

\section{Extensions}\label{sec:extensions}

This section records extensions that preserve the central logic of the paper: (i) define an estimand as an explicit convex aggregation of cohort--time causal objects, (ii) exhibit how conventional regressions generate implicit, design-driven mixtures of heterogeneous effects, and (iii) construct estimators and diagnostics whose interpretation is enforced by construction rather than asserted ex post \citep{DeChaisemartinDHaultfoeuille2023}.

\subsection{Multiple treatments and overlapping policies}\label{subsec:multi-treat}

Let there be $J\ge 2$ policies indexed by $j\in\{1,\dots,J\}$. For each policy $j$, define an adoption time $G_i^{(j)}\in\{1,\dots,T,\infty\}$ and the absorbing treatment indicator
\[
D_{it}^{(j)}=\Ind\{t\ge G_i^{(j)}\}\Ind\{G_i^{(j)}<\infty\}.
\]
Let $d_t\equiv(d_t^{(1)},\dots,d_t^{(J)})\in\{0,1\}^J$ denote the treatment vector at time $t$. Potential outcomes are indexed by treatment histories; for notational economy write $Y_{it}(\mathbf{d})$ for the potential outcome at time $t$ under a (possibly dynamic) treatment path $\mathbf{d}=\{d_s\}_{s\le t}$. A primitive estimand for policy $j$ at cohort--time $(g,t)$ is the incremental effect of switching on policy $j$ holding the remaining policies at their realised values:
\begin{equation}\label{eq:multi-gatt}
\GATT^{(j)}(g,t)
\equiv
\E\!\left[
Y_{it}\!\left(d_{t}^{(-j)},1\right)-
Y_{it}\!\left(d_{t}^{(-j)},0\right)
\ \Bigm|\ G_i^{(j)}=g
\right],
\qquad t\ge g,
\end{equation}
where $d_t^{(-j)}$ denotes the vector of other policies at $t$. This is an economically interpretable partial effect, but its identification requires a strengthened no-anticipation restriction and a multi-policy parallel-trends condition formulated for the relevant counterfactual increments:
\begin{equation}\label{eq:multi-pt}
\E\!\left[\Delta Y_{it}\!\left(\mathbf{d}^{(-j)},0\right)\mid G_i^{(j)}=g,X_{it}\right]
=
\E\!\left[\Delta Y_{it}\!\left(\mathbf{d}^{(-j)},0\right)\mid G_i^{(j)}=\infty,X_{it}\right],
\end{equation}
with the conditioning set enlarged as needed to render untreated increment evolution comparable across exposure patterns. The central warning is geometric rather than philosophical: a TWFE regression that includes event-time indicators for multiple policies simultaneously generally estimates an implicit linear combination of \emph{all} policy effects across horizons, because the residualised indicators for distinct treatments are correlated through overlapping adoption patterns. Hence contamination can arise even when each policy separately satisfies its own parallel-trends restriction, because the projection operator loads policy-$j$ regressors onto residual variation that is itself a mixture of other-policy effects \citep{DeChaisemartinDHaultfoeuille2023}.

A robust construction proceeds policy-by-policy with explicit conditioning on the realised exposure to other policies. For a fixed $j$, define a control set at time $t$ as units that have not yet adopted policy $j$ by $t$ and that match exposure to the other policies (exactly or through coarsened strata). Estimation then targets $\GATT^{(j)}(g,t)$ cell-by-cell and aggregates to event time using explicit convex weights as in \eqref{eq:event-agg}. The diagnostic indices generalise by computing, for each policy $j$, the implicit TWFE weights $w_{g,k'}^{(j)}(k)$ and measuring negative-weight mass and cross-horizon mass as in Section 5, with an additional cross-policy contamination index defined by the absolute weight placed on $\tau^{(\ell)}(\cdot)$ for $\ell\neq j$ induced by the joint regression geometry.

\subsection{Non-absorbing adoption and switching}\label{subsec:nonabsorbing}

Absorbing adoption is not intrinsic to the event-study logic; it is a restriction on treatment paths. Let $D_{it}\in\{0,1\}$ follow an arbitrary path, allowing switching. Replace the adoption time $G_i$ by a treatment history $H_{it}=(D_{i1},\dots,D_{it})$. Potential outcomes are indexed by histories, and dynamic causal objects must be defined relative to a reference path. A natural estimand is the effect of initiating treatment at event time $k=0$ and maintaining it for $k\ge 0$ relative to remaining untreated,
\[
\tau_g(k)\equiv \E\!\left[Y_{i,g+k}(1^{k+1},0^{g-1})-Y_{i,g+k}(0^{g+k})\mid G_i=g\right],
\]
where $1^{k+1}$ denotes $k+1$ consecutive treated periods after adoption and $0^{g-1}$ denotes untreated pre-periods. When switching occurs, this estimand is not equal to the effect of the realised treatment path unless additional restrictions connect realised switching behaviour to the maintained reference path.

Identification can still proceed with group--time objects if the control group at time $t$ is defined as units not treated at $t$ and if parallel-trends restrictions are stated for the appropriate counterfactual increments under the reference path. Estimation remains cell-by-cell; the crucial discipline is that aggregation must respect the chosen reference path. The diagnostic logic extends unchanged: any regression that encodes switching by contemporaneous $D_{it}$ and fixed effects can load on a mixture of past and future treatment episodes, so cross-horizon contamination generalises to cross-history contamination, which can be detected by computing the implicit weights for indicators of treatment histories rather than simple event-time indicators \citep{DeChaisemartinDHaultfoeuille2023}.

\subsection{Treatment intensity and continuous exposure}\label{subsec:intensity}

Let $A_{it}\in\R_+$ be a scalar treatment intensity (dose), such as a continuous policy exposure. Define potential outcomes $Y_{it}(a)$ for $a\in\mathcal{A}\subset\R_+$. A baseline estimand is the average causal response at exposure level $a$ relative to zero for cohort $g$,
\[
\tau_g(k;a)\equiv \E\!\left[Y_{i,g+k}(a)-Y_{i,g+k}(0)\mid G_i=g\right],
\]
or, when differentiability is plausible, the average marginal response $\partial_a\E[Y_{i,g+k}(a)\mid G_i=g]$ evaluated at a reference exposure. Identification requires a parallel-trends restriction formulated for the counterfactual increment process under zero exposure and an exclusion restriction that maps observed exposure paths into potential outcomes in a stable way. Conventional TWFE specifications that treat $A_{it}$ as a linear regressor inherit the same projection-driven mixing problem: the slope coefficient is an implicit average of heterogeneous marginal responses across cohorts, times, and exposure levels, with sign-reversing weights possible once fixed effects are partialled out \citep{DeChaisemartinDHaultfoeuille2023}.

A robust alternative is to discretise exposure into bins that define mutually exclusive exposure states and then apply the group--time logic to each bin relative to a baseline, with explicit convex aggregation across cohorts and horizons. When continuous exposure is retained, orthogonal-score methods of Section 7 can be applied to estimate average causal responses as linear functionals, with the Riesz representer corresponding to the appropriate exposure-tilting functional \citep{ChernozhukovNeweySingh2022}. In both cases, the diagnostics generalise by computing the projection-induced weights for exposure indicators or basis expansions of $A_{it}$, and then measuring negative-weight mass and cross-horizon mass as in Section 5, now interpreted as properties of the exposure design rather than of binary adoption.

These extensions show that the central distinction is not between ``TWFE'' and ``alternatives'' as software choices, but between estimands that are explicit convex functionals of well-defined causal objects and estimands that are implicit, design-driven mixtures whose interpretation fails under heterogeneity, overlap of treatments, switching, or intensity variation \citep{DeChaisemartinDHaultfoeuille2023}.

\clearpage
\appendix

\section{Appendix: Proof sketches and additional results}

\subsection{Proof sketch for Proposition \ref{prop:contamination}}\label{subsec:proofsketch-contamination}

Write \eqref{eq:twfe-es} in stacked form. Let $D\in\R^{nT\times (|\mathcal{K}|-1)}$ collect the event-time indicators $\{D_k\}_{k\in\mathcal{K}\setminus\{k_0\}}$, let $X$ collect the unit and time fixed effects (and any additional controls), and let $M_X\equiv I-X(X'X)^{-1}X'$ be the residual-maker. Define the residualised event-time design $Z\equiv M_XD$ and write $Z=[Z_k\ \ Z_{-k}]$ where $Z_k$ is the column associated with event time $k$ and $Z_{-k}$ contains the remaining event-time columns. By Frisch--Waugh--Lovell,
\begin{equation}\label{eq:fwl-k}
\hat\beta_k
=
\big(Z_k'M_{Z_{-k}}Z_k\big)^{-1}Z_k'M_{Z_{-k}}Y,
\qquad
M_{Z_{-k}}\equiv I-Z_{-k}(Z_{-k}'Z_{-k})^{-1}Z_{-k}',
\end{equation}
so $\hat\beta_k$ is the slope coefficient from regressing $Y$ on the residualised regressor $Z_k$ after partialling out the remaining residualised event-time indicators. Equivalently, there exists a coefficient vector $a_k$ such that
\begin{equation}\label{eq:zk-orth}
\tilde Z_k \equiv M_{Z_{-k}}Z_k = Z_k - Z_{-k}a_k,
\qquad\text{and}\qquad
\hat\beta_k = \frac{\tilde Z_k'Y}{\tilde Z_k'Z_k}.
\end{equation}

Under absorbing adoption and no anticipation, decompose the observed outcome as
\begin{equation}\label{eq:Y-decomp}
Y_{it}
=
Y_{it}(\infty)
+
\sum_{g=1}^T\Ind\{G_i=g\}\sum_{k'\ge 0}\tau_g(k')\,\Ind\{t-g=k'\}
+
\varepsilon_{it},
\end{equation}
with $\E[\varepsilon_{it}\mid X_{it},G_i]=0$. Stacking \eqref{eq:Y-decomp} yields
\[
Y = Y(\infty)+\sum_{k'} D_{k'}\tau(k')+\varepsilon,
\]
where the sum ranges over all (relevant) event times and $\tau(k')$ is the vector collecting the cohort-specific effects at horizon $k'$. Substituting into \eqref{eq:zk-orth} and taking probability limits gives
\begin{equation}\label{eq:plim-mix}
\plim\,\hat\beta_k
=
\sum_{k'} \underbrace{\frac{\tilde Z_k' D_{k'}}{\tilde Z_k' D_k}}_{\textstyle \sum_g w_{g,k'}(k)}\,\tau(k')
\quad
\text{(grouped by horizons)},
\end{equation}
and regrouping the numerator by cohort and horizon yields the representation \eqref{eq:weights}. Hence cross-horizon contamination occurs whenever there exists $k'\neq k$ such that
\begin{equation}\label{eq:nonorth}
\tilde Z_k' D_{k'} \neq 0.
\end{equation}
Condition \eqref{eq:nonorth} is generic under staggered adoption because the event-time indicators are functions of the common adoption partition and are therefore not mutually orthogonal after residualisation by unit and time fixed effects. In particular, $M_X$ removes cohort means and time means, producing residualised indicators $Z_k$ that take both positive and negative values across cohort--time cells; the subsequent projection $M_{Z_{-k}}$ preserves sign changes in $\tilde Z_k$ unless the design is degenerate (for example, a single cohort or a single event-time regressor). Therefore $\tilde Z_k$ typically correlates with indicators for other horizons, implying non-zero weights on $k'\neq k$ in \eqref{eq:weights}.

Negative weights arise because the weights in \eqref{eq:weights} are ratios of inner products involving $\tilde Z_k$, and $\tilde Z_k$ changes sign across cohort--time cells. Specifically, for any cohort--horizon cell $(g,k')$ the implied weight contribution is proportional to the average of $\tilde Z_k$ over that cell. If $\tilde Z_k$ assigns negative mass to some treated cells (or to some not-yet-treated cells that enter the projection through $M_X$ and $M_{Z_{-k}}$), then the corresponding $w_{g,k'}(k)$ is negative. This sign reversal is entirely geometric: it is induced by the fixed-effect residualisation and the linear dependence structure among event-time indicators, and it does not require a failure of Assumption \ref{ass:pt}. Heterogeneity in $\tau_g(k)$ then ensures that these non-convex weights translate into distorted probability limits and potentially spurious pre-trends, as formalised in the TWFE heterogeneity analyses of \citet{SunAbraham2021} and surveyed and generalised in \citet{DeChaisemartinDHaultfoeuille2023}.

\subsection{Implementation notes}\label{subsec:implementation}

\paragraph{Design diagnostics as functions of $(G_i,T,\mathcal{K})$.}
Fix an event window $\mathcal{K}\subset\mathbb{Z}$ and baseline period $k_0\in\mathcal{K}$. Construct the stacked event-time design matrix $D\in\R^{nT\times (|\mathcal{K}|-1)}$ with columns
\[
D_k \equiv \big(\Ind\{t-G_i=k\}\Ind\{G_i<\infty\}\big)_{(i,t)\in[n]\times[T]},
\qquad k\in\mathcal{K}\setminus\{k_0\}.
\]
Let $X$ denote the fixed-effect design for unit and time effects (and any additional controls that are to be partialled out). Form the residual-maker $M_X\equiv I-X(X'X)^{-1}X'$ and the residualised event-time design
\[
Z\equiv M_XD.
\]
For each horizon $k$, let $e_k$ be the selector for column $k$ in the chosen ordering, and define the coefficient-weight vector
\begin{equation}\label{eq:pi-impl}
\pi(k)\equiv Z(Z'Z)^{-1}e_k,
\qquad\text{so that}\qquad
\hat\beta_k=\pi(k)'Y.
\end{equation}
The vector $\pi(k)$ is computable without $Y$ once $Z$ is known. To obtain cohort--horizon weights, average $\pi_{it}(k)$ over the cohort--time cells and map cells to event time $k'=t-g$:
\begin{equation}\label{eq:w-impl}
w_{g,k'}(k)
\equiv
\sum_{t:\,t-g=k'} \E\!\left[\pi_{it}(k)\mid G_i=g\right],
\end{equation}
which is implemented by sample analogues replacing the conditional expectation by the cohort mean. The negative-weight risk index and cross-horizon contamination index are then
\[
\mathcal{N}(k)=\sum_{g}\sum_{k'}\abs{w_{g,k'}(k)}\Ind\{w_{g,k'}(k)<0\},
\qquad
\mathcal{C}(k)=\sum_{g}\sum_{k'}\abs{w_{g,k'}(k)}\Ind\{k'\neq k\}.
\]
Because $D$ is a deterministic function of $(G_i,T,\mathcal{K})$ and $M_X$ depends only on the fixed-effect structure, these diagnostics are design properties and can be reported prior to estimating treatment effects.

\paragraph{Sensitivity regions as finite-dimensional optimisation.}
Let $\theta=\tau(k)$ be a target event-time estimand. Under a restriction class $\delta\in\Delta(\mathcal{R})$ (Section \ref{sec:sensitivity}), the identified set is $\Theta(\mathcal{R})=[\underline{\theta}(\mathcal{R}),\overline{\theta}(\mathcal{R})]$ with endpoints defined by \eqref{eq:bounds}. In implementation, $\delta$ is indexed on a finite set of cohort--time cells, so the optimisation is finite-dimensional. Under the curvature-bounded class \eqref{eq:RRclass}, the constraints are linear inequalities in $\delta$ when expressed in differences (bounds on $\delta$ and on $\Delta^2\delta$), so the computation reduces to a linear programme once the mapping $\delta\mapsto\theta(\delta)$ is written in linear form for the chosen estimator. More general shape restrictions yield convex programmes of comparable dimensionality. Sample analogues of the bound computations substitute estimated reduced-form quantities and incorporate sampling uncertainty through the critical-value construction used for uniformly valid confidence regions in Theorem \ref{thm:uniform} \citep{RambachanRoth2023}.

\subsection*{Appendix A.3.\ \ Geometry of Restricted Deviations and the Identified Set}\label{app:A3}

\begin{figure}[htbp]
\centering
\begin{tikzpicture}[scale=1.1, >=Stealth]

    \draw[->, thick] (-4.5,0) -- (5.5,0) node[above right] {Event Time $k$};
    \draw[->, thick] (0,-3.5) -- (0,3.5) node[above] {Deviation $\delta_{g,t}$};
    
    \foreach \x in {-4,-3,-2,-1,1,2,3,4}
        \draw (\x,0.1) -- (\x,-0.1);
    \node[below] at (-1, -0.1) {$-1$};
    \node[below] at (0, -0.1) {$0$};
    \node[below] at (1, -0.1) {$1$};

    \draw[dashed, thick, black!60] (-4.2,0) -- (5.2,0) node[below right, font=\footnotesize] {Parallel Trends ($\delta=0$)};

    \fill[gray!20] (-4.2, -0.4) rectangle (0, 0.4);
    \draw[gray, thick] (-4.2, 0.4) -- (0, 0.4);
    \draw[gray, thick] (-4.2, -0.4) -- (0, -0.4);
    
    \draw[<->, thin] (-2.5, 0) -- (-2.5, 0.4) node[midway, right, font=\scriptsize] {$B$};
    \node[font=\footnotesize, text=gray!80!black] at (-2, 0.8) {Pre-trend constraint};

    
    \draw[thick, blue!70!black, domain=0:4.5, samples=100] plot (\x, {0.4 + 0.12*\x*\x});
    \draw[thick, blue!70!black, domain=0:4.5, samples=100] plot (\x, {-0.4 - 0.12*\x*\x});
    
    \fill[blue!10, opacity=0.7, domain=0:4.5, variable=\x] 
        (0, 0.4) -- plot ({\x}, {0.4 + 0.12*\x*\x}) -- (4.5, -2.83) 
        -- plot[domain=4.5:0] ({\x}, {-0.4 - 0.12*\x*\x}) -- cycle;

    
    \draw[<->, blue!80!black] (3, 1.48) -- (3, 2.4) node[midway, right, font=\footnotesize] {Widening due to $\Gamma$};
    
    \node[blue!80!black, font=\bfseries] at (2.5, 0.8) {Identified Set $\Theta(\mathcal{R})$};
    
    \draw[red, thick, smooth] plot coordinates {(-3.5, 0.1) (-2, -0.2) (-1, 0.1) (0, 0.2) (1, 0.3) (2, 0.8) (3, 0.6) (4, 1.5)};
    \node[red, right, font=\footnotesize] at (4, 1.5) {Permissible Violation};

    \node[anchor=north] at (-2, -3) {\textsc{Pre-Treatment}};
    \node[anchor=north] at (2, -3) {\textsc{Post-Treatment}};
    \draw[dotted] (0, -3) -- (0, 0);

\end{tikzpicture}
\caption{\textbf{The Geometry of Restricted Violations.} The identified set for the bias $\delta_{g,t}$ (shaded blue) expands over event time. The width at $k<0$ is governed by the pre-trend fit tolerance $B$. The expansion rate for $k>0$ is governed by the smoothness parameter $\Gamma$, which limits how explicitly the post-treatment trend can deviate (curvature) relative to the pre-trend.}
\label{fig:sensitivity-cone}
\end{figure}
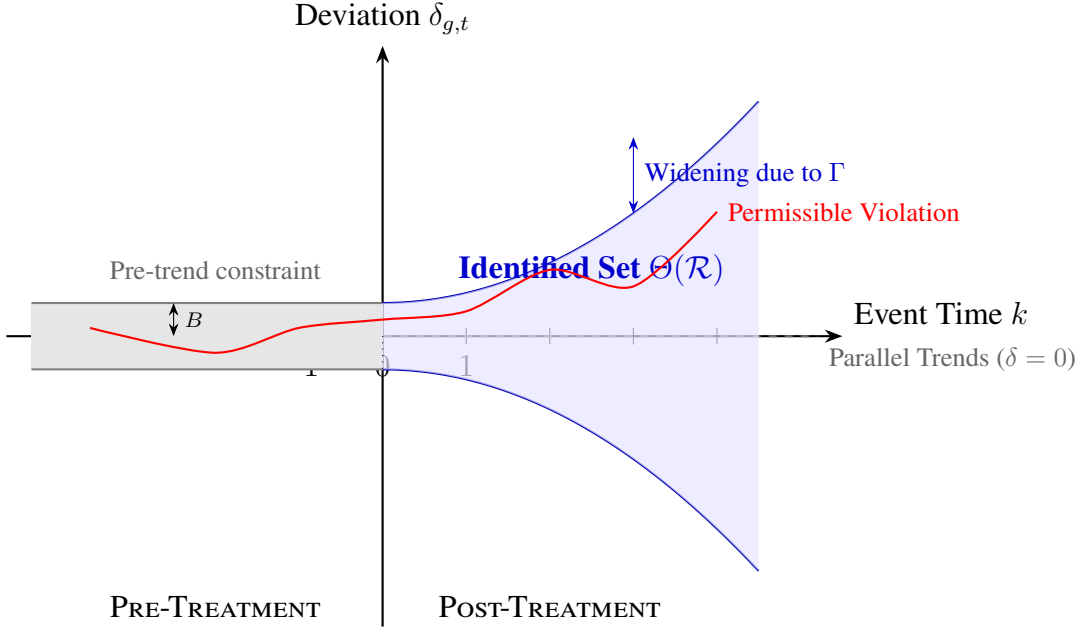

Section \ref{sec:sensitivity} defines the deviation process $\delta_{g,t}$ in \eqref{eq:delta} and the restriction class $\Delta(\mathcal{R})$ that replaces the sharp condition $\delta_{g,t}\equiv 0$ with calibrated, economically interpretable bounds. Figure \ref{fig:sensitivity-cone} provides a geometric interpretation of the restriction class in the event-time coordinate $k=t-g$ for a fixed cohort $g$.

The horizontal axis is event time $k$, with $k<0$ denoting pre-treatment periods and $k\ge 0$ denoting post-treatment periods. The vertical axis plots the deviation $\delta_{g,t}$, interpreted as the cohort-specific wedge between the untreated increment for cohort $g$ and the untreated increment for the control group (never-treated or not-yet-treated, as defined in Section \ref{sec:sensitivity}). The dashed line at zero corresponds to the baseline parallel-trends restriction. In pre-treatment periods, the shaded grey band represents a tolerance region $\abs{\delta_{g,t}}\le B$ for $k<0$, which formalises the idea that pre-trend fit is not judged by informal visual inspection but by an explicit bound that can be calibrated from sampling uncertainty in estimated pre-period coefficients.

For post-treatment periods, the admissible set expands as event time increases. The blue region depicts a smoothness-restricted envelope indexed by $\Gamma$, which limits how rapidly the deviation process can bend away from its pre-treatment behaviour. In discrete time, this is naturally formulated as a bound on the second difference $\abs{\Delta^2 \delta_{g,t}}\le \Gamma$, which constrains curvature in the deviation path and therefore ties permissible post-period drift to a disciplined continuation of pre-period drift rather than permitting arbitrary breaks at $k=0$. The figure therefore represents a nested family of models: as $(B,\Gamma)$ increase, the admissible deviation set grows and the identified set $\Theta(\mathcal{R})$ in \eqref{eq:identified-set} widens accordingly.

The red trajectory illustrates a single admissible deviation path within the restriction class. Each admissible $\delta$ induces a corresponding value of the target estimand $\theta=\tau(k)$ through the mapping from deviations to counterfactual trends described in Section \ref{sec:sensitivity}. The identified set $\Theta(\mathcal{R})$ is the image of $\Delta(\mathcal{R})$ under that mapping, and the endpoints of $\Theta(\mathcal{R})$ are obtained by solving the lower- and upper-bound optimisation problems in \eqref{eq:bounds}. Figure \ref{fig:sensitivity-cone} is therefore not decorative: it is the diagrammatic analogue of the optimisation defining $\Theta(\mathcal{R})$, and it clarifies why credible inference targets coverage of the entire set rather than a single point estimate when parallel trends is relaxed.

\subsection{Derivations for calibration mappings}\label{app:calibration_derivations}

\paragraph{Setup.}
Let $\theta$ denote the target scalar estimand (for example an average post-period effect), and let $\widehat{\theta}$ be the baseline estimator. Let $\mathcal{V}(B,\Gamma,\Delta)$ denote the admissible set of violations under restriction class $\mathcal{R}$, where $\Delta$ abbreviates $\Delta(\mathcal{R})$. Define the identified set for $\theta$ as
\begin{equation}
\mathcal{I}(B,\Gamma,\Delta)
\;:=\;
\Bigl\{
\theta(v): v\in \mathcal{V}(B,\Gamma,\Delta)
\Bigr\},
\label{eq:identified_set_def}
\end{equation}
where $\theta(v)$ denotes the estimand implied by violation path $v$.

\paragraph{Generic monotonicity requirement.}
A calibration mapping $\widehat{\Delta}(\mathcal{R})=\phi_{\mathcal{R}}(\cdot)$ is monotone in diagnostics if, for two datasets (or diagnostic realisations) $d_1,d_2$,
\begin{equation}
d_1 \preceq d_2
\quad\Longrightarrow\quad
\phi_{\mathcal{R}}(d_1)\le \phi_{\mathcal{R}}(d_2),
\label{eq:mono_requirement}
\end{equation}
where $\preceq$ denotes coordinatewise ordering in the diagnostics (larger pre-trend magnitudes, larger drift, more placebo rejections).

\subsubsection{Monotonicity of the canonical \texorpdfstring{$\Delta(\mathcal{R})$}{Delta(R)} rule}
Recall the canonical calibration rule
\begin{equation}
\widehat{\Delta}(\mathcal{R})
\;=\;
c_{\mathcal{R}}\cdot A_{\mathrm{pre}},
\qquad
A_{\mathrm{pre}}=\max_{\ell\in\mathcal{L}_{\mathrm{pre}}}|\widehat{\beta}_{\ell}|.
\label{eq:delta_rule_app}
\end{equation}
For any fixed $c_{\mathcal{R}}\ge 0$, monotonicity in $\{\widehat{\beta}_{\ell}\}$ follows because $A_{\mathrm{pre}}$ is monotone in absolute magnitudes:
if $|\widehat{\beta}_{\ell}^{(1)}|\le |\widehat{\beta}_{\ell}^{(2)}|$ for all $\ell\in\mathcal{L}_{\mathrm{pre}}$, then
\begin{equation}
A_{\mathrm{pre}}^{(1)}
=
\max_{\ell}|\widehat{\beta}_{\ell}^{(1)}|
\le
\max_{\ell}|\widehat{\beta}_{\ell}^{(2)}|
=
A_{\mathrm{pre}}^{(2)}
\quad\Longrightarrow\quad
\widehat{\Delta}^{(1)}(\mathcal{R})\le \widehat{\Delta}^{(2)}(\mathcal{R}).
\label{eq:delta_mono}
\end{equation}

\subsubsection{Monotonicity of the data-driven \texorpdfstring{$B$}{B} rule}
For $\widehat{B}=\kappa_{B}A_{\mathrm{pre}}$ with $\kappa_{B}\ge 0$, the same argument yields
\begin{equation}
A_{\mathrm{pre}}^{(1)}\le A_{\mathrm{pre}}^{(2)}
\quad\Longrightarrow\quad
\widehat{B}^{(1)}\le \widehat{B}^{(2)}.
\label{eq:B_mono}
\end{equation}
For the holdout definition,
\begin{equation}
\widehat{B}_{\mathrm{holdout}}
=
\min\Bigl\{b\ge 0:\;
g(b)\le \bar{g}
\Bigr\},
\qquad
g(b):=\max_{\ell\in\mathcal{L}_{2}}|\widehat{\beta}_{\ell}(b)|,
\qquad
\bar{g}:=q_{0.95}\bigl(\{|\widehat{\beta}_{\ell}|:\ell\in\mathcal{L}_{1}\}\bigr),
\label{eq:B_holdout_app}
\end{equation}
monotonicity follows under the sufficient condition that $g(b)$ is nonincreasing in $b$ (the admissible perturbation enlarges with $b$ and can only reduce the residual pre-trend discrepancy on holdout):
\begin{equation}
b_1\le b_2
\quad\Longrightarrow\quad
g(b_2)\le g(b_1).
\label{eq:g_nonincreasing}
\end{equation}
Under \eqref{eq:g_nonincreasing}, the feasible set $\{b:g(b)\le \bar{g}\}$ is an upper set, hence the minimum exists (possibly $0$) and is unique.

\subsubsection{Monotonicity and interpretation of the \texorpdfstring{$\Gamma$}{Gamma} mapping}
Recall
\begin{equation}
\widehat{\mathcal{D}}
=
\max_{\ell\in\mathcal{L}_{\mathrm{pre}}\setminus\{\min\mathcal{L}_{\mathrm{pre}}\}}
\left|\widehat{\beta}_{\ell}-\widehat{\beta}_{\ell-1}\right|,
\qquad
\widehat{\Gamma}
=
100\cdot \frac{\widehat{\mathcal{D}}}{|\widehat{\tau}|+\varepsilon_{\tau}}.
\label{eq:Gamma_map_app}
\end{equation}
For fixed $\widehat{\tau}$ and $\varepsilon_{\tau}>0$, $\widehat{\Gamma}$ is monotone in $\widehat{\mathcal{D}}$:
\begin{equation}
\widehat{\mathcal{D}}^{(1)}\le \widehat{\mathcal{D}}^{(2)}
\quad\Longrightarrow\quad
\widehat{\Gamma}^{(1)}\le \widehat{\Gamma}^{(2)}.
\label{eq:Gamma_mono}
\end{equation}
The units interpretation follows directly: $\widehat{\Gamma}$ equals the maximal per-period drift in pre-trends as a percentage of the baseline effect scale $|\widehat{\tau}|+\varepsilon_{\tau}$.

\subsubsection{Sufficient conditions for identified set expansion}
The calibration inputs are useful only if enlarging $(B,\Gamma,\Delta)$ weakly enlarges the violation set and hence the identified set.
Assume the following nesting conditions for admissible violations:
\begin{equation}
B_1\le B_2,\;\Gamma_1\le \Gamma_2,\;\Delta_1\le \Delta_2
\quad\Longrightarrow\quad
\mathcal{V}(B_1,\Gamma_1,\Delta_1)\subseteq \mathcal{V}(B_2,\Gamma_2,\Delta_2).
\label{eq:V_nesting}
\end{equation}
A sufficient set of primitive conditions for \eqref{eq:V_nesting} is that $\mathcal{V}(B,\Gamma,\Delta)$ is defined by constraints of the form
\begin{equation}
\|v\|_{\star}\le B,
\qquad
\|Dv\|_{\diamond}\le \Gamma\cdot s_{\Gamma},
\qquad
\|P_{\mathcal{R}}v\|_{\circ}\le \Delta,
\label{eq:constraint_template}
\end{equation}
for seminorms $\|\cdot\|_{\star},\|\cdot\|_{\diamond},\|\cdot\|_{\circ}$, a linear difference operator $D$, a restriction-class projection (or linear map) $P_{\mathcal{R}}$, and a scale factor $s_{\Gamma}>0$.
Under \eqref{eq:constraint_template}, each feasible set is an intersection of sublevel sets, each of which is nested in its bound, implying \eqref{eq:V_nesting}.

Given \eqref{eq:V_nesting}, identified set expansion holds:
\begin{equation}
\mathcal{V}(B_1,\Gamma_1,\Delta_1)\subseteq \mathcal{V}(B_2,\Gamma_2,\Delta_2)
\quad\Longrightarrow\quad
\mathcal{I}(B_1,\Gamma_1,\Delta_1)\subseteq \mathcal{I}(B_2,\Gamma_2,\Delta_2),
\label{eq:I_expansion}
\end{equation}
because $\mathcal{I}(B,\Gamma,\Delta)$ is the image of $\mathcal{V}(B,\Gamma,\Delta)$ under the map $v\mapsto \theta(v)$.

\subsubsection{Monotonicity of the sensitivity region}
Let $\mathsf{C}(B,\Gamma,\Delta)\in\{0,1\}$ indicate whether the target conclusion is unchanged under $(B,\Gamma,\Delta)$ (for example, a sign conclusion or rejection decision).
Assume $\mathsf{C}$ is monotone decreasing in the size of the identified set, meaning
\begin{equation}
\mathcal{I}(B_1,\Gamma_1,\Delta_1)\subseteq \mathcal{I}(B_2,\Gamma_2,\Delta_2)
\quad\Longrightarrow\quad
\mathsf{C}(B_2,\Gamma_2,\Delta_2)\le \mathsf{C}(B_1,\Gamma_1,\Delta_1).
\label{eq:C_mono}
\end{equation}
Then the robustness region
\begin{equation}
\mathcal{S}
:=
\{(B,\Gamma,\Delta):\mathsf{C}(B,\Gamma,\Delta)=1\}
\label{eq:S_region}
\end{equation}
is a down-set in $(B,\Gamma,\Delta)$ under coordinatewise order: if $(B,\Gamma,\Delta)\in\mathcal{S}$, then any $(B',\Gamma',\Delta')$ with $0\le B'\le B$, $0\le \Gamma'\le \Gamma$, $0\le \Delta'\le \Delta$ also lies in $\mathcal{S}$.

\subsubsection{Placebo and holdout selection as minimality problems}
Let $\pi(B,\Gamma,\Delta)$ denote a placebo rejection rate and $h(B,\Gamma,\Delta)$ a holdout discrepancy.
Selection in Box~6.1 corresponds to the constrained minimality problem
\begin{equation}
(\widehat{B},\widehat{\Gamma},\widehat{\Delta})
\in
\arg\min_{(B,\Gamma,\Delta)\in\mathcal{G}}
\;\;
w_{B}B+w_{\Gamma}\Gamma+w_{\Delta}\Delta
\quad\text{s.t.}\quad
\pi(B,\Gamma,\Delta)\le \alpha,
\;\;
h(B,\Gamma,\Delta)\le \bar{h},
\label{eq:min_problem}
\end{equation}
over a finite grid $\mathcal{G}$, with weights $w_{B},w_{\Gamma},w_{\Delta}>0$ and threshold $\bar{h}$ defined from the pre-period reference split.
Under grid finiteness, existence is immediate; under monotonicity of $\pi$ and $h$ in $(B,\Gamma,\Delta)$, the feasible set is an upper set and the minimiser is attained on its lower boundary.

\clearpage
\bibliography{references}

\begin{thebibliography}{23}
\newcommand{\enquote}[1]{``#1''}
\expandafter\ifx\csname natexlab\endcsname\relax\def\natexlab#1{#1}\fi

\bibitem[\protect\citeauthoryear{Abadie, Angrist, and Frandsen}{Abadie et~al.}{2025}]{AbadieAngristFrandsen2025}
\textsc{Abadie, A., J.~Angrist, and B.~Frandsen} (2025): \enquote{Harvesting Differences-in-Differences and Event-Study Evidence,} Working Paper 34550, National Bureau of Economic Research.

\bibitem[\protect\citeauthoryear{Arkhangelsky, Athey, Hirshberg, Imbens, and Wager}{Arkhangelsky et~al.}{2021}]{ArkhangelskyAtheyHirshbergImbensWager2021}
\textsc{Arkhangelsky, D., S.~Athey, D.~A. Hirshberg, G.~W. Imbens, and S.~Wager} (2021): \enquote{Synthetic Difference-in-Differences,} \emph{American Economic Review}.

\bibitem[\protect\citeauthoryear{Athey and Imbens}{Athey and Imbens}{2022}]{AtheyImbens2022}
\textsc{Athey, S. and G.~W. Imbens} (2022): \enquote{Design-based Analysis in Difference-in-Differences Settings with Staggered Adoption,} \emph{Journal of Econometrics}.

\bibitem[\protect\citeauthoryear{Baker, Larcker, and Wang}{Baker et~al.}{2022}]{BakerLarckerWang2022}
\textsc{Baker, A.~C., D.~F. Larcker, and C.~C.~Y. Wang} (2022): \enquote{How Much Should We Trust Staggered Difference-in-Differences Estimates?} \emph{Journal of Financial Economics}, 144, 370--395.

\bibitem[\protect\citeauthoryear{Callaway and Sant'Anna}{Callaway and Sant'Anna}{2021}]{CallawaySantAnna2021}
\textsc{Callaway, B. and P.~H.~C. Sant'Anna} (2021): \enquote{Difference-in-Differences with Multiple Time Periods,} \emph{Journal of Econometrics}, 225, 200--230.

\bibitem[\protect\citeauthoryear{Chernozhukov, Newey, and Singh}{Chernozhukov et~al.}{2022}]{ChernozhukovNeweySingh2022}
\textsc{Chernozhukov, V., W.~K. Newey, and R.~Singh} (2022): \enquote{Automatic Debiased Machine Learning of Causal and Structural Effects,} \emph{Econometrica}, 90, 967--1027.

\bibitem[\protect\citeauthoryear{de~Chaisemartin and D'Haultf{\oe}uille}{de~Chaisemartin and D'Haultf{\oe}uille}{2023}]{DeChaisemartinDHaultfoeuille2023}
\textsc{de~Chaisemartin, C. and X.~D'Haultf{\oe}uille} (2023): \enquote{Two-way Fixed Effects and Differences-in-Differences with Heterogeneous Treatment Effects: A Survey,} \emph{The Econometrics Journal}, 26, C1--C30.

\bibitem[\protect\citeauthoryear{Gardner}{Gardner}{2021}]{Gardner2021}
\textsc{Gardner, J.} (2021): \enquote{Two-stage difference-in-differences,} Tech. rep., Working Paper.

\bibitem[\protect\citeauthoryear{Goodman-Bacon}{Goodman-Bacon}{2021}]{GoodmanBacon2021}
\textsc{Goodman-Bacon, A.} (2021): \enquote{Difference-in-differences with variation in treatment timing,} \emph{Journal of Econometrics}, 225, 254--277.

\bibitem[\protect\citeauthoryear{Lee}{Lee}{2025}]{Lee2025}
\textsc{Lee, W.} (2025): \enquote{A Simple Approach to Staggered Difference-in-Differences with Spillovers,} Tech. rep., Working Paper.

\bibitem[\protect\citeauthoryear{Manski}{Manski}{2003}]{Manski2003}
\textsc{Manski, C.~F.} (2003): \emph{Partial Identification of Probability Distributions}, Springer.

\bibitem[\protect\citeauthoryear{Masten and Poirier}{Masten and Poirier}{2021}]{MastenPoirier2021}
\textsc{Masten, M.~A. and A.~Poirier} (2021): \enquote{Salvaging Falsified Instrumental Variable Models,} \emph{Econometrica}, 89, 1449--1469.

\bibitem[\protect\citeauthoryear{Newey}{Newey}{1994}]{Newey1994}
\textsc{Newey, W.~K.} (1994): \enquote{The Asymptotic Variance of Semiparametric Estimators,} \emph{Econometrica}, 62, 1349--1382.

\bibitem[\protect\citeauthoryear{Rambachan and Roth}{Rambachan and Roth}{2023}]{RambachanRoth2023}
\textsc{Rambachan, A. and J.~Roth} (2023): \enquote{A More Credible Approach to Parallel Trends,} \emph{The Review of Economic Studies}, 90, 2555--2591.

\bibitem[\protect\citeauthoryear{Roth}{Roth}{2024}]{Roth2024}
\textsc{Roth, J.} (2024): \enquote{Interpreting Event-Studies from Recent Difference-in-Differences Methods,} .

\bibitem[\protect\citeauthoryear{Roth, Koles{\'a}r, and Montiel~Olea}{Roth et~al.}{2025}]{RothKolesarMontielOlea2025}
\textsc{Roth, J., M.~Koles{\'a}r, and J.~L. Montiel~Olea} (2025): \enquote{Evaluating Counterfactual Policies Using Instruments,} Tech. rep., Working Paper.

\bibitem[\protect\citeauthoryear{Roth and Sant'Anna}{Roth and Sant'Anna}{2023}]{RothSantAnna2023}
\textsc{Roth, J. and P.~H.~C. Sant'Anna} (2023): \enquote{When is Parallel Trends Sensitive to Functional Form?} \emph{Econometric Theory}.

\bibitem[\protect\citeauthoryear{Roth, Sant{'}Anna, Bilinski, and Poe}{Roth et~al.}{2023}]{RothSantAnnaBilinskiPoe2023}
\textsc{Roth, J., P.~H.~C. Sant{'}Anna, A.~Bilinski, and J.~Poe} (2023): \enquote{Whats Trending in Difference-in-Differences? A Synthesis of the Recent Econometrics Literature,} \emph{Journal of Econometrics}, conditionally accepted; available as arXiv:2201.01194.

\bibitem[\protect\citeauthoryear{Sant'Anna and Zhao}{Sant'Anna and Zhao}{2025}]{SantAnnaZhao2025}
\textsc{Sant'Anna, P. H.~C. and J.~Zhao} (2025): \enquote{Efficient Difference-in-Differences and Event Study Estimators,} Tech. rep., Cowles Foundation Discussion Paper / arXiv.

\bibitem[\protect\citeauthoryear{Sun and Abraham}{Sun and Abraham}{2021}]{SunAbraham2021}
\textsc{Sun, L. and S.~Abraham} (2021): \enquote{Estimating Dynamic Treatment Effects in Event Studies with Heterogeneous Treatment Effects,} \emph{Journal of Econometrics}, 225, 175--199.

\bibitem[\protect\citeauthoryear{Wing, Yozwiak, Hollingsworth, Freedman, and Simon}{Wing et~al.}{2024}]{WingEtAl2024}
\textsc{Wing, C., M.~Yozwiak, A.~Hollingsworth, S.~Freedman, and K.~Simon} (2024): \enquote{Designing Difference-in-Difference Studies with Staggered Treatment Adoption: Key Concepts and Practical Guidelines,} \emph{Annual Review of Public Health}, 45, 485--505.

\bibitem[\protect\citeauthoryear{Wooldridge}{Wooldridge}{2021}]{Wooldridge2021}
\textsc{Wooldridge, J.~M.} (2021): \enquote{Two-Way Mundlak Regression for Panel Data Analysis,} \emph{Journal of Econometrics}, 225, 534--548.

\bibitem[\protect\citeauthoryear{Zhu}{Zhu}{2024}]{Tominaga2024}
\textsc{Zhu, S.} (2024): \enquote{Recent Developments in Event Study Estimation,} Director: Valeria Gargiulo.

\end{thebibliography}
\end{document}